\documentclass[10pt]{article}
\ifx\macrosloaded\relax\endinput\else\let\macrosloaded\relax\fi
\newtheorem{theorem}{Theorem}[section]

\newtheorem{Prop}{Proposition}[section]

\newtheorem{Def}{Definition}[section]

\catcode`\@=11
\newcommand{\eqinsec}{\relax\@addtoreset{equation}{section}}
\renewcommand{\theequation}{\ifx\showlabels\iftrue\the\id\else\thesection.\arabic{equation}\fi}
\catcode`\@=13
\newcounter{supeq}
\newenvironment{subeq}
\stepcounter{equation}
\def\theequation{\ifx\showlabels\iftrue\the\id\else\thesection.\arabic{equation}\fi}
\newtoks\id
\newcommand{\eqlabel}[1]{\label{#1}\global\id={(#1)}} 

\newcommand{\tr}{\mbox{tr}}
\newcommand{\be}{\begin{equation}}
\newcommand{\eeq}{\end{equation}}
\newcommand{\bea}{\begin{eqnarray}}
\newcommand{\eea}{\end{eqnarray}}
\newcommand{\beaa}{\begin{eqnarray*}}
\newcommand{\eeaa}{\end{eqnarray*}}
\newcommand{\bseq}{\begin{subeq}}
\newcommand{\eseq}{\end{subeq}}
\newcommand{\ba}{\begin{array}}
\newcommand{\ea}{\end{array}}
\newcommand{\eql}{\eqlabel}

\def \rectangle#1#2{\hbox{\vrule\vbox to #2
{\hrule\hbox to #1{\hfil}\vfil\hrule}\vrule}}
\newcommand{\edd}{\end{document}}

\renewcommand{\c}{\cdot}
\newcommand{\NI}{\noindent}

\newcommand{\Lb}{\underline{L}}

\newcommand{\Si}{\Sigma}
\newcommand{\ga}{\gamma}
\newcommand{\Ga}{\Gamma}

\newcommand{\GGa}{{\bf \Gamma}}

\newcommand{\rr}{\mbox{${\bf R}$}}

\newcommand{\ggg}{\mbox{${\bf g}$}}

\newcommand{\dd}{\mbox{${\bf D}$}}

\newcommand{\omm}{\mbox{${\bf\om}$}}
\newcommand{\oomm}{\mbox{${\bf\oom}$}}

\newcommand{\lap}{\mbox{$\bigtriangleup$}}
\newcommand{\lapp}{\mbox{$\bigtriangleup  \mkern-13mu / \,$}}
\newcommand{\nab}{\mbox{$\nabla$}}
\newcommand{\nabb}{\mbox{$\nabla \mkern-13mu /$\,}}

\newcommand{\partialb}{\mbox{$\partial \mkern-10mu /$\,}}

\newcommand{\ddb}{\mbox{$\dd \mkern-13mu /$\,}}

\newcommand{\pr}{\partial}
\newcommand{\hot}{\widehat{\otimes}}

\newtheorem{Le}{Lemma}[section]

\newcommand{\Lie}{\mbox{$\cal L$}}

\newcommand{\nn}{\nonumber}

\newcommand{\chib}{\underline{\chi}}

\newcommand{\de}{\delta}
\newcommand{\De}{\Delta}
\newcommand{\ep}{\epsilon}
\newcommand{\xib}{\underline{\xi}}
\newcommand{\chih}{\hat{\chi}}

\newcommand{\chibh}{\underline{\hat{\chi}}}

\newcommand{\und}[1]{\underline{#1}}

\newcommand{\Nb}{\und{N}}

\newcommand{\Cb}{\und{C}}

\newcommand{\n}{n}

\newcommand{\ub}{{\underline{u}}}
\renewcommand{\c}{\cdot}

\newcommand{\M}{{\cal M}}

\renewcommand{\aa}{\underline{\alpha}}
\newcommand{\bb}{\underline{\beta}}
\renewcommand{\a}{\alpha}
\renewcommand{\b}{\beta}
\newcommand{\dual}{\mbox{}^{\star}\!}

\newcommand{\si}{\sigma}

\newcommand{\ro}{\rho}

\newcommand{\ze}{\zeta}
\newcommand{\divv}{\mbox{div}\mkern-19mu /\,\,\,\,}

\newcommand{\curll}{\mbox{curl}\mkern-19mu /\,\,\,\,}

\newcommand{\om}{\omega}
\newcommand{\oom}{\Omega}

\newcommand{\omb}{\underline{\omega}}

\newcommand{\etab}{\underline{\eta}}
\newcommand{\la}{\lambda}

\newcommand{\dddd}{{\bf D} \mkern-13mu /\,}

\def\frac#1#2{{{#1}\over{#2}}}

\newcommand{\ML}{\!\!\!\!\!\!\!\!\!}

\begin{document}
\title{LOCAL AND GLOBAL ANALYTIC SOLUTIONS FOR A CLASS OF CHARACTERISTIC PROBLEMS OF THE EINSTEIN VACUUM EQUATIONS IN THE ``DOUBLE NULL FOLIATION GAUGE"}
\author{Giulio CACIOTTA , Francesco NICOL\`{O}
\footnote{Dipartimento di Matematica, Universit\`{a} degli Studi di Roma
``Tor Vergata", Via della Ricerca Scientifica, 00133-Roma, Italy}
\\Universit\`{a} degli studi di Roma ``Tor Vergata"\date{\today}}\maketitle

\begin {abstract}
\NI {The main goal of this work consists in showing that the analytic solutions for a class of characteristic problems for the Einstein vacuum equations have an existence region much larger than the one provided by the Cauchy-Kowalevski theorem due to the intrinsic hyperbolicity of the Einstein equations. To prove this result we first describe a geometric way of writing the vacuum Einstein equations for the characteristic problems we are considering, in a gauge characterized by the introduction of a double null cone foliation of the spacetime. Then we prove that the existence region for the analytic solutions can be extended to a larger region which depends only on the validity of the apriori estimates for the Weyl equations, associated to the ``Bel-Robinson norms". In particular if the initial data are sufficiently small we show that the analytic solution is global. Before showing how to extend the existence region we describe the same result in the case of the Burger equation, which, even if much simpler, nevertheless requires analogous logical steps required for the general proof. Due to length of this work, in this paper we  mainly concentrate on the definition of the gauge we use and on writing in a ``geometric" way the Einstein equations, then we show how the Cauchy-Kowalevski theorem is adapted to the characteristic problem for the Einstein equations and we describe how the existence region can be extended in the case of the Burger equation. Finally we describe the structure of the extension proof in the case of the Einstein equations. The technical parts of this last result is the content of a second paper.}
\end{abstract}
\newpage
\tableofcontents
\section{Introduction}\label{S.0}
In this paper we prove a result about the existence region of the analytic solution of a class of characteristic problems, namely those whose ``initial data" are given on a null hypersurface consisting of the union of a truncated outgoing null cone and of a truncated incoming cone intersecting the previous one along a   surface diffeomorphic to $S^2$. This class of characteristic problems has been studied by different authors, for instance H.Muller Zum Hagen, \cite{Muller},  H.Muller Zum Hagen and H.J.Seifert, \cite{MullerSeifert}, Christudoulou and Muller Zum Hagen, \cite{Ch-MZ},  Dossa in a series of papers, see \cite{Dossa} and references therein, but, in particular,  we recall the anticipating work by A.Rendall, \cite{rendall:charact}, where a thorough examination has been done to show how to obtain initial data satisfying the costraint equations and the harmonic gauge conditions and, subsequently, a way of obtaining a local existence result is presented. Recently following, but largely improving the A.Rendall result, we suggest to the reader attention the paper by Y. Choquet-Bruhat, P.Y.Chrusciel, J.M. Martin-Garcia, ``The Cauchy problem on a characteristic cone for the Einstein equations in arbitrary dimensions", to appear on Annales Henry Poincar\`e .  In that paper the authors prove a local existence result for the characteristic problem with initial data on a null cone,  using again the harmonic gauge and proving in a very detailed way how the initial data constraints have to be satisfied and how, relying on the Dossa results, the local existence result can be proved. Moreover the nature of the characteristic problem, initial data on the null cone, adds the extra problems of the ``tip of the cone" they solve completely. 

\NI In the present paper our goal is to show that the real analytic solutions of the class of characteristic problems we are considering have, due to the ÒhyperbolicityÓ of the Einstein equations, a larger existence region than the one we can obtain by the application of the Cauchy-Kowalevski theorem. More precisely we prove that the extension of the analyticity existence region depends only on a finite number of derivatives of the initial data, namely on some appropriate Sobolev norms; moreover if we assume the initial data small in these norms we can prove the global existence of the analytic solutions. Some analogous results have been obtained in the past by S.Alinhac, G.Metivier, where they proved the propagation of the analiticity for hyperbolic systems of p.d.e., see [A-M] and references therein; more recent results can be found in S.Spagnolo, ``Propagation of analyticity for a class of nonlinear hyperbolic equations", \cite{Spagn},
and references therein. To prove this result we present a strategy analogous to that used by S.Klainerman and one of the authors (F.N.) in \cite{Kl-Ni:rew}, for the Burger equation in a non characteristic problem. Clearly, as the Einstein equations are much more complicated, the extension to the present case is significantly more difficult. Nevertheless some general aspects are borrowed from that toy model and suggests the nature of the different technical problems we have to deal. 

\NI As it will appear clearly in the rest of this introduction, in the bulk of this paper and of the following one, preliminary to the existence proof of the analytic solutions a detailed examination of the gauge used and how the contraints are satisfied is needed. This part, crucial to the development of the existence proof, is, in our opinion, of interest in itself and new in many aspects. To it a greater part of the present paper is devoted.
Let us  summarize the various steps of our approach. 
\smallskip

{\bf i)} As discussed in detail in Subsection \ref{SS4.1} and stated before the crucial ingredient to extend the analiticity region is the hyperbolicity of the p.d.e. equations we are considering. The region where the analitical solution exists is the region where the apriori  energy estimates can be proved. Therefore the control of energy norms is the key step. To achieve these norms and their apriori estimates is trivial in the Burger case, but much more delicate in the Einstein equations, even more if we look for energy type estimates valid (for small initial data) everywhere. It is, in fact, well known that the standard energy norms associated to the vacuum Einstein equation in the harmonic gauge are very difficult to use to obtain a global solution even in the noncharacteristic problem, see \cite{L-R}, and no results are at our disposal for the characteristic case. It turns out that an efficient strategy to achieve a global existence result is the one introduced first by D. Christoudoulou and S. Klainerman in [Ch-Kl] and subsequently modified by S.Klainerman and F. Nicol\`o, see [Kl-Ni2]. In this approach the ``energy type" norms to bound are those associated to the Bel-Robinson tensor, quadratic in the Riemann tensor. The control of these norms is strictly tied to the control of the connection coefficients\footnote{Sometimes called Ricci coefficients.} of the spacetime and the equations which control these last quantities are the so called ``structure equations", see \cite{Sp:Spivak}, Vol 2,  which have the form of elliptic Hodge systems or of transport equations along the null directions.
\smallskip

{\bf ii)} This strongly suggests the use of a foliation already introduced in \cite{Kl-Ni:book}, Chapter 3, the ``double null cone foliation". The main differences are that first, the structure equations were used there to obtain good estimates for the various $L^p$ integral norms while here to show how the Einstein equations can be expressed as a subset of these structure equations. More precisely, as in the non characteristic case one can foliate the spacetime with a family of spacelike hypersurfaces and write the Einstein equations as a system of first order equations for the (riemannian) metric and the second fundamental form adapted to this foliation, here we assume the spacetime foliated by a family of outgoing  cones and incoming (truncated) cones and write the Einstein equations as a set of first order equations involving the metric adapted to these cones and the connection coefficients, basically, the first derivatives of this metric. The complete detailed description of the procedure to write the Einstein equations in the way we are sketching here is given in subsections \ref{SS2.3} and \ref{SS3.2}.

\NI The second difference is that in \cite{Kl-Ni:book} a local solution was already assumed to exist and proved in the more standard harmonic gauge, here even the local existence is proved in the ``double null cone foliation gauge". Therefore in this approach we will never use any foliation made by spacelike hypersurfaces, the derivatives of the various unknown functions of our equations are always done with respect to the angular variables and to the $\ub$ and $u$ variables, the affine parameters of the null geodesics generating the outgoing or incoming cones.
\smallskip

{\bf iii)}  Exactly as in the non characteristic case the choice of the spacelike hypersurface foliation is basically equivalent to the choice of a gauge, in the present case specifying the ``double null cone foliation" is just the choice of the gauge and the quantities $\oom$ and $X$ which appear in the expression of our metric in the adapted coordinates,
\bea
{\ggg}=-2{{\oom}}^2(dud\ub\!+\!d\ub du)\!+\!\ga_{ab}(X^adu+d\om^a)(X^bdu+d\om^b)\ ,\eql{met0int}
\eea
play the role of the lapse function and the shift vector. Analogously to them they will have to satisfy some differential equations. In subsection \ref{SS2.3} a very extended discussion about this gauge is given.
\smallskip

{\bf iv)} As we are dealing with a characteristic problem it is expected that the initial data cannot be given in a complete free way, but they have to satisfy some constraints. In the case of the Einstein equations the situation is more complicated as even in the non characteristic case the initial data cannot be given in a free way.\footnote{See for instance the detailed discussion in \cite{Friedrich:Rendall}.} Therefore our initial data have to satisfy two different kind of constraints, those due to the nature of the Einstein equations, the analogous of the constraints equations for the second fundamental form $k_{ij}$ and those associated to the gauge choice namely, in the present case, the equations for $\oom$ and $X$. 
In our presentation, more geometric than the one using, for instance, the harmonic gauge, the distinction between these two kinds of constraints is completely clear and it is natural that in these equations no transverse (to the cones) derivatives appear. 

\NI Finally to show that the solutions of our equations are in fact solutions of the Einstein equations we have to prove, exactly as in the non characteristic case,  that the (Einstein) constraint equations once satisfied by the initial data are satisfied everywhere. This is proved in subsection \ref{SS3.2}.
\smallskip

{\bf v)} As we want to prove that our analytic solutions can be extended to the whole spacetime\footnote{A slightly imprecise statement which, nevertheless, should be clear.} we have first  to provide a local analitic solution of the characteristic problem. This is discussed in Section \ref{S3} where we adapt the Cauchy-Kowalevski theorem to the characteristic problem following G.F.D.Duff, \cite{Duff} and H.Friedrich, \cite{friedrich:cauchychar}.
\smallskip

{\bf vi)} In Section \ref{S4} the central part of our program is described and partially proved. In subsection \ref{SS4.1} an analogous result is proved for the Burger equation using, and somewhat extending, a previous result of S.Klainerman and one of the present authors (F.N.), \cite{Kl-Ni:rew}. Although the problem, in that case, is much simpler some of the basic ideas can be borrowed and transported. In the Burger equation case we prove that, due to the hyperbolicity of the equation, some apriori estimates hold for the Sobolev energy norms (with $s=2$) up to a  time $T$, depending only on these norms. Then it is proved that in the region of analiticity provided by the Cauchy-Kowalevski theorem it is possible to show that all the derivatives satisfy some appropriate estimates such that the series describing the analitic solutions have a convergence radius depending on the initial analitic data and on the Sobolev norms associated to the first derivatives, but independent from the point (of the analiticity region) around which we perform the series expansion. This is proved controlling, with a delicate inductive argument, the norms of all order derivatives. Once this result is achieved the analiticity region can be extended and as, again, the series convergence radius depends only on the initial data and on the Sobolev norms (bounded, via the apriori estimates, in terms of the Sobolev initial data norms) it can be proved that the procedure can be repeated in the whole time interval where the apriori estimates hold, obtaining the final result.
\smallskip

\NI In the second part of the section, subsection \ref{SS4.2}, we discuss how this approach can be implemented in the case of the Einstein equations. This requires a lot of technical work to which a subsequent paper is devoted. There we discuss in a very detailed way the problems one encounters and how we have solved them. We recall here some of these problems and give in subsection \ref{SS4.2} an extended discussion of how they are faced. 
\smallskip

\NI 1) The first problem is the position of the initial data, in subsection \ref{SS3.2} it was already discussed the constraints they have to satisfy, here we have to show how they can be given on the whole initial null hypersurface and not only in a portion of it, a generalization of what has been done in \cite{Ca-Ni:char}, where we were interested only on Sobolev initial data.

\NI 2) The local existence result has been proved in Section \ref{S3}, the strategy to extend the analiticity region is to repeat the inductive mechanism which allows to control the norms of all derivatives again in an uniform way. This is the more complicated technical part; to achieve it we have to use the transport  equations for the connection coefficients and the hyperbolicity of the Einstein equations, more precisely the a priori estimates for the integral norms of the Bel-Robinson tensor. The main lemma needed to prove our result, Lemma \ref{L6.1a}, is stated in subsection \ref{SS4.2} while its proof and the subsequent steps to prove our result are written in the subsequent paper.
\smallskip

\NI To summarize this discussion we are convinced that to satisfy our goal of extending as much as possible the analiticity region for the Einstein equations our gauge choice is the most convenient, even more as it seems so naturally intertwined to the characteristic problem. Moreover this formalism is perfectly suited to control the integral norms of the Bel-Robinson tensor and prove that, for small initial data, they can be bounded in the whole spacetime.
\medskip

\NI{\bf Acknowledgments:} {\em The idea of trying to propagate as much as possible the analiticity of the solutions of the Einstein equations was suggested, years ago, to one of the author (F.N.) by S.Klainerman and was implemented together on the ``toy model" of the Burger equation in \cite{Kl-Ni:rew}; moreover both authors are pleased to thank S.Klainerman one for the long and rich collaboration and (G.C.) for the useful period spent in the Math. Department of Princeton University and for the long and profitable conversations he had with him.}
 \section{The characteristic problem for the vacuum Einstein equations, assuming the spacetime foliated by outgoing and incoming null cones}\label{S.1}

In this section and in the following one we present a way of writing the Einstein equations suited to study and solve the class of characteristic problems we are investigating. The basic idea is to assume that the spacetime we are constructing is foliated by a family of outgoing and incoming null cones, a foliation used in \cite{Kl-Ni:book}  which we believe very appropriate to study the characteristic problems for the Einstein equations.

\NI We will see that, with obvious differences, our approach is similar to the one used for the non characteristic problem when the spacetime is foliated by three dimensional hypersurfaces and the evolution part of the Einstein equations,\footnote{Therefore apart from the constraint equations.} are first order equations in terms of the Riemannian metric of the hypersurfaces, $g_{ij}$, and their extrinsic curvature $k_{ij}$.

\NI Therefore the various steps required to accomplish our goal are in order
\smallskip

i) Define the class of characteristic problems we are considering.
\smallskip

ii) Define the gauge we use.
\smallskip

iii) Identify the evolution equations in the coordinates associated to the gauge.
\smallskip

iv) Identify the constraints equations in this formalism.
\smallskip

v) the conservation of the constraints: Once steps i), ii), iii), iv) are clearly done, we  show how the analytic solutions can be obtained in this characteristic case with the appropriate Cauchy-Kowalevski approach, that exactly as in non characteristic case we can define our analytic solution as a solution of a ``reduced problem" and subsequently that, once the constraints are satisfied from the initial data, they are satisfied in all the existence region so that the analytic solution is really a solution of the (vacuum) Einstein equations.

\NI In the following subsections and in Section \ref{S3} we will concentrate on steps i),...v).

\subsection{The class of characteristic problems}
As we said in the introduction we are considering the case of initial data on a null hypersurface consisting of the union of a truncated outgoing null cone $C_0\equiv C(\la_0)$, see later for the ``cone" definition and also \cite{Ca-Ni:char}, and of a truncated incoming cone $\Cb_0\equiv\Cb(\nu_0)$ intersecting the previous one along an $S^2$ surface. Moreover we expect that analogous results can be easily obtained when the initial hypersurface is made by the intersection of two null hyperplanes. More delicate is when looking for the solutions of the Einstein equations inside an outgoing cone where we give the initial data. We believe that this problem, solved for the local existence in the previous work by by Y. Choquet-Bruhat, P.Y.Chrusciel, J.M. Martin-Garcia, quoted in the introduction, can be also faced with our technique, in our double null cone gauge. 

\subsection{The ``double null foliation gauge"}\label{SS2.2}
Let us recall what is the meaning of a ``geometric" gauge choice in the non characteristic problem associated to a spacetime foliation. In that case, see also \cite{Kl-Ni:book}, Lemma 1.3.2 , we can write the metric in the following way
\bea
\ggg(\c,\c)=-\Phi^2dt^2+g_{ij}(X^idt+dx^i)(X^jdt+dx^j)\eql{noncmet}
\eea
where $g_{ij}$ are the components of the riemannian metric induced on the generic spacelike hypersurface of the foliation. The unit normal to the generic hypersurface is 
\bea
N=\frac{1}{\sqrt{\Phi^2+|X|^2}}\left(\frac{\pr}{\pr t}-X^j\frac{\pr}{\pr x^j}\right)\ .
\eea
The coordinates $\{x^i\}\ ,i\in\{1,2,3\}$ define a point $p$ of a generic hypersurface  labeled by a parameter $\tau$, $p\in\Si_{\tau}$, and the diffeomorphism $\Psi_{N}(\de)$ associated to the vector field $N$ send $p$ to a point $q\in \Si_{\tau+\de}$ with the same $\{x^i\}$ coordinates.

\NI Up to now, although we have defined the coordinates $\{t,x^i\}$, the ``gauge" is not yet defined as we have not specified the lapse and the shift functions $\Phi$ and $X$. They are defined completely once that we define the hypersurfaces which foliate the spacetime. For instance in the choice used in \cite{C-K:book}, the hypersurfaces are assumed labeled by the time coordinate $\tau=t$ and moreover they are assumed ``maximal" which means that the trace of their second fundamental form is identically zero. The first statement implies that $X=0$ and the second one that the lapse function $\Phi$ satisfies the hyperbolic equation
\[\lap\Phi=|k|^2\Phi\ .\]
 Similar considerations can be done for the ``gauge" associated to the ``CMC foliation" used by L.Andersson and V.Moncrief, \cite{An-Mon1}.
\smallskip

\NI Following these ideas the ``geometric" gauge we are choosing is associated to a double null cone foliation, see \cite{Kl-Ni:book} and later on for its definition. The spacetime we are going to construct is foliated by a family of null outgoing cones $C(\la)$ and a family of null incoming cones $\Cb(\nu)$, more precisely a portion of null cones, see later on for details. We denote $S(\la,\nu)$ their intersections,
\bea
S(\la,\nu)=C(\la)\cap\Cb(\nu)\ 
\eea
which are two-dimensional surfaces diffeomorphic\footnote{Here this is part of the assumptions we are doing about the foliation, later on we prove that this is true where the solution exists.} to $S^2$ and with $N$ and $\Nb$ the equivariant vector fields whose associated diffeomorphisms $\Phi$ and $\underline{\Phi}$ send the $S(\la,\nu)$ surfaces to the analogous surfaces on the outgoing or the incoming cones respectively,
\bea
&&\Phi(\de)[S(\la,\nu)]= S(\la,\nu+\de)\nn\\
&&\underline{\Phi}(\de)[S(\la,\nu)]= S(\la+\de,\nu)\ .\eql{diff1}
\eea
 Let us also define the map $\Psi(\la,\nu)$ we will use extensively later on,
\bea
\Psi(\la,\nu):S_0\ni p_0\rightarrow q=\Psi(\la,\nu)(p_0)={\Phi}(\nu-\nu_0)(\underline\Phi(\la-\la_0)(p_0)\in S(\la,\nu)\ .\ \ 
\eea
Once we have specified the foliation of the spacetime we can define two coordinates adapted to our gauge, namely the parameter $\la$ and $\nu$ which determine the incoming or the outgoing cones. More precisely we denote these coordinates $u$ and $\ub$ and the outgoing and incoming cones $C(\la)$ and $\Cb(\nu)$ are, respectively,
\bea
C(\la)=\{p\in\M|u(p)=\la\}\ \ ,\ \ \Cb(\nu)=\{p\in\M|\ub(p)=\nu\}
\eea
where $\M$ denotes the spacetime. 

\NI As the coordinates $u,\ub$ allow to define the null cones it is important to know how to connect them to an arbitrary set of coordinates. This is provided by the eikonal equation, 
\bea
g^{\mu\nu}\pr_{\mu}w\pr_{\nu}w=0   \eql{eik}
\eea 
where $g^{\mu\nu}$ are the components of the inverse of the metric tensor written in an arbitrary set of coordinates $x^{\mu}$. Then $u=u(x)$ and $\ub=\ub(x)$ are solution of the eikonal equations with appropriate initial data\footnote{Different choice of the ``initial data" for \ref{eik} on the external outgoing and incoming cones give rise to different null cone foliations.} such that their level hypersurfaces are, at least locally, the null cones of our foliation. It is well known, see for instance \cite{Kl-Ni:book}, that the vector fields 
\[L=-g^{\mu\nu}\pr_{\nu}u\frac{\pr}{\pr x^{\mu}}\ \ ,\ \ \Lb=-g^{\mu\nu}\pr_{\nu}\ub\frac{\pr}{\pr x^{\mu}}\]
are the tangent vector field of the null geodesics generating the outgoing and the incoming cones, satisfying
\[D_{L}L=0\ \ ,\ \ D_{\Lb}\Lb=0\ .\]
Their scalar product defines the scalar function $\oom$,
\bea
\ggg(L,\Lb)=-(2\oom)^{-2}
\eea
and the vector fields $e_3,e_4$,
\bea
e_3=2\oom\Lb\ \ ,\ \ e_4=2\oom L\ ,
\eea
satisfy the relation $g(e_3,e_4)=-2$. Finally the equivariant vector fields previously introduced, $N,\Nb$,\footnote{It is immediately to check that they satisfy eqs. \ref{diff1}.} have the following expressions
\bea
N=\oom e_4\ \ ,\ \ \Nb=\oom e_3\ .
\eea
Once introduced the coordinates $u,\ub$ we are left with defining the remaining two coordinates, we have to interpret as angular coordinates, which allow to specify a point on each surface $S(\la,\nu)$. The procedure is similar to the one envisaged previously in the non characteristic case, to define them we have to define a map which sends a point $p$ of ``angular coordinates" $\{\om^a\}\ ,\ a\in\{1,2\}$ belonging to the intersection of the ``initial cones" $S(\la_0,\nu_0)=C(\la_0)\cap\Cb(\nu_0)$ to a point $q\in S(\la,\nu)$ to which the same coordinates are assigned. The map we choose is made in the following way: first we move, starting from a point $p\in S(\la_0,\nu_0)$ of angular coordinates $\{\om^a\}$,  along the $\Cb(\nu_0)$ cone using the integral curves of the vector field $\Nb$ up to a point $q'\in S(\la,\nu_0)$, then applying the diffeomorphism generated by the vector field $N$ we move ``inside" the region of the spacetime up to the point  $q\in(\la,\nu)$ and assigne to this point the same angular coordinates $\{\om^a\}$; formally, for any $a$,
\bea
\om^a(\Psi(\la,\nu)(p))\equiv\om^a({\Phi}(\nu-\nu_0)\underline\Phi(\la-\la_0)(p))=\om^a(p)\ .\eql{omdef}
\eea
Once we have introduced the coordinates relative to our gauge we have still to write, in these coordinates, an explicit expression for the metric tensor. This can be done defining a null moving frame, $\{e_4,e_3,e_A\}, \ A\!\in\!\{1,2\}$ adapted to this double null foliation where
\bea
e_A=e_A^a\frac{\pr}{\pr\om^a}, \ A\!\in\!\{1,2\}\ ,
\eea
are orthonormal vector fields tangent at each point $p\in\M$ to the surface $S(\la,\nu)$ containing $p$ and  $e_3\ ,\ e_4$
are null vector fields orthogonal to the ${e_A}$'s, outgoing and incoming, respectively, which, basically, means ``tangent" to the null hypersurfaces $C(\la)$ and $\Cb(\nu)$.\footnote{$e_4$ is, at same time, tangent and normal to $C(\la)$ and analogously $e_3$ with respect to $\Cb(\nu)$.} 
Moreover, recalling the meaning of the vector fields $N$ and $\Nb$, it follows that
\bea
N=\frac{\pr}{\pr\ub}\ \ \mbox{and}\ \ e_4=\frac{1}{\oom}\frac{\pr}{\pr\ub}\ .
\eea
The explicit expression of $\Nb$, and therefore of $e_3$ is somewhat different as in a curved spacetime these two vector fields do not commute; it can be proved, see \cite{Kl-Ni:book} that $\Nb$ must have the following expression
\bea
\Nb=\frac{\pr}{\pr u}+X^a\frac{\pr}{\pr\om^a}\ \ \mbox{and therefore}\ \ e_3=\frac{1}{\oom}\left(\frac{\pr}{\pr u}+X^a\frac{\pr}{\pr\om^a}\right)\ .
\eea
where the property and the equation that the vector field $X$
has to satisfy will be discussed later on.\footnote{The definition of the coordinates $\om^a$, eq. \ref{omdef}, implies that 
$X=0$ on $\Cb_0$.}
\smallskip

\NI Once we have the explicit expression of the null frame in the adapted\footnote{Here with adapted we mean both adapted to the ``gauge" we are defining, both to the leaves $S(\la,\nu)$ of the null cones.} coordinates we can write the metric tensor in these coordinates obtaining
\bea
{\ggg}=-2{{\oom}}^2(dud\ub\!+\!d\ub du)\!+\!\ga_{ab}(X^adu+d\om^a)(X^bdu+d\om^b)\eql{met0}
\eea
where $\ga_{ab}$ are the components of the induced metric on $S(\la,\nu)$. 
\smallskip

\NI{\bf Remark:} {\em
Observe  that this is not the more general metric we can write ``adapted" to the foliation, in fact there are only seven metric components different from zero. This follows as we have chosen the coordinate $\ub$ such that the $N$ vector field was $\frac{\pr}{\pr\ub}$. We could nevertheless define the coordinates in such a way that 
$N=\frac{\pr}{\pr\ub}+Y^a\frac{\pr}{\pr\om^a}$,
in this case repeating the previous argument the metric turns out to be, written in the same $u,\ub$ coordinates,
$$
{\ggg}=-2{{\oom}}^2(dud\ub\!+\!d\ub du)\!+\!\ga_{ab}(X^adu+Y^ad\ub+d\om^a)(X^bdu+Y^bd\ub+d\om^b)\eql{met1}
$$
with ten components different from zero.\footnote{Nevertheless not all independent due to the nature of the double null cone foliation.} This last expression is the one analogous to the metric \ref{noncmet} written for the non characteristic problem
$$
\ggg(\c,\c)=-\Phi^2dt^2+g_{ij}(X^idt+dx^i)(X^jdt+dx^j)
$$
while the previous expression \ref{met0} corresponds to the one used for instance in \cite{C-K:book} associated to the ``maximal" foliation, namely
$$
\ggg(\c,\c)=-\Phi^2dt^2+g_{ij}dx^idx^j\eql{noncmet0}\ .$$}
Comparing the metric expression \ref{met0} with the one in \ref{noncmet} we see that $\oom$ plays basically the role of the lapse function and $X$ of the shift vector. Exactly as in that case we expect that the gauge choice is completed once we are able to determine these functions. In the non characteristic case this could be done in different ways, for instance in \cite{Kl-Ni:book} the shift vector was imposed equal to zero and the lapse function had to satisfy an elliptic equation. In other cases, see for instance \cite{C-K:book} or \cite{An-Mon1},\footnote{In their case the spacetime is spatially compact and the reduced equations are the evolution equations (2.5a), (2.5b) together with the elliptic equations (2.8a), (2.8b) relative to $N$ and $X$ which specify the gauge; given a solution of this set, the constraint equations are proved to hold everywhere in the spacetime once they are assumed for the initial data.} both quantities had to satisfy some differential equations. In the present characteristic case we will see that both $\oom$ and $X^a$ have to satisfy some first order differential equations. We will obtain them in the next sections after we write the Einstein equations in these coordinates. To do it we use in a systematic way the structure equations for a Lorentzian manifold. 

\subsection{The Einstein equations in the $\{\la,\nu,\theta,\phi\}$ coordinates}\label{SS2.3}
In this section and in the following section we fulfill the first goal of this paper, namely we write the Einstein equations in a way suited to the class of characteristic problems we are considering and connected to the gauge choice discussed in the previous subsection. We recall shortly the main properties of our approach.

\NI i) As we said before we assume the spacetime foliated by null cones, outgoing and incoming as prescribed by the chosen gauge. Therefore the Einstein equations become evolution equations for the metric components $\ga_{ab}$ of the two dimensional leaves $S(\la,\nu)$ which foliate the null cones. Moreover as in the non characteristic case the choice of the gauge will imply that the lapse function, $\oom$, and the shift vector $X$ have to satisfy some equations. 

\NI ii) As in the analogous formulation of the non characteristic case we express the Einstein equations as a system of first order equations; we define the family of first order equations which express the Einstein equations and separate them in two groups, in the first one we collect those equations which can be interpreted as the evolution part of the Einstein equations\footnote{Together with the equations for $\oom$ and $X$.} and in the second one those which can be interpreted as constraint equations. This is the central goal of this subsection.

\NI iii) The third aspect to remark is that we write our equations in terms of the coordinates $u\equiv\la$, $\ub\equiv\nu$, $\om^a$, where $\nu$ and $\la$ are, as defined in the previous subsection, affine parameters for the null geodesics along the null outgoing and incoming cones and we do not use the more standard time coordinate, $t$.\footnote{The simplest analogy is solving the two dimensional homogeneous linear wave equation written as: $$\frac{\pr^2}{\pr\la\nu}u=0\ .$$} With this choice our equations have a more geometric ``flavour" as all the quantities we introduce are connected in a direct way to the geometric properties of our foliation and the equations are transport equations along the null cones. This approach and this formalism turn out to be very appropriate, as already discussed in \cite{Kl-Ni:book}, to obtain apriori estimates for ``energy-type" quantities. This, moreover, will allow us to prove the main goal of this work, namely, as discussed in the introduction, that the analytic solution of the characteristic problem has a much larger existence region depending on the ``hyperbolic" apriori estimates, a region which can be unbounded if the initial data have $H^s$ norms, with appropriate $s$, sufficiently small.
\smallskip

\NI To write the Einstein equations in a way satisfying i) ii) and iii) we use intensively the structure equations, see for instance, \cite{Sp:Spivak}, Vol 2, adapted to a Lorentzian manifold.
\subsubsection{The structure equations}
We recall some general aspects of the structure equations. We denote the null orthonormal frame in the following way:
\bea
&\{e_{(\alpha)}\}=\{e_{\alpha}\}=\{e_{(1)},e_{(2)},e_{(3)},e_{(4)}\}
\eea
where, 
\bea
e_{(3)}=2\oom\Lb,\ \ e_{(4)}=2\oom L\ \ ,\ \ e_{(1)}=e_{\theta},\ \ e_{(2)}=e_{\phi}
\eea
and\bea
\nn\\
&\{\theta^{(\alpha)}\}=\{\theta^{\alpha}\}=\{\theta^{(1)},\theta^{(2)},
\theta^{(3)},\theta^{(4)}\}
\eea
 are the corresponding forms satisfying
\bea
\theta^{(\a)}(e_{(\b)})=\delta^{\a}_{\b}
\eea
and it follows that
\bea
\theta^{(3)}_\mu= -\frac{1}{2}g_{\mu\nu}{e_4}^\nu\ ,\ \theta^{(4)}_\mu= -\frac{1}{2}g_{\mu\nu}{e_3}^\nu\ .
\eea
We define 
\bea
&&\dd_{e_{\a}}e_{\b}\equiv \GGa^{\ga}_{\a\b}e_{\gamma}\nn\\
&&\rr(e_{\a}e_{\b})e_{\ga}\equiv \rr^{\de}_{\ga\a\b}e_{\de} \eql{Gadef}
\eea
where $\dd$ is the connection of the spacetime associated to the Lorentz metric $g$,
$\dd_{e_{\a}}e_{\b}$ is the covariant derivative of the vector field $e_{\b}$ in the direction
$e_{\a}$ and $\rr$ is the Riemann tensor (here the first greek letters are ``names" and do not denote components),\footnote{Obviously choosing a coordinate basis
$\{\frac{\partial}{\partial x^{0}},\frac{\partial}{\partial x^{1}},\frac{\partial}{\partial
x^{2}},\frac{\partial}{\partial x^{3}}\}$ the $\GGa^{\ga}_{\a\b}$ defined in \ref{Gadef} are just the
usual Christoffel symbols, $\Ga^{\mu}_{\nu\ro}$, $\dd_{\frac{\partial}
{\partial x^{\mu}}} \equiv D_{\mu}$ is the covariant derivative with respect to
${\partial_\mu}$ and 
$\rr^{\de}_{\ga\a\b}=R^{\mu}_{\nu\ro\si}e^{\nu}_{\ga}e^{\ro}_{\a}e^{\si}_{\b}
\theta^{\de}_{\mu}$ .}
\[\rr(e_{\a}e_{\b})e_{\ga}=\dd_{e_{\a}}(\dd_{e_{\b}}{e_{\ga}})-
\dd_{e_{\b}}(\dd_{e_{\a}}{e_{\ga}})-\dd_{[e_{\a},e_{\b}]}{e_{\ga}}\]
Defining the following one and two forms:
\bea
\omm^{\a}_{\b}&\equiv&\GGa^{\a}_{\ga\b}\theta^{\ga}\nn\\
\oomm^{\a}_{\b}&\equiv&\frac{1}{2}\rr^{\a}_{\b\ga\de}\theta^{\ga}\wedge\theta^{\de}
\eea
we have the following result, see \cite{Sp:Spivak}, Vol.2, whose proof is in the appendix to this section.
\begin{Prop}\label{P2.1}
$\omm^{\a}_{\b}$ and $\oomm^{\a}_{\b}$ satisfy the following structure equations
\bea
&&d\theta^{\a}=-\omm^{\a}_{\ga}\wedge\theta^{\ga}\\
&&d\om^{\de}_{\ga}=-\om^{\de}_{\si}\wedge\om^{\si}_{\ga}+\Omega^{\de}_{\ga}\ ,
\eea
called, respectively, the first and the second structure equations.
\end{Prop}

\NI The knowledge of a null orthonormal frame in a whole region is equivalent to knowing the metric in that region, therefore the first set of structure equations can be thought as ``first order equations" for the metric components. Viceversa the one forms $\om^\a_\b$ are connected to the first derivatives of the (components of the) moving frame and, therefore play the role of the first derivatives of the metric components; the second set of structure equations represent first order equations for these first derivatives.\footnote{Observe that the second group of structure equations depends, through $\Omega^\a_\b$, also on the Riemann tensor components. This could suggest that expressing these equations as partial differential equations for the various components, to have a closed system of equations one should also consider the Bianchi equations for the Riemann tensor and the Riemann components as independent variables. Although this could be done, see for instance \cite{friedrich:cauchychar}, this is not what we do, as we discuss in great detail later on.}

\NI  To have an explicit expression for the structure equations  we  recall the definition of the ``connection coefficients" (sometimes called ``Ricci coefficients"). In the defined null orthonormal frame they have the following expressions:
\bea
\chi_{AB}&=&\ggg(\dd_{{{e_A}}}{e_4},e_{B})\ \ \ ,\ \chib_{ab}\ =\ \ggg(\dd_{{{e_A}}}{e_3},e_{B})\nn\\
\xib_{A}&=&\frac{1}{2}\ggg(\dd_{{e_3}}{e_3},e_{A})\ \ \ ,\ 
\xi_{A}=\frac{1}{2}\ggg(\dd_{{e_4}}{e_4},e_{A})\nn\\
\omb&=&\frac{1}{4}\ggg(\dd_{{e_3}}{e_3},{e_4})\ \ \ ,\ 
\om=\frac{1}{4}\ggg(\dd_{{e_4}}{e_4},{e_3})\nn\\
\etab_{A}&=&\frac{1}{2}\ggg(\dd_{e_{4}}{e_3},{{e_A}})\ \ \ ,\ 
\eta_{A}=\frac{1}{2}\ggg(\dd_{e_{3}}{e_4},{{e_A}})\nn\\
\ze_{A}&=&\frac{1}{2}\ggg(\dd_{e_{A}}{e_4},{e_3})\nn
\eea
They are 2-covariant tensors, vectors and scalar functions defined on the two dimensional surfaces $S(\la,\nu)$. 

\NI In terms of these quantities the one forms, $\om^{\b}_{\a}$, have the following expressions:
\bea
&&\om^4_3=0\ \ ,\ \ \om^4_4=-2\om\theta^4+2\omb\theta^3-\ze_a\theta^A\ \ ,\ \ \om^4_A=\etab_A\theta^4+\frac{1}{2}\chib_{BA}\theta^B\nn\\
&&\om^3_3=2\om\theta^4-2\omb\theta^3+\ze_a\theta^A\ \ ,\ \ \om^3_4=0\ \ ,\ \ \om^3_A=\eta_A\theta^3+\frac{1}{2}\chi_{BA}\theta^B\nn\\
&&\om^A_4=2\eta_A\theta^3+\chi_{BA}\theta^B\ \ ,\ \ \om^A_3=2\etab_A\theta^4+\chib_{BA}\theta^B
\eea
The one forms $\om^A_B$ have a different expression which do not depend on the connection coefficients we have introduced,
\bea 
\om^A_B=\ggg(\ddb_4e_B,e_A)\theta^4+\ggg(\ddb_3e_B,e_A)\theta^3+\ggg(\dd_Ce_B,e_A)\theta^C\ .
\eea
It is a long, but simple task to write the structure equations in terms of the metric components and the connection coefficients. The first set of structure equations 
\beaa
d\theta^{\a}(e_\b,e_\ga)=-\omm^{\a}_{\de}\wedge\theta^{\de}(e_\b,e_\ga)
\eeaa
becomes:
\bea
&&\ML\ML\{\a,\b,\ga\}=\{A,B,4\}: \ \ \frac{\pr}{\pr\nu}\ga_{ab}=2\oom\chi_{ab}\nn\\
&&\ML\ML\{\a,\b,\ga\}=\{A,B,3\}: \ \ \frac{\pr}{\pr\la}\ga_{ab}+\Lie_X\ga_{ab}\!=2\oom\chib_{ab}\nn\\
&&\ML\ML\{\a,\b,\ga\}=\{A,4,3\}: \ \ \frac{\pr}{\pr\nu}X^c=4\oom^2\ze_Ce_C^c\eql{Istruct1}\\
&&\ML\ML\{\a,\b,\ga\}=\{4,4,3\}: \ \ \frac{\pr}{\pr\nu}\log\oom=-2\oom\om\nn\\
&&\ML\ML\{\a,\b,\ga\}=\{3,3,4\}: \ \ \left(\frac{\pr}{\pr\la}+X^a\frac{\pr}{\pr\om^a}\right)\log\oom=-2\oom\omb\nn
\eea
\bea
&&\ML\ML\{\a,\b,\ga\}=\{4,A,4\}: \ \ \nabb_A\log\oom=\etab_A+\ze_A\nn\\
&&\ML\ML\{\a,\b,\ga\}=\{3,A,3\}: \ \ \nabb_A\log\oom=\eta_A-\ze_A\nn\\
&&\ML\ML\{\a,\b,\ga\}=\{4,A,3\}: \ \ \ \xib_A=0\nn\\
&&\ML\ML\{\a,\b,\ga\}=\{3,A,4\}: \ \ \ \xi_A=0\ .\eql{Istruct2}
\eea
{\bf Remark:} {\em 
To obtain equations \ref{Istruct1} and \ref{Istruct2} we have used only the relation $e_3=2\oom\Lb\ ,\ e_4=2\oom L$. Recall that the structure equations for a moving (null orthonormal) frame do not imply that $\{e_3,e_A\}, A\in\{1,2\}$ or $\{e_4,e_A\}, A\in\{1,2\}$ are integrable distributions. In our present case, with the previous definition of $e_3,e_4$, this is true.}
\smallskip

\NI As our purpose is to write the Einstein equations as a system of first order equations for the component of the metric and their derivatives, equations \ref{Istruct1} of the first set of structure equations tell us exactly that we have to look for first order partial differential equations for the connection coefficients $\{\chi,\chib,\ze,\om,\omb\}$. They correspond in fact to the derivatives along the inward (or outward) direction with respect to the null cones outgoing or incoming, of the metric components $\ga_{ab}$, of the lapse function and of the shift vector. 
Once we have written the appropriate partial differential equations for these connection coefficients we will show that  we reduce to a closed set of equations in the metric components and in these connection coefficients. To find the appropriate equations 
we rely on the second set of the structure equations,
\beaa
(d\om^{\de}_{\ga}+\om^{\de}_{\si}\wedge\om^{\si}_{\ga})(e_{\a},e_{\b})=\Omega^{\de}_{\ga}(e_{\a},e_{\b})\ .
\eeaa
These equations can be written in a more explicit way in terms of the connection coefficients. One has, nevertheless, to remember that the structure equations are  identities, valid in a generic manifold with a (Riemannian or Lorentzian) metric. They can also be seen as ``integrability conditions" for the existence of a moving frame in the whole manifold. One has to remark that in their right hand side the terms $\Omega^{\de}_{\ga}(e_{\a},e_{\b})$ appear which can be written in terms of the various components of the Riemann tensor. Therefore these equations can be seen as identities defining the Riemann tensor in terms of first derivatives of the connection coefficients, but if we impose some condition on the Riemann tensor they become partial differential equations to solve with respect to the connection coefficients. The condition to impose to the Riemann tensor, or more precisely to $\Omega^{\de}_{\ga}$, are that the vacuum Einstein equations have to be satisfied, namely that the corresponding Ricci tensor is identically zero. Therefore we have to look to the explicit expression of these equations under this condition.

\NI It is a long, but standard and certainly not new\footnote{What is certainly more uncommon is that we do not use  a subset of these equations to get some norm estimates for various terms assuming we already have a solution, but to solve them, which requires a delicate choice of the subset.}  to realize that the structure equations have the expressions written in the following where we indicate also the Ricci components ($=0$) to which they are associated.\footnote{Remind that we used the relations between $\eta,\etab$ and $\ze$, the expression of $\om$ and $\omb$ in terms of $\oom$ and the fact that  $\xi=\xib=0$, obtained from the first set of structure equations, see for instance \cite{Kl-Ni:book}, Chapter 3.} 
\bea
&&\ML\ML[{\bf R}(e_4,e_4)\!=\!0:]\nn\\
&&\ML\ML\dd_4\tr\chi+\frac{1}{2}(\tr\chi)^2+2\om \tr\chi+|\chih|^2=0\nn\\
&&\nn\\
&&\ML\ML[{\bf R}(e_4,e_A)\!=\!0:]\nn\\
&&\ML\ML\dddd_4\zeta+\zeta\chi+\tr\chi\zeta-\divv\chi+\nabb\tr\chi+\ddb_4\nabb\log\oom=0\nn\\
&&\nn\\
&&\ML\ML[\de_{AB}{\bf R}(e_A,e_B)\!=\!0:]\nn\\
&&\ML\ML\dd_4\tr\chib+\frac{1}{2}\tr\chi \tr\chib-2\om \tr\chib+\chih\cdot\chibh+2\divv(\ze\!-\!\nabb\log\oom)\!-\!2|\ze\!-\!\nabb\log\oom|^2=2\ro\nn\\
&&\nn\\
&&\ML\ML[\widehat{{\bf R}(e_A,e_B)}\!=\!0:]\nn\\
&&\ML\ML\dddd_4\chibh\!+\!\frac{1}{2}\tr\chi\chibh\!+\!\frac{1}{2}\tr\chib\chih\!-\!2\om\chibh\!+\!\nabb\hot(\ze\!-\!\nabb\!\log\oom)
\!-\!(\ze\!-\!\nabb\!\log\oom)\hot(\ze\!-\!\nabb\!\log\oom)\!=\!0\nn\\
&&\nn\\
&&\ML\ML[{\bf R}(e_3,e_4)\!=\!0:]\nn\\
&&\ML\ML\dd_4\omb\!-\!2\om\omb\!-\!\ze\c\nabb\log\oom\!-\!\frac{3}{2}|\ze|^2\!+\!\frac{1}{2}|\nabb\log\oom|^2\!+\frac{1}{2}\!\big({\bf
K}\!+\!\frac{1}{4}\tr\chi\tr\chib\!-\!\frac{1}{2}\chih\c\chibh\big)\!=\!0\nn\\
&&\nn\\
&&\ML\ML[{\bf R}(e_3,e_3)\!=\!0:]\nn\\
&&\ML\ML\dd_3{\tr\chib}+\frac{1}{2}(\tr\chib)^2+2\omb \tr\chib+|\chibh|^2=0 \eql{2.1ff}\\
&&\nn\\
&&\ML\ML[{\bf R}(e_3,e_A)\!=\!0:]\nn\\
&&\ML\ML\dddd_3\zeta+\tr\chib\zeta+\zeta\chib+\divv\chib-\nabb\tr\chib-\ddb_3\nabb\log\oom=0\nn\\
&&\nn\\
&&\ML\ML[\de_{AB}{\bf R}(e_A,e_B)\!=\!0:]\nn\\
&&\ML\ML\dddd_3\tr\chi\!+\!\tr\chib\tr\chi\!-\!2\omb\tr\chi\!-\!2\divv(\ze+\nabb\log\oom)\!-\!2|\ze+\nabb\log\oom|^2\!+\!2{\bf K}=0\nn\\
&&\nn\\
&&\ML\ML[\widehat{{\bf R}(e_A,e_B)}\!=\!0:]\nn\\
&&\ML\ML\dddd_3\chih\!+\!\frac{1}{2}\tr\chib\chih\!+\!\frac{1}{2}\tr\chi\chibh\!-\!2\omb\chih
\!-\!\nabb\hot(\zeta\!+\!\nabb\!\log\oom)\!-\!(\zeta\!+\!\nabb\!\log\oom)\hot(\zeta\!+\!\nabb\!\log\oom)=0\nn\\
&&\nn\\
&&\ML\ML{{\bf R}(e_4,e_3)}\!=\!0:]\nn\\
&&\ML\ML\dd_3\om\!-\!2\omb\om\!+\!\ze\c\nabb\log\oom\!-\!\frac{3}{2}|\ze|^2\!+\!\frac{1}{2}|\nabb\log\oom|^2\!+\frac{1}{2}\!\big({\bf
K}\!+\!\frac{1}{4}\tr\chib\tr\chi\!-\!\frac{1}{2}\chibh\c\chih\big)\!=\!0\nn
\eea
where  $\dd_3\!=\!\dd_{e_3}, \dd_4\!=\!\dd_{e_4}$,  
$\dddd_3, \dddd_4$ their projection on the tangent spaces $TS(\la,\nu)$,
$\nabb$ the covariant derivatives associated to the metric $\ga$ induced by $\ggg$ on the surfaces $S(\la,\nu)$, $\bf K$ is the
curvature of these $S(\la,\nu)$ surfaces and \[\widehat{{\bf R}(e_A,e_B)}={\bf R}(e_A,e_B)-2^{-1}\de_{AB}{\bf R}(e_A,e_B)\ .\]
 
 \NI In the appendix we write the general form of the second set of structure equations and show how from them, the first set and our gauge choice equations \ref{2.1ff} follow. 
 \smallskip
 
 \NI We have now the explicit expression for the first set and the second set of the structure equations, namely \ref{Istruct1} and \ref{2.1ff}. They are 24 equations, ten from the first set and fourteen from the second one while we have only sixteen unknown functions $\oom, X, \ga, \chi,\chib,\ze,\om,\omb$. 
The fact that there are more equations that unknown functions is not, in the present case, a real difficulty as the structure equations are automatically satisfied in a Lorentzian manifold and, therefore, also in a vacuum Einstein manifold. Therefore we have to  choose a subset between them which forms a complete set of equations for the  sixteen unknown functions and then prove, as expected, that the remaining equations play the analogous role of the standard constraint equations and are automatically satisfied once they are imposed on the initial data. The nature of the constraint equations is discussed in detail in subsection \ref{SS3.2}.
The sixteen equations we choose are:
\bea
&&\ML\ML\frac{\partial{\ga}}{\partial\la}-2{\oom}\ \!{\chib}+\Lie_{X}{\ga}=0\nn\\
&&\ML\ML\frac{\partial\log{\oom}}{\partial\la}+{{\nabb}}_{X}\log{\oom}+{2{\oom}}\ \!{\omb}=0\nn\\
&&\ML\ML\dddd_3\tr\chi\!+\!\tr\chib\tr\chi\!-\!2\omb\tr\chi\!-\!2\divv(\ze+\nabb\log\oom)\!-\!2|\ze+\nabb\log\oom|^2\!+\!2{\bf K}=0\nn\\
&&\ML\ML\dddd_3\chih\!+\!\frac{1}{2}\tr\chib\chih\!+\!\frac{1}{2}\tr\chi\chibh\!-\!2\omb\chih
\!-\!\nabb\hot(\zeta\!+\!\nabb\!\log\oom)\!-\!(\zeta\!+\!\nabb\!\log\oom)\hot(\zeta\!+\!\nabb\!\log\oom)=0\nn\\
&&\ML\ML\dddd_3\zeta+\tr\chib\zeta+\zeta\chib+\divv\chib-\nabb\tr\chib-\ddb_3\nabb\log\oom=0\eql{subsetstrinc}\\
&&\ML\ML\dd_3\om\!-\!2\omb\om\!+\!\ze\c\nabb\log\oom\!-\!\frac{3}{2}|\ze|^2\!+\!\frac{1}{2}|\nabb\log\oom|^2\!+\frac{1}{2}\!\big({\bf
K}\!+\!\frac{1}{4}\tr\chib\tr\chi\!-\!\frac{1}{2}\chibh\c\chih\big)\!=\!0\nn
\eea
\bea
&&\ML\ML\frac{\partial{X}}{\partial\nu}+4{\oom}^2{Z}=0\nn\\
&&\ML\ML\dddd_4\tr\chib+\tr\chi\tr\chib -2\om\tr\chib+2\divv(\ze\!-\!\nabb\log\oom)\!-\!2|\ze\!-\!\nabb\log\oom|^2+\!2{\bf K}=0\ \ \ \eql{subsetstrout}\\
&&\ML\ML\dddd_4\chibh\!+\!\frac{1}{2}\tr\chi\chibh\!+\!\frac{1}{2}\tr\chib\chih\!-\!2\om\chibh\!+\!\nabb\hot(\ze\!-\!\nabb\!\log\oom)
\!-\!(\ze\!-\!\nabb\!\log\oom)\hot(\ze\!-\!\nabb\!\log\oom)\!=\!0\nn\\
&&\ML\ML\dd_4\omb\!-\!2\om\omb\!-\!\ze\c\nabb\log\oom\!-\!\frac{3}{2}|\ze|^2\!+\!\frac{1}{2}|\nabb\log\oom|^2\!+\frac{1}{2}\!\big({\bf
K}\!+\!\frac{1}{4}\tr\chi\tr\chib\!-\!\frac{1}{2}\chih\c\chibh\big)\!=\!0\ .\nn
\eea
{\bf Remarks:} {\em

\NI i) The equations \ref{subsetstrinc} and  \ref{subsetstrout} are appropriate, as we will see in the following, to apply Cauchy-Kowalevski theorem and find analytic solutions. 
To write them as first order p.d.e. equations for the tensor components requires still some more work due to the presence of the Gauss curvature $K$ which depends on the second angular variables of the metric $\ga$. This will be discussed in the next subsection.
\smallskip

\NI ii) The second important remark is that once we have solved equations \ref{subsetstrinc} and  \ref{subsetstrout}, looking at \ref{2.1ff} our analytic solution is such that \[{\bf R}(e_A,e_B)=0\ ,\ {\bf R}(e_A,e_4)=0\ ,\ {\bf R}(e_3,e_4)=0\ \]
and we have still to prove that the remaining Ricci equations are satisfied. This is discussed in detail in subsection \ref{SS3.2} where we show that the remaining equations to be satisfied have to be considered as ``constraint equations".
\smallskip

\NI iii) Observe that in this approach there are ten independent connection coefficients, $\chi,\chib,\ze,\om,\omb$. They, basically, correspond to
the second fundamental form $k_{ij}$ of the ``maximal foliation gauge" or of the ``CMC foliation gauge". The difference is that $k$ has only six
components. This is due to the fact that in those cases the foliation is made by only one family of hypersurfaces $\Si_t$ while here there are both
the $C(\la)$ and the $\Cb(\nu)$ null hypersurfaces. If we consider only the $\{C(\la)\}$ foliation, $\chib,\ze,\omb$  are the $S$-tensors corresponding to  $k$ ($\chi,\ze,\om$ in the opposite case). In both situations there are six components, as expected.}
\medskip

\NI Equations \ref{subsetstrinc}, \ref{subsetstrout} are perfectly defined as tensorial equations, but, to consider them as p.d.e. equations, they  have to be written as equations for the tensor components. In this case they do not maintain exactly the same expressions. There are many ways to rewrite these equations as standard partial differential equations whose unknown are the components of the various tensors involved, for instance one could choose a Fermi transported null orthonormal frame\footnote{Nevertheless this is possible only with respect to a null direction, but not simultaneously to both.} as was done in \cite{Kl-Ni:book}; here we present a more general approach using the diffeomorphism $\Psi(\la,\nu)$ previoulsly introduced to map, via the pullback associated to $\Psi(\la,\nu)$, these equations on a manifold $S_0\times R^2$ where the equations become equations for the various components in the angular variables and in the variables $\la$ and $\nu$, which are just the parameters of the diffeomprphism $\Psi(\la,\nu)$. Here we state the result and its detailed proof is
in the appendix.
\begin{Prop}\label{Prop2.2}
In the coordinates $\{\la,\nu,\theta,\phi\}$ associated to the ``double null foliation gauge", equations \ref{subsetstrinc} and \ref{subsetstrout} written for the various  of  metric and  connection coefficients components have the following expression:
{\em
\bea
&&\ML\ML\frac{\partial{\ga}}{\partial\la}-2{\oom}\ \!{\chib}+\Lie_{X}{\ga}=0\nn\\
&&\ML\ML\frac{\partial\log{\oom}}{\partial\la}+{{\nabb}}_{X}\log{\oom}+{2{\oom}}\ \!{\omb}=0\nn\\
&&\ML\ML\frac{\partial{\tr\chi}}{\partial\la}+{\oom}{\tr\chib}{\tr\chi}-2{\oom}{\omb}{\tr\chi}+{\nabb}_{\! X}{\tr\chi}
-2{\oom}{\divv}{(\ze+\nabb\log\oom)}-2{\oom}|{\ze+\nabb\log\oom}|^2 +2{\oom}{\bf K}\!=\!0\nn\\
&&\ML\ML\frac{\partial\hat{{\chi}}}{\partial\la}-\frac{{\oom}{\tr\chib}}{2}\hat{{\chi}}+\frac{{\oom}{\tr\chi}}{2}\hat{{\chib}}
+\frac{\partial\log{\oom}}{\partial\la}\hat{{\chi}}+({\nabb}_{\! X}\log{\oom})\hat{{\chi}}
-{\oom}(\hat{{\chi}}\c\hat{{\chib}})\ga-{\oom}\ \!{\nabb}\hot{(\ze+\nabb\log\oom)}\nn\\
&&\ML\ML-{\oom}{(\ze+\nabb\log\oom)}\hot{(\ze+\nabb\log\oom)}+\Lie_{{X}}\hat{{\chi}}\!=\!0\eql{subsetstrinc2}\\
&&\ML\ML\frac{\partial{\ze}}{\partial\la}+{\oom}\ \!{\tr\chib}{\ze}+{\oom}{\divv}{\chib}-{\oom}{\nabb}{\tr\chib}
-\frac{\partial{\nabb}{\log\oom}}{\partial\la}+{\oom}{\nabb}\log{\oom}\!\c\!{\chib}+{\Lie_{ X}{\ze}}-\Lie_{X}{\nabb}\log{\oom}\!=\!0\nn\\
&&\ML\ML\frac{\partial{{\om}}}{\partial\la}\!+\!{\nabb}_{\!X}{\om}\!-\!2{\oom}\ \!\!{{\omb}}\
\!{{\om}}\!-\!\frac{3}{2}{{\oom}}|{\ze}|^2\!+\!{{\oom}}{\ze}\!\c\!{\nabb}\log{\oom}
\!+\!\frac{1}{2}{\oom}|{\nabb}\log{\oom}|^2\!+\!\frac{1}{2}{\oom}\!\left({{\bf K}}
\!+\!\frac{1}{4}{\tr\chi}{\tr\chib}\!-\!\frac{1}{2}\hat{{\chi}}\!\c\!\hat{{\chib}}\!\right)=0\nn
\eea
\bea
&&\ML\ML\frac{\partial{X}}{\partial\nu}+4{\oom}^2{Z}=0\nn\\
&&\ML\ML\frac{\partial{\tr\chib}}{\partial\nu}+{\oom}{\tr\chi}{\tr\chib}
-2{\oom}{\om}{\tr\chib}-2{\oom}{\divv}{(-\ze+\nabb\log\oom)}-2{\oom}|-\ze+\nabb\log\oom|^2+2{\oom}{\bf K}\!=\!0\nn\\
&&\ML\ML\frac{\partial\hat{{\chib}}}{\partial\nu}-\frac{{\oom}{\tr\chi}}{2}\hat{{\chib}}
+\frac{{\oom}{\tr\chib}}{2}{\chih}+\frac{\partial\log{\oom}}{\partial\nu}\hat{{\chib}}-{\oom}(\hat{{\chib}}\c\hat{{\chi}})\ga
-{\oom}{\nabb}\hot{(-\ze+\nabb\log\oom)}\nn\\
&&\ML\ML-{\oom}(-\ze+\nabb\log\oom)\hot(-\ze+\nabb\log\oom)\!=\!0\eql{subsetstrout2}\\
&&\ML\ML\frac{\partial{{\omb}}}{\partial\nu}-2{\oom}\ \!\!{{\om}}\ \!{{\omb}}-\frac{3}{2}{{\oom}}|{\ze}|^2
\!-{{\oom}}{\ze}\c{\nabb}\log{\oom}\!+\!\frac{1}{2}{\oom}|{\nabb}\log{\oom}|^2
\!+\!\frac{1}{2}{\oom}\!\left({{\bf K}}\!+\!\frac{1}{4}{\tr\chib}{\tr\chi}
\!-\!\frac{1}{2}\hat{{\chib}}\c\hat{{\chi}}\!\right)=0\ .\nn
\eea}
\end{Prop}
Equations \ref{subsetstrinc2}, \ref{subsetstrout2} are not yet a system of first order equations. In fact
$\bf K$, the curvature of the two dimensional surfaces $S(\la,\nu)$, 
depends on  second tangential derivatives of $\ga$. Moreover in the (transport) evolution equations along $\Cb(\nu)$ for $\tr\chi$, $\chih$, $\ze$ and in the (transport) evolution equations along $C(\la)$ for $\tr\chib$ and $\chibh$ the second derivatives of $\log\oom$ with respect to the angular variables are present.

\NI To have a real first order system we define some new independent variables and their evolution equations, namely:
\bea
v_{\c\c\c}=\partialb_{\c}\ga(\c,\c)\ ,\  w_{\c\c}=\partialb_{\c}X(\c)\  \ ,\ \  \psi_{\c}=\partialb_{\c}\log\oom\ \ .\eql{3.13wq}
\eea
Their evolution equations are obtained deriving the evolution equations of $\ga(\c,\c)$, $X(\c)$ and
$\log\oom$, the second one in the outgoing direction, the other two in the incoming one. The unknown function $w_{\c,\c}$ is
introduced as in the evolution equation for $v$ the second tangential derivatives of $X$ appear. These equations do not contain more
than first derivatives of the previous unknown variables and this transforms the system of equations into a larger system of first order
equations. It is a matter of computation, which we report in the appendix to this section, to obtain the following evolution equations for $\psi_a$, ${v}_{adb}$ and for $w_{ab}$:
\bea
&&\ML\frac{\partial}{\partial\la}\psi_a=-2\oom\partial_a\omb-2\oom\omb\psi_a-(\nabb_aX)^c\psi_c-X^c\nabb_c\psi_a\nn\\
&&\ML\frac{\partial}{\partial\la}v_{cba}=-(\pr_cX^d)v_{dab}-\pr_Xv_{cab}+\partial_cw_{ab}+\partial_cw_{ba}+2\oom\partial_c\chib_{ab}+2\oom\psi_c\chib_{ab}\eql{3.18d}\\
&&\ML\frac{\partial w_{ab}}{\partial\nu}=-8\oom^2\psi_a\ze_b-4\oom^2\partial_a\ze_b+2\oom\psi_a\chi_{bc}X^c+2\oom(\partial_a\chi_{bc})X^c\ .\nn
\eea
We write now the final system of first order equations for the various tensors components omitting the indices to simplify the notations,
\bea
&&{\ML\ML}\frac{\partial\ga}{\partial\om}-v=0\ \ ,\ \ \frac{\partial\log\oom}{\partial\om}-\psi=0\ \ ,\ \ \frac{\partial\hat{X}}{\partial\om}-w=0\nn\\
&&{\ML\ML}\frac{\partial{\ga}}{\partial\la}-2{\oom}\ \!{\chib}+\Lie_{X}{\ga}=0\nn\\
&&{\ML\ML}\frac{\partial\log{\oom}}{\partial\la}+\psi(X)+{2{\oom}}\ \!{\omb}=0\nn\\
&&{\ML\ML}\frac{\partial v}{\partial\la}+\nabb_Xv+(\partialb X)\c v-S(\partialb\!\otimes\!w)-2\oom\partialb\!\otimes\!\chib-2\oom\ \!\psi\!\otimes\!\chib=0\eql{5.37l}\\
&&{\ML\ML}\frac{\partial\psi}{\partial\la}+\nabb_X\psi+2\oom\omb\psi+\psi(\nabb X)+2\oom\partialb\omb=0\nn\\
&&{\ML\ML}\frac{\partial{\tr\chi}}{\partial\la}+{\oom}{\tr\chib}{\tr\chi}-2{\oom}{\omb}{\tr\chi}+{\nabb}_{\! X}{\tr\chi}
-2{\oom}{\divv}{(\ze+\psi)}-2{\oom}|{\ze+\psi}|^2 +2{\oom}{\bf K}\!=\!0\nn\\
&&{\ML\ML}\frac{\partial\hat{{\chi}}}{\partial\la}+\Lie_{{X}}\hat{{\chi}}-\frac{{\oom}{\tr\chib}}{2}\hat{{\chi}}+\frac{{\oom}{\tr\chi}}{2}\hat{{\chib}}
-{2\oom}\ \!{\omb}\ \!\chih
-{\oom}(\hat{{\chi}}\c\hat{{\chib}})\ga-{\oom}\ \!{\nabb}\hot{(\ze+\psi)} -{\oom}(\ze+\psi)\hot(\ze+\psi)\!=\!0\nn\\
&&{\ML\ML}\frac{\partial{\ze}}{\partial\la}+{\Lie_{X}{\ze}}+{\oom}\ \!{\tr\chib}{\ze}+{\oom}{\divv}{\chibh}-\frac{1}{2}{\oom}{\partialb}{\tr\chib}
\left.+2\oom\omb\psi+2\oom\partialb\omb\right.+{\oom}\psi\!\c\!{\chib}\!=\!0\nn\\
&&{\ML\ML}\frac{\partial{{\om}}}{\partial\la}\!+\!{\partialb}_{\! X}{\om}\!-\!2{\oom}\ \!\!{{\omb}}\
\!{{\om}}\!-\!\frac{3}{2}{{\oom}}|\ze|^2\!+\!\frac{1}{4}{{\oom}}\ze\!\c\!\psi\!+\!\frac{1}{2}{\oom}|\psi|^2
\!+\!\frac{1}{2}{\oom}\!\left({{\bf K}}\!+\!\frac{1}{4}{\tr\chi}{\tr\chib}\!-\!\frac{1}{2}\hat{{\chi}}\!\c\!\hat{{\chib}}\!\right)=0\nn\\
&&\nn\\
&&{\ML\ML}\frac{\partial\hat{X}}{\partial\nu}+4{\oom}^2{\ze}=0\nn\\
&&{\ML\ML}\frac{\partial w}{\partial\nu}+8\oom^2\psi\!\otimes\!\ze+4\oom^2\partialb\!\otimes\!\ze-2\oom\psi\!\otimes\!(\chi\!\c\!X)
-2\oom(\partialb\!\otimes\!\chi)\!\c\!X=0\nn\\
&&{\ML\ML}\frac{\partial{\tr\chib}}{\partial\nu}+{\oom}{\tr\chi}{\tr\chib}
-2{\oom}{\om}{\tr\chib}+2{\oom}{\divv}\ze\!-\!2{\oom}{\divv}\psi-2{\oom}|\ze\!-\!\psi|^2+2{\oom}{\bf K}\!=\!0\eql{2.70gql}\\
&&{\ML\ML}\frac{\partial\hat{{\chib}}}{\partial\nu}-\frac{{\oom}{\tr\chi}}{2}\hat{{\chib}}
+\frac{{\oom}{\tr\chib}}{2}{\chih}-2\oom\om\chibh-{\oom}(\hat{{\chib}}\c\hat{{\chi}})\ga+{\oom}{\nabb}\hot(\ze\!-\!\psi)
-{\oom}(\ze\!-\!\psi)\hot(\ze\!-\!\psi)\!=\!0\nn\\
&&{\ML\ML}\frac{\partial{{\omb}}}{\partial\nu}-2{\oom}\ \!\!{{\om}}\ \!{{\omb}}-\frac{3}{2}{{\oom}}|\ze|^2
\!-{{\oom}}\ze\!\c\!\psi\!+\!\frac{1}{2}{\oom}|\psi|^2
\!+\!\frac{1}{2}{\oom}\!\left({{\bf K}}\!+\!\frac{1}{4}{\tr\chib}{\tr\chi}
\!-\!\frac{1}{2}\hat{{\chib}}\!\c\!\hat{{\chi}}\!\right)=0\nn
\eea
where $\hat{X}$ is the covariant vector $\hat{X}_a=\ga_{ab}X^b$, $S$ means symmetrization, $V\hot W$ is twice the traceless part of the symmetric
tensorial product $S(V\otimes W)$, $\bf K$ has to be thought as a function of $\ga$, $v$ and $\partialb\!v$.\footnote{$\partialb$ is the ordinary partial
derivative with respect to the angular variables, $\om^a$, and $\nabb$ is the Levi-Civita connection with respect to $\ga$.}

\subsection{The first order system of equations as solutions of the vacuum Einstein characteristic problem, the constraint problem.}\label{SS3.2}

The first order system of p.d.e. equations \ref{5.37l}, \ref{2.70gql} describes a characteristic problem  which can be solved via the
Cauchy-Kowalevski theorem (its characteristic version, as discussed later on in Section \ref{S3}), giving the inital data on the two null hypersurfaces $C_0$ and $\Cb_0$.\footnote{Which initial data can be given freely and which constrained is a delicate point we discuss in detail in subsection \ref{SS4.2}.} As we said before, see remark ii) after equations \ref{subsetstrout}, the equations we want to solve are not all the equations associated to $\rr_{\mu\nu}=0$. Therefore we have to determine under which conditions a solution of the equations \ref{5.37l}, \ref{2.70gql} is a solution of the Einstein equations. Observe that, looking at the structure equations, apart from equations \ref{5.37l}, \ref{2.70gql}, the following equations have to be satisfied by the vacuum Einstein equations: 
\bea
&&\ \ \ \ \ \ \ \ \ \ \ \ \ \ \ \frac{\partial{\ga}}{\partial\nu}-2{\oom}{\chi}=0\ \ ,\ \
\frac{\partial\!\log{\oom}}{\partial\nu}+{2{\oom}}{\om}=0\nn\\ 
&&\ML\ML\ \ \ \mbox{${\bf R}(e_4,e_A)\!=\!0$ :}\ \
\frac{\partial{\ze}}{\partial\nu}+{\oom}\
\!{\tr\chi}{\ze}-{\oom}{\divv}{\chi}+{\oom}{\nabb}{\tr\chi}-2\oom\om\psi-2\oom\nabb\om-{\oom}\psi\!\c\!{\chi}\!=\!0\nn\\
&&\ML\ML\ \ \ \mbox{${\bf R}(e_4,e_4)\!=\!0$ :}\ \
\frac{\partial{\tr\chi}}{\partial\nu}+\frac{{\oom}{\tr\chi}}{2}{\tr\chi}+2{\oom}{\om}{\tr\chi}+{\oom}|\hat{{\chi}}|^2\!=\!0\eql{2.81ff}\\
&&\ML\ML\ \ \ \mbox{${\bf R}(e_3,e_3)\!=\!0$ :}\ \
\frac{\partial{\tr\chib}}{\partial\la}+\frac{{\oom}{\tr\chib}}{2}{\tr\chib}+{\partialb}_{\!X}{\tr\chib}
+2{\oom}{\omb}{\tr\chib}+{\oom}|\hat{{\chib}}|^2\!=\!0\ .\nn
\eea
The first two equations of \ref{2.81ff} are at the ``level" of equations for the metric components and follow from the first set of the structure equations.
This is somewhat analogous to what happens in the maximal foliation gauge used by D.Christodoulou and
S.Klainerman, \cite{C-K:book}, where, once we impose $\tr k\!=\!0$ on
$\Si_0$, one has to prove that $\tr k$ remains equal zero on any $t$-constant hypersurface, justifying the definition of the maximal foliation gauge. In
other words proving that the first two equations, once satisfied on $C_0$, are satisfied on any $C(\la)$ shows that we are in the double null foliation gauge.

\NI The remaining three equations have to be satisfied to make the components of the Ricci tensor, ${\bf R}(e_4,e_A)$, ${\bf R}(e_4,e_4)$ and
${\bf R}(e_3,e_3)$, identically zero. Here it is appropriate to introduce the notion of ``signature" for the various functions involved:\footnote{This
was introduced by D.Christodoulou and S.Klainerman, \cite{C-K:book}, for the null components of the Riemann tensor.}
\begin{Def}
We call ``signature" of the various connection coefficients the number of times the null vector $e_4$ appears in their definition minus the
number of times $e_3$ is present. Each derivative along $C_0$ increases the signature by one and viceversa for each derivative along $\Cb_0$\ .
\end{Def}
Observe that
the last three equations in \ref{2.81ff}\ are at the level of connection coefficients and have signature $+1$, $+2$ and $-2$. This can be
interpreted as the indication that in these equations there are no ``derivatives" with respect to the transverse directions,
$e_3$ for $C(\la)$ and $e_4$ for $\Cb(\nu)$. These equations have, therefore, to be seen as constraint equations and we have to prove that, if satisfied
from the initial data on $C_0$ and $\Cb_0$, they are satisfied on each outgoing or incoming cone, respectively.\footnote{Observe
that the effect of a coordinate choice and of the choice of the system of equations make the set of equations \ref{2.81ff} asymmetric with
respect to the $\la$, $\nu$ interchange. It is also easy to see that we have a certain arbitrariness in choosing the first order system, for
instance one could interchange the role of the ``$\nu$" and ``$\la$" directions.}
This is the content of the following lemma which connects the solutions of \ref{5.37l}, \ref{2.70gql} to the solutions of the Einstein equations, 
\begin{Le}\label{L2.1}
Let $\Psi=(\ga_{ab},\log\oom, X_a,v_{c,ab},w_{ba},\chi_{ab},\ze_a,\om,\chib_{ab},\omb)$ be a solution of the first order
system made by equations \ref{5.37l}, \ref{2.70gql}.

\NI If \ $\{\ga_{ab},\log\oom,\tr\chi,\ze\}$ are a solution of the first four equations of \ref{2.81ff} on $C_0$ and $\tr\chib$ is a
solution of the last one on $\Cb_0$, it follows that they are solutions of the same equations on any cone $C(\la)$ and
$\Cb(\nu)$ respectively.                
\end{Le}
{\bf Proof:} We show that there exist first order transport equations along $\la$ for the left hand sides of the first
four equations of \ref{2.81ff} and a transport equation along $\nu$ for the left hand side of the last equation. Therefore, if these expressions
are zero on $C_0$ or on $\Cb_0$, they are identically zero for all $\la,\nu$ values. A way to obtain these transport equations is just a
long computation using equations \ref{5.37l}, \ref{2.70gql}. We write only the proof for the second equation of the first line of
\ref{2.81ff}. 
\bea
\frac{\partial}{\partial\la}\!\left(\!\frac{\partial\!\log{\oom}}{\partial\nu}+{2{\oom}}{\om}\!\right)
=\frac{\partial}{\partial\nu}\!\left(\frac{\partial\!\log{\oom}}{\partial\la}\right)+2\!\left(\frac{\partial\!\log{\oom}}{\partial\la}\right)\!\oom\om
+2\oom\frac{\partial\om}{\partial\la}\eql{3.23a}
\eea
Using again equations \ref{5.37l}, \ref{2.70gql}, the terms in the right hand side become
\bea
&&\frac{\partial}{\partial\nu}\!\left(\frac{\partial\!\log{\oom}}{\partial\la}\right)
=\frac{\partial}{\partial\nu}\left(-2\oom\omb-\partial_X\!\log\oom\right)\nn\\
&&=-2\!\left(\frac{\partial}{\partial\nu}\log\oom\right)\!\oom\omb-2\oom\frac{\partial\omb}{\partial\nu}
-\frac{\partial X^a}{\partial\nu}\partial_a\!\log\oom-\partial_X\!\!\left(\frac{\partial\log\oom}{\partial\nu}\right)\\
&&2\!\left(\frac{\partial\!\log{\oom}}{\partial\la}\right)\!\oom\om=2\oom\om\!\left(-2\oom\omb-\partial_X\!\log\oom\right)
=-4\oom^2\om\omb-2\oom\om\partial_X\!\log\oom\ .\nn 
\eea
Substituting in \ref{3.23a} and denoting ${\cal I}=\left(\frac{\partial\!\log{\oom}}{\partial\nu}+{2{\oom}}{\om}\right)$  we obtain
\bea
\frac{\partial{\cal I}}{\partial\la}=-\nabb_X{\cal I}-2\oom\omb{\cal I}
+2\oom\left[\left(\frac{\partial\!\om}{\partial\la}+\partial_{\!X}\om\right)-\frac{\partial\omb}{\partial\nu}
-\frac{1}{2\oom}\frac{\partial X^a}{\partial\nu}\partial_a\!\log\oom\right]
\eea
and, using again equations \ref{5.37l}, \ref{2.70gql}, the term in the square bracket is identically zero. Therefore 
$\cal I$ satisfies the following equation
\bea
\frac{\partial{\cal I}}{\partial\la}+\nabb_X{\cal I}+2\oom\omb{\cal I}=0
\eea
which implies that $\cal I$ is equal to zero on every outgoing cone $C(\la)$ provided it is set equal zero on $C_0$.
\smallskip

\NI To complete the proof of the lemma at the level of the metric components, let us consider the first equation of \ref{2.81ff}.
Proceeding as before and denoting
\[{\cal I}_{ab}=\frac{\partial\ga_{ab}}{\partial\nu}-2\oom\chi_{ab}\ ,\]
a long, but straightforward computation shows that ${\cal I}_{ab}$ satisfies the equation:
\bea
\frac{\partial{\cal I}_{ab}}{\partial\la}-\oom\tr\chib{\cal I}_{ab}+\Lie_X{\cal I}_{ab}=0\eql{3.26c}
\eea
which implies again that if $\frac{\partial\ga_{ab}}{\partial\nu}-2\oom\chi_{ab}=0$ on $C_0$ then this relation holds on any $C(\la)$.
To complete Lemma \ref{L2.1} the same result has to be proved for the remaining equations in \ref{2.81ff}. If we proceed as before the
computation will turn out very long and laborious. This can be avoided observing that this result 
follows by a straitghforward application of the Bianchi equations. In fact let us consider the Lorentzian manifold with metric
\beaa
\ggg(\c,\c)=|X|^2d\la^2-2\oom^2(d\la d\nu+d\nu d\la)-X_a(d\la d\om^a+d\om^ad\la)+\ga_{ab}d\om^ad\om^b,\ \ \ \ 
\eeaa
where $X,\oom,\ga$ satisfies equations \ref{5.37l}, \ref{2.70gql}. From these equations and the result just stated it follows that the components of
$\Psi$, $\{\chi_{ab},\ze_a,\om,\chib_{ab},\omb\}$, can be interpreted as the connection coefficients associated
to this metric. Therefore, as they satisfy equations \ref{5.37l}, \ref{2.70gql} it follows that, see equations \ref{2.1ff}, the 
null components ${\bf R}(e_A,e_B), {\bf R}(e_3,e_4)$ and ${\bf R}(e_3,e_A)$ of the Ricci tensor are identically zero. To prove the remaining part of Lemma
\ref{L2.1} amounts to prove that also ${\bf R}(e_4,e_4)$, ${\bf R}(e_A,e_4)$ and ${\bf R}(e_3,e_3)$ are identically zero, provided
they are equal to zero on the initial hypersurface. To prove this result  we use the
contracted Bianchi equations. In fact from them one  deduces the following identities
\bea
D^{\mu}R_{\mu\nu}-\frac{1}{2}D_{\nu}R=0\ . \eql{3.27t}
\eea
Denoting $\{e_3,e_4,e_A\}$, $A\in\{1,2\}$, a null orthonormal frame and writing
\[g^{\mu\nu}=-\frac{1}{2}(e_3^{\mu}e_4^{\nu}+e_4^{\mu}e_3^{\nu})+\sum_Ae_A^{\mu}e_A^{\nu}\ ,\]
equation \ref{3.27t} can be written, mutiplying it with $e_4$ and $e_B$ respectively,
\bea
&&\ML-\frac{1}{2}(D_3R_{\mu\nu})e_4^{\mu}e_4^{\nu}+\sum_A(D_AR_{\mu\nu})e_A^{\mu}e_4^{\nu}-\frac{1}{2}(D_4R_{\mu\nu})e_A^{\mu}e_A^{\nu}=0\nn\\
&&\ML-\frac{1}{2}(D_4R_{\mu\nu})e_3^{\mu}e_B^{\nu}-\frac{1}{2}(D_3R_{\mu\nu})e_4^{\mu}e_B^{\nu}+\sum_A(D_AR_{\mu\nu})e_A^{\mu}e_B^{\nu}+\frac{1}{2}(D_BR_{\mu\nu})e_4^{\mu}e_3^{\nu}\nn\\
&&\ML-\sum_A\frac{1}{2}(D_BR_{\mu\nu})e_A^{\mu}e_A^{\nu}=0\ .
\ \eql{3.28t}
\eea
Rewriting these equations as transport equations for the various null Ricci components, from the first set of structure equations for the null frame, see for instance \cite{Kl-Ni:book}, Chapter 3:
\beaa
\dd_{A}e_B&=&\nabb_{A}e_B+\frac{1}{2}\chi_{AB}e_3+\frac{1}{2}\chib_{AB}e_4\nn\\
\dd_{A}e_3&=&\chib_{AB}e_B+\ze_{A}e_3\ \ \ ,\ \ \ \dd_{A}e_4=\chi_{AB}e_B-\ze_{A}e_4\nn\\
\dd_{3}e_A&=&\dddd_3 e_A+\eta_{A}e_3\ \ \ ,\ \ \ \dd_{4}e_A=\dddd_4 e_A+\etab_{A}e_4\eql{struct1}\\
\dd_{3}e_3&=&(\dd_3\log\oom)e_3\ \ \ ,\ \ \ \dd_{3}e_4=-(\dd_3\log\oom)e_4+2\eta_{B}e_{B}\nn\\
\dd_{4}e_4&=&(\dd_4\log\oom)e_4\ \ \ ,\ \ \ \dd_{4}e_3=-(\dd_4\log\oom)e_3+2\etab_{B}e_{B}\nn
\eeaa
and recalling that all the null Ricci components with signature $-1$ and $0$ are already equal to zero, equations \ref{3.28t} become:
\bea
&&-\left(\!\frac{\partial}{\partial\la}+\partial_X\!\right)\!{\bf R}(e_4,e_4)-\tr\chib{\bf R}(e_4,e_4)+4\omb {\bf R}(e_4,e_4)+2\nabb_A{\bf R}(e_A,e_4)\nn\\
&&\ \ \ -2\big(\eta_B-g(\nabb_Ae_A,e_B)+\nabb_B\log\oom\big){\bf R}(e_B,e_4)=0\eql{3.29t}\\
&&-\left(\!\frac{\partial}{\partial\la}+\partial_X\!\right)\!{\bf R}(e_B,e_4)-\tr\chib{\bf R}(e_A,e_4)-\chib(e_A,e_B){\bf R}(e_B,e_4)-4\omb{\bf R}(e_B,e_4)\nn\\
&&\ \ \ +g(\ddb_3e_A,e_B){\bf R}(e_B,e_4)=0\ .\nn
\eea
From the second equation and the assumed initial conditions it follows that ${\bf R}(e_B,e_4)=0$, which, substituted in the first equation, implies that
${\bf R}(e_4,e_4)=0$. The proof that ${\bf R}(e_3,e_3)=0$ goes in the same way, is somewhat simpler and we do not report here. Once Lemma \ref{L2.1} is proved it follows that $\Psi$ is a solution of the vacuum Einstein equations. 
\smallskip

\NI {\bf Remark:} {\em The previous discussion makes cristal clear, in the Einstein equations characteristic problem,  which are the  equations we have to consider as evolution equations and which have to be interpreted as constraint equations, which is enough to satisfy on the ``initial data" to have them satisfied everywhere, The first ones are equations \ref{5.37l}, \ref{2.70gql} while the ``constraint equations" are equations \ref{2.81ff} which do not involve inward (outward) derivatives of the initial data.

\NI One has also to remark that if we consider only the equations \ref{5.37l}, \ref{2.70gql} and we do not care about initial data satisfying \ref{2.81ff}, we are still considering a well defined characteristic problem whose solutions nevertheless do not define an Einstein vacuum spacetime. Nevertheless as this is a characteristic problem in itself, even in this more general case the initial data have to satisfy some constraints, namely the initial data associated to $\{\ga,\oom,v,\psi,\chi,\ze,\om\}$ have to satisfy the constraints prescribed by eqs \ref{5.37l} on $\Cb_0\equiv\Cb(\nu_0)$, while they are given in a free way on $C_0=C(\la_0)$ and the opposite has to be imposed for the initial data of $\{{\hat X},w,\chib,\omb\}$.}
\medskip

\NI In conclusion one has to recognize that, in some sense,  the characteristic problem for the Einstein equations has two kind of constraint equations that the initial data have to satisfy, the first one connected to the more general problem \ref{5.37l}, \ref{2.70gql} and the second one to the requirement that also equations \ref{2.81ff} have to be satisfied.
We summarize this discussion in the following theorem,
\begin{theorem}\label{T3.1}
Let $\Psi=(\ga_{ab},\log\oom, X_a,v_{c,ab},\psi_a,w_{ba},\chi_{ab},\ze_a,\om,\chib_{ab},\omb)$ be a solution, in a region, $\{(\la,\nu)|(\la,\nu)\in
[0,\overline{\la}]\times[0,\overline{\nu}]\}$, of the characteristic first order Cauchy problem made by equations \ref{5.37l}, \ref{2.70gql}
with the initial data on the null hypersurface ${\cal S}=C_0\cup\Cb_0$ satisfying on $\cal S$, beside equations  \ref{5.37l}, \ref{2.70gql}, considered as equations on $\Cb_0$ and $C_0$ respectively, the constraint equations, see \ref{2.81ff}, 
\bea
&&\ML\ML\mbox{\bf On $C_0$\ :}\ \ \ 
\frac{\partial{\ga}}{\partial\nu}-2{\oom}{\chi}=0\ \ ,\ \ \frac{\partial\!\log{\oom}}{\partial\nu}+{2{\oom}}\om=0\nn\\
&&\ \ \ \ \frac{\partial{{\tr}\chi}}{\partial\nu}+\frac{{\oom}{{{tr}}\chi}}{2}{\tr\chi}+2{\oom}{\om}{\tr\chi}+{\oom}|\hat{{\chi}}|^2\!=\!0\nn\\
&&\ \ \ \ \frac{\partial{\ze}}{\partial\nu}+{\oom}\ \!{\tr\chi}{\ze}-{\oom}{\divv}{\chi}+{\oom}{\partialb}\!{\tr\chi}
+\frac{\partial{\partialb}{\log\oom}}{\partial\nu}-{\oom}{\partialb}\!\log{\oom}\!\c\!{\chi}\!=\!0\nn\\
&&\nn\\
&&\ML\ML\mbox{\bf On $\Cb_0$:}\ \ \  \frac{\partial{\tr\chib}}{\partial\la}+\frac{{\oom}{\tr\chib}}{2}{\tr\chib}
+\partial_X\!{\tr\chib}+2{\oom}{\omb}{\tr\chib}+{\oom}|\hat{{\chib}}|^2\!=\!0\ ,
\eea
then in the same region the metric tensor
\bea
\ggg(\c,\c)=|X|^2d\la^2-2\oom^2(d\la d\nu+d\nu d\la)-X_a(d\la d\om^a+d\om^ad\la)+\ga_{ab}d\om^ad\om^b\ \ \ \ \eql{2.84hbm}
\eea
is a solution of the Einstein vacuum equations.
\end{theorem}  
\smallskip

\NI Summarizing we collect here all the constraint equations the initial data have to satisfy, namely equations \ref{5.37l}, \ref{2.70gql} and \ref{2.81ff},
\bea
&&\ML\mbox{\bf On $C_0$\ :}\nn\\
&&\ML\partialb\ga-v=0\ \ ,\ \ \partialb\!X-w=0\ \ ,\ \ \partialb\!\log\oom-\psi=0\nn\\
&&\ML\frac{\partial{\ga}}{\partial\nu}-2{\oom}\chi=0\ \ ,\ \ \frac{\partial{X}}{\partial\nu}+4{\oom}^2{Z}=0\ \ ,\ \
\frac{\partial\!\log{\oom}}{\partial\nu}+{2{\oom}}\om=0\nn\\
&&\ML\frac{\partial{\tr\chi}}{\partial\nu}+\frac{{\oom}{{{tr}}\chi}}{2}{\tr\chi}+2{\oom}{\om}{\tr\chi}+{\oom}|\hat{{\chi}}|^2\!=\!0\nn\\
&&\ML\frac{\partial{\ze}}{\partial\nu}+{\oom}\ \!{\tr\chi}{\ze}-{\oom}{\divv}{\chi}+{\oom}{\partialb}\!{\tr\chi}
+\frac{\partial{\partialb}{\log\oom}}{\partial\nu}-{\oom}{\partialb}\!\log{\oom}\!\c\!{\chi}\!=\!0\eql{5.37la}\\
&&\ML\frac{\partial{\tr\chib}}{\partial\nu}+{\oom}{\tr\chi}{\tr\chib}
-2{\oom}{\om}{\tr\chib}+2{\oom}{\divv}{\ze}-2\oom\lapp\log\oom-2{\oom}|\ze-\partialb\log\oom|^2+2{\oom}{\bf K}\!=\!0\nn\\
&&\ML\frac{\partial\hat{{\chib}}}{\partial\nu}-\frac{{\oom}{\tr\chi}}{2}\hat{{\chib}}
+\frac{{\oom}{\tr\chib}}{2}{\chih}-2\oom\om\chibh-{\oom}(\hat{{\chib}}\c\hat{{\chi}})\ga+{\oom}{\nabb}\hot{\ze}-\oom\nabb\hot\partialb\!\log\oom\nn\\
&&\ML-{\oom}(-\ze+\partialb\!\log\oom)\hot(-\ze+\partialb\!\log\oom)\!=\!0\nn\\
&&\ML\frac{\partial{{\omb}}}{\partial\nu}\!-\!2{\oom}\ \!\!{{\om}}\ \!{{\omb}}\!-\!     
\frac{3}{2}{{\oom}}|{\ze}|^2\!+\!{{\oom}}{\ze}\!\c\!{\partialb}\!\log{\oom}\!+\!\frac{1}{2}{\oom}|\partialb\!\log{\oom}|^2
\!+\!\frac{1}{2}{\oom}\!\left(\!{{\bf K}}\!+\!\frac{1}{4}{\tr\chib}{\tr\chi}\!-\!\frac{1}{2}\hat{{\chib}}\!\c\!\hat{{\chi}}\!\right)\!=\!0\nn
\eea
\bea
&&\ML\mbox{\bf On $\Cb_0$:}\nn\\
&&\ML\frac{\partial{\ga}}{\partial\la}-2{\oom}\ \!{\chib}+\Lie_{X}{\ga}=0\ \ ,\ \
\frac{\partial\log{\oom}}{\partial\la}+\partial_X\!\log\oom+{2{\oom}}\ \!{\omb}=0\nn\\
&&\ML\frac{\partial{\tr\chib}}{\partial\la}+\frac{{\oom}{\tr\chib}}{2}{\tr\chib}
+\partial_X\!{\tr\chib}+2{\oom}{\omb}{\tr\chib}+{\oom}|\hat{{\chib}}|^2\!=\!0\nn\\ 
&&\ML\frac{\partial{\ze}}{\partial\la}+{\oom}\
\!{\tr\chib}{\ze}+{\oom}{\divv}{\chib}-{\oom}{\partialb\!}{\tr\chib}
-\frac{\partial({\partialb\!}{\log\oom})}{\partial\la}+{\oom}{\partialb}{\log\oom}\!\c\!{\chib}+{\Lie_{ X}{\ze}}
-\Lie_{X}{\partialb}{\log\oom}\!=\!0\ \nn\\
&&\ML\frac{\partial{\tr\chi}}{\partial\la}+{\oom}{\tr\chib}{\tr\chi}-2{\oom}{\omb}{\tr\chi}+{\partial}_{\! X}\!{\tr\chi}
-2\oom\divv\ze-2\oom\lapp\log\oom-2{\oom}|\ze+\nabb\oom|^2 +2{\oom}{\bf K}\!=\!0\nn\\
&&\ML\frac{\partial\hat{{\chi}}}{\partial\la}-\frac{{\oom}{\tr\chib}}{2}\hat{{\chi}}+\frac{{\oom}{\tr\chi}}{2}\hat{{\chib}}
+\frac{\partial\log{\oom}}{\partial\la}\hat{{\chi}}+({\partial}_{\! X}\!\log{\oom})\hat{{\chi}} 
-{\oom}(\hat{{\chi}}\c\hat{{\chib}})\ga-{\oom}\ \!{\nabb}\hot{\ze}-{\oom}\ \!{\nabb}\hot\ \!{\partialb}\!\log\oom\nn\\
&&\ML-{\oom}(\ze+{\partialb}\!\log\oom)\hot(\ze+{\partialb}\!\log\oom)+\Lie_{{X}}\hat{{\chi}}\!=\!0\eql{2.84q}\\
&&\ML\frac{\partial{{\om}}}{\partial\la}\!+\!\nabb_X\om-2{\oom}\ \!\!{{\omb}}\ \!{{\om}}\!-\!\frac{3}{8}{{\oom}}|\zeta|^2
\!-{{\oom}}{\ze}\!\c\!{\partialb}\!\log{\oom}\!+\!\frac{1}{2}{\oom}|{\partialb}\!\log\oom|^2
\!+\!\frac{1}{2}{\oom}\!\left({{\bf K}}\!+\!\frac{1}{4}{\tr\chib}{\tr\chi}
\!-\!\frac{1}{2}\hat{{\chib}}\c\hat{{\chi}}\!\right)\!=\!0\ .\nn
\eea
{\bf Remark:}{\em The implementation of the initial conditions, namely the way of obtaining initial data 
satisfying the constraint equations, \ref{5.37la}, \ref{2.84q} with appropriate (Sobolev) regularity and asymptotic behaviour has been discussed in
\cite{Ca-Ni:char}. Here we will have basically to repeat the same argument, but imposing the analiticity, this makes this problem more complicated and how we solve it will be discussed later on. Next section, Section \ref{S3}, is devoted to find a local analytic solution for the system of equations \ref{5.37l}, \ref{2.70gql}. To do it  we rewrite them in a more compact notation and show how, following  G.F.D.Duff, \cite{Duff} and H.Friedrich, \cite{friedrich:cauchychar}, we can apply the Cauchy-Kowalevski theorem to this characteristic case.}

\section{The analytic solution of the characteristic problem 
via the Cauchy-Kowalevski theorem.}\label{S3}
The method we use to obtain a real analytic solution of the characteristic problem defined by the system of equations
\ref{5.37l}, \ref{2.70gql}  with initial data satisfying equations \ref{5.37la}, \ref{2.84q} is a variant of the Cauchy-Kowalevski method. The 
adaptation of the Cauchy-Kowalevski theorem to characteristic problems has been developed by G.F.D.Duff, \cite{Duff}, for the linear case and,
subsequently, by H.Friedrich, \cite{friedrich:cauchychar}, for the non linear problem. Friedrich result is suited to the present case,
therefore we just recall the main lines of the proof, a straightforward adaptation of his result.

\NI The system of equations \ref{5.37l}, \ref{2.70gql} can be written in a much more compact form in the following way:
\bea
&&\frac{\partial {\bf V}}{\partial\la}={\bf F}({\bf V},{\bf W},\partialb{\bf V},\partialb{\bf W})\nn\\
&&\frac{\partial {\bf W}}{\partial\nu}={\bf G}({\bf V},{\bf W},\partialb{\bf V})\eql{4.1dd}
\eea
where ${\bf V}$ and ${\bf W}$ are vector functions valued in $R^{18}$ and $R^{10}$ respectively, defined by
\bea
&&{\bf V}=\{V^s\}=\{\ga,\oom,v,\psi;\om,\ze,\chi\}\ \ ,\ \ {\bf W}=\{W^t\}=\{X,w,\omb,\chib\}\nn\\
&&s\in\{1,...,18\}\ \ ,\ \ t\in\{1,...,10\}\ .
\eea
The initial data are assigned on the union of the null hypersurfaces $C_0$ and $\Cb_0$. They have to be analytic functions satisfying the costraint equations \ref{5.37la}, \ref{2.84q}; this is possible as it is proved later on and has been proved in \cite{Ca-Ni:char} for initial data belonging to a suitable Sobolev space.\footnote{The situation is somewhat simpler here with respect to \cite{Ca-Ni:char} due to the fact that we are considering a local problem and we do not have to worry about the asymptotic behaviour of the initial data, but only require that the initial data be analytic.} Let us denote the  initial data ${\bf V}_0={\bf V}_0(\nu,\om^a),{\bf W}_0={\bf W}_0(\la,\om^a)$ on $C_0$ and $\Cb_0$ respectively.\footnote{The initial data defined here, ${\bf V}_0={\bf V}_0(\nu,\om^a),{\bf W}_0={\bf W}_0(\la,\om^a)$ satisfy the constraint equations \ref{2.81ff}, the remaining ones are automatically obtained using the Cauchy-Kowalevski method.}

\NI The existence of a local real analytic solution of the system \ref{4.1dd} with initial data ${\bf V}_0={\bf V}_0(\nu,\om^a),{\bf W}_0={\bf
W}_0(\la,\om^a)$ is proved in the following theorem:
\begin{theorem}\label{T4.1}
the system of equations \ref{4.1dd} with initial data ${\bf V}_0={\bf V}_0(\nu,\om^a),{\bf W}_0={\bf
W}_0(\la,\om^a)$
 admits a unique real analytic solution in a
region $(\la,\nu)\in [0,\overline{\la}]\times[0,\overline{\nu}]$ whose size is determined by the initial data.
\end{theorem}
{\bf Proof:} 
System \ref{4.1dd} can be rewritten for the new unkown functions
\[({\bf V}-{\bf V}_0)(\la,\nu,\om^a)\ \mbox{and}\ \  ({\bf W}-{\bf W}_0)(\la,\nu,\om^a)\] we denote again ${\bf V}$ and ${\bf W}$. It is easy
to show that it has the form:
\bea
&&\frac{\partial V^s}{\partial\la}=F^{s,a}_{s'}\frac{\partial V^{s'}}{\partial\om^a}+{\tilde F}^{s,a}_{t'}\frac{\partial W^{t'}}{\partial\om^a}+f^s\eql{4.3c}\\
&&\frac{\partial W^t}{\partial\nu}={\tilde G}^{t,a}_{s'}\frac{\partial V^{s'}}{\partial\om^a}+g^t\ \ ,\eql{4.3d}
\eea
with 
$\om^a\in\{\theta,\phi\}$ and we sum over repeated indices. With these new ${\bf V}, {\bf W}$, the initial data are 
\bea
{\bf V_0}=(0,...,0)\ \ ,\ \ {\bf W_0}=(0,...,0)
\eea
and the coefficients $F^{s,a}_{s'},{\tilde F}^{s,a}_{t'},f^s,{\tilde G}^{t,a}_{s'},g^t$ depend, besides ${\bf V}$ and $\bf W$, on the original data ${\bf V_0}$ and ${\bf W_0}$ and through them on the coordinates $\{x^\mu\}$. 

\NI Following Friedrich, \cite{friedrich:cauchychar}, the proof is basically made by two main steps. The first one provides a recursive mechanism to get 
all the derivatives in $\nu,\la,\om^a$ for $V$ and $W$ observing that, as initial data, we have all the derivatives in $\nu,\om^a$ for $V$ and in $\la,\om^a$ for $W$. The remaining mixed derivatives are obtained through the equations \ref{4.3c} and \ref{4.3d}. In the second and more delicate step we prove the convergence  of the formal power series we have obtained. 
\smallskip

\NI {\bf i) the recursive determination of the derivatives:} From equation \ref{4.3d} we control $\pr_{\nu}W$ and $\pr_{\nu}\nabb^qW$ for any $q\geq 0$. From equation \ref{4.3c} we control $\pr_{\la}V$ and $\pr_{\la}\nabb^qV$ for any $q\geq 0$. Deriving equation \ref{4.3c} with respect to $\nu$ we control  $\pr_{\nu}\pr_{\la}\nabb^qV$, deriving equation \ref{4.3d} with respect to $\la$ we control $\pr_{\nu}\pr_{\la}\nabb^qW$. Deriving with respect to $\la$ equation \ref{4.3c} we control $\pr^2_{\la}\nabb^qV$ and deriving with respect to $\nu$ equation \ref{4.3d} we control $\pr^2_{\nu}\nabb^qW$. Iterating the procedure we obtain all the mixed derivatives.
\smallskip

\NI{\bf Remark:} {\em It should be clear that this procedure which allows to obtain formal power series for $V$ and $W$ both on $C_0$ and on $\Cb_0$ satisfies also the constraint equations for $V$ on $\Cb_0$ and $W$ on $C_0$ as discussed in remark before Theorem \ref{T3.1}.}
\medskip

\NI {\bf ii) The convergence of the formal series}
\medskip

\NI The functions $F^{s,a}_{s'},{\tilde F}^{s,a}_{t'},{\tilde G}^{t,a}_{s'},f^s,g^t$ depend on the analytic initial data ${\bf V}_0$ and ${\bf W}_0$ and on the unknown functions $\bf V$ and $\bf W$. More specifically, looking at the explicit expression of system \ref{5.37l}, \ref{2.70gql}, 
they are polynomials in $\bf V$ and $\bf W$ and can be written as
\bea
&&F^{s,a}_{s'}=\sum_{\a\b}F^{s,a}_{s';\a\b}(x^\mu)W^{\a}V^{\b}\ \ ,\ \ {\tilde F}^{s,a}_{t'}=\sum_{\a\b}{\tilde
F}^{s,a}_{t';\a\b}(x^\mu)W^{\a}V^{\b}\nn\\ &&{\tilde G}^{t,a}_{s'}=\sum_{\a\b}{\tilde G}^{t,a}_{s';\a\b}(x^\mu)W^{\a}V^{\b}\eql{4.4zs}\\
&&f^s=\sum_{\a\b}f^s_{\a\b}(x^\mu)W^{\a}V^{\b}\ \ ,\ \ g^t=\sum_{\a\b}g^t_{\a\b}(x^\mu)W^{\a}V^{\b}\ ,\nn
\eea
with
\bea
&&|\a|=0\ \ ,\ \ |\b|\leq 2\ \ \mbox{for}\ \ F\ \ ,\ \ |\a|=0\ \ ,\ \ |\b|\leq 1\ \ \mbox{for}\ \ \tilde{F}\nn\\
&&|\a|=0\ \ ,\ \ |\b|\leq 2\ \ \mbox{for}\ \ \tilde{G}\eql{4.5r}\\
&&|\a|\leq 1\ \ ,\ \ |\b|\leq 3\ \ \mbox{for}\ \ f\ \ ,\ \ |\a|\leq 1\ \ ,\ \ |\b|\leq 3\ \ \mbox{for}\ \ g\ \ .\nn
\eea
As we assumed that the initial data ${\bf V}_0,{\bf W}_0$ are real analytic, the functions $F^{s,a}_{s';\a\b}(x^\mu),{\tilde F}^{s,a}_{t';\a\b}(x^\mu),{\tilde G}^{t,a}_{s';\a\b}(x^\mu),f^s_{\a\b}(x^\mu),g^t_{\a\b}(x^\mu)$ are real analytic\footnote{$\{x^{\mu}\}=\{x^1,x^2,x^3,x^4\}=\{\la,\nu,\om^a\}$.} and can be written as power
series in $\{x^{\mu}\}$,
\bea
&&F^{s,a}_{s';\a\b}(x)=\sum_{\ga}F^{s,a}_{s';\a\b\ga}x^{\ga}\ \ ,\ \ {\tilde F}^{s,a}_{t';\a\b}(x)=\sum_{\ga}{\tilde F}^{s,a}_{t';\a\b\ga}x^{\ga}\nn\\
&&{\tilde G}^{t,a}_{s';\a\b}(x)=\sum_{\ga}{\tilde G}^{t,a}_{s';\a\b\ga}x^{\ga}\eql{4.6y}\\
&&f^s_{\a\b}(x)=\sum_{\ga}f^s_{\a\b\ga}x^{\ga}\ \ ,\ \ g^t_{\a\b}(x)=\sum_{\ga}g^t_{\a\b\ga}x^{\ga}\nn .
\eea
Due to the real analyticity we can assume that their convergence radius be $> R$,\footnote{As we are concerned about a local solution we consider a compact portion of $C_0\cup\Cb_0$.} for a given $R>0$ and that in
$B_R(0)\subset R^4$ 
the coefficients of the expansions \ref{4.6y} satisfy
\bea
|F^{s,a}_{s';\a\b\ga}|\ ,\  |{\tilde F}^{s,a}_{t';\a\b\ga}|\ ,\  |{\tilde G}^{t,a}_{s';\a\b\ga}|\ ,\  |f^s_{\a\b\ga}|\ ,\  |g^t_{\a\b\ga}|\leq
\frac{M}{R^{|\ga|}}\ \ .
\eea
Let us define the function $H(\theta_{\mu}x^{\mu})=H(y)$, 
with $\theta_1>1,\theta_2=\theta_3=\theta_4=1$:
\bea
H(\theta_{\mu}x^{\mu})\equiv\frac{M}{1-\frac{\theta_{\mu}x^{\mu}}{R}}=\frac{MR}{R-\theta_{\mu}x^{\mu}}=
\sum_{\ga}\frac{M}{R^{|\ga|}}\frac{|\ga|!}{\ga_1!\ga_2!\ga_3!\ga_4!}\theta_1^{\ga_1}x^{\ga}=\sum_{\ga}H_{\ga}x^{\ga}\ .\ \ \ \ 
\eea
It follows from the previous definitions that
\bea
|F^{s,a}_{s';\a\b\ga}|\ ,\  |{\tilde F}^{s,a}_{t';\a\b\ga}|\ ,\  |{\tilde G}^{t,a}_{s';\a\b\ga}|\ ,\  |f^s_{\a\b\ga}|\ ,\  |g^t_{\a\b\ga}|\leq
H_{\ga}
\eea
which means that the function $H(\theta_{\mu}x^{\mu})$ majorizes the functions $F^{s,a}_{s';\a\b}(x)$, ${\tilde F}^{s,a}_{t';\a\b}(x)$, ${\tilde
G}^{t,a}_{s';\a\b}(x)$ ,$f^s_{\a\b}(x)$ , $g^t_{\a\b}(x)$. We define now the functions
$\hat{F}^{s,a}_{s'}\ ,\ \hat{{\tilde F}}{}^{s,a}_{t'}\ ,\ \hat{\tilde G}{}^{t,a}_{s'}\ \ ,\ \  \hat{f}^s\ \ ,\ \ \hat{g}^t$ as the power
series \ref{4.4zs} with the coefficient functions $F^{s,a}_{s';\a\b}(x^\mu)...$ substituted by the function $H(\theta_{\mu}x^{\mu})$ times a  matrix
$P^{s,a}_{s';\a\b}...$ with non negative real coefficients and which defines the same polynomial $r({\bf V},{\bf W})=\sum_{\a\b}{\tilde
p}_{\a\b}W^{\a}V^{\b}$, ${\tilde p}_{\a\b}>0$  , $|\a|\leq 1\ ,\ |\b|\leq 3$, for all indices $s,s',t,a$.
Therefore\footnote{The matrix elements ${P}^{s,a}_{s'}, {\tilde P}^{s,a}_{t'}, {\tilde Q}^{t,a}_{s'},{p}^s,{q}^t$ are all zero or one.}
\bea
&&\hat{F}^{s,a}_{s'}=H(\theta_{\mu}x^{\mu})\sum_{\a\b}P^{s,a}_{s';\a\b}W^{\a}V^{\b}=H(\theta_{\mu}x^{\mu})r({\bf V},{\bf W}){P}^{s,a}_{s'}\nn\\
&&\hat{\tilde F}{}^{s,a}_{t'}=H(\theta_{\mu}x^{\mu})\sum_{\a\b}{\tilde P}^{s,a}_{t';\a\b}W^{\a}V^{\b}=H(\theta_{\mu}x^{\mu})r({\bf V},{\bf W}){\tilde P}^{s,a}_{t'}\nn\\
&&\hat{\tilde G}{}^{t,a}_{s'}=H(\theta_{\mu}x^{\mu})\sum_{\a\b}{\tilde Q}^{t,a}_{s';\a\b}W^{\a}V^{\b}
=H(\theta_{\mu}x^{\mu})r({\bf V},{\bf W}){\tilde Q}^{t,a}_{s'}\eql{4.4zz}\\
&&\hat{f}^s=H(\theta_{\mu}x^{\mu})\sum_{\a\b}{p}^s_{\a\b}W^{\a}V^{\b}=H(\theta_{\mu}x^{\mu})r({\bf V},{\bf W}){p}^s\nn\\
&&\hat{g}^t=H(\theta_{\mu}x^{\mu})\sum_{\a\b}{q}^t_{\a\b}W^{\a}V^{\b}=H(\theta_{\mu}x^{\mu})r({\bf V},{\bf W}){q}^t\ .\nn
\eea
Let us consider the following system of equations:
\bea
&&\frac{\partial \hat{V}^s}{\partial\la}=\hat{F}^{s,a}_{s'}\frac{\partial \hat{V}^{s'}}{\partial\om^a}
+\hat{\tilde F}{}^{s,a}_{t'}\frac{\partial\hat{W}^{t'}}{\partial\om^a}+\hat{f}^s(x,\hat{V},\hat{W})\nn\\
&&\frac{\partial \hat{W}^t}{\partial\nu}=\hat{\tilde G}{}^{t,a}_{s'}\frac{\partial\hat{V}^{s'}}{\partial\om^a}
+\hat{g}^t(x,\hat{V},\hat{W})\ \ ,\eql{4.12dw}
\eea 
the following lemma holds:
\begin{Le}\label{L3.1w}
The solution $\hat{\bf V},\hat{\bf W}$ of system \ref{4.12dw} with initial conditions $\hat{\bf V}_0=\hat{\bf W}_0$ majorizing the initial conditions
${\bf V}_0={\bf W}_0=0$, is majorizing $\bf V$ and $\bf W$, solution of system  \ref{4.3c} \ref{4.3d} with initial conditions ${\bf V}_0={\bf W}_0=0$.
\end{Le}
\NI{\bf Proof:} The first step is to write formal power expansions for $\bf V$ and $\bf W$; {This is a formal expansion around a generic point of $S_0\equiv S(\la_0,\nu_0)$; redefining $\la$ and $\nu$ as $\la-\la_0$, $\nu-\nu_0$ we consider it as an expansion around $(\la,\nu)=(0,0)$. The convergence of the formal expansion can be repeated for all the angular coordinates $\{\om^a\}$ of the points of $S_0$ so that, finally, we obtain an analytic solution in a neighbourhood of $S_0$.} Therefore, redefining also $\om^a$ as $(\om^a-\om_0^a)$ we have:
\bea
\ML\ML V^s(x)=\sum_{j,k,\underline{a}}v^s_{jk\underline{a}}\la^j\nu^k\om_1^{a_1}\om_2^{a_2}\
\ ,\ \ \ W^t(x)=\sum_{j,k,\underline{a}}w^t_{jk\underline{a}}\la^j\nu^k\om_1^{a_1}\om_2^{a_2}\eql{4.13s}
\eea
Observe that the initial conditions imply $v^s_{0k\underline{a}}=w^t_{j0\underline{a}}=0$ and that, plugging the formal expansions in  \ref{4.3c}, \ref{4.3d}, we obtain recursively, as explained before, all the coefficients $v^s_{jk\underline{a}}$ and $w^t_{jk\underline{a}}$. We proceed in the same way
expanding the solution $\hat{\bf V},\hat{\bf W}$ of system \ref{4.12dw},
\bea
\ML\ML \hat{V}^s(x)=\sum_{j,k,\underline{a}}\hat{v}^s_{jk\underline{a}}\la^j\nu^k\om_1^{a_1}\om_2^{a_2}\
\ ,\ \ \ \hat{W}^t(x)=\sum_{j,k,\underline{a}}\hat{w}^t_{jk\underline{a}}\la^j\nu^k\om_1^{a_1}\om_2^{a_2}\eql{4.14w} 
\eea
and we observe that, due to the definition of the functions $\hat{F}^{s,a}_{s'},\hat{\tilde F}{}^{s,a}_{t'},\hat{\tilde
G}{}^{t,a}_{s'},\hat{f}^s,\hat{g}^t$, \ref{4.4zz}, it follows that $\hat{v}^s_{jk\underline{a}},\hat{w}^t_{jk\underline{a}}\geq 0$ and that the formal solution $(\hat{\bf V},\hat{\bf W})$, \ref{4.14w}, majorizes the (formal) solution $({\bf V},{\bf W})$, \ref{4.13s}, namely the following inequalities hold, for all index values,
\bea
\hat{v}^s_{jk\underline{a}}\geq |{v}^s_{jk\underline{a}}|\ \ ,\ \ \hat{w}^t_{jk\underline{a}}\geq |{w}^t_{jk\underline{a}}|  \ .
\eea
The proof is achieved if we can find a real analytic solution of the system \ref{4.12dw} with initial conditions such that the
coefficients of its power expansion satisfy $\hat{v}^s_{0k\underline{a}}\geq 0\ ,\hat{w}^t_{j0\underline{a}}\geq 0$; in this case, for an
appropriate radius $R$, the expansions \ref{4.13s} describe a real analytic solution of system
\ref{4.3d} with initial conditions ${\bf V}_0={\bf W}_0=0$. 

\NI To find a solution of the system \ref{4.12dw} with initial conditions $\hat{\bf V}_0=\hat{\bf W}_0$ such that
\bea
\hat{v}^s_{0k\underline{a}}\geq 0\ ,\hat{w}^t_{j0\underline{a}}\geq 0\ ,\eql{4.16t}
\eea
we make the following ansatz:
\bea
&&\hat{V}^s(x^{\mu})=\tilde{V}^s(y=\theta_{\mu}x^{\mu})=\tilde{V}(y)\nn\\
&&\hat{W}^t(x^{\mu})=\tilde{W}^t(y=\theta_{\mu}x^{\mu})=\tilde{W}(y)\ .\eql{4.16y}
\eea
Each equation for $\hat{V}^s$ and $\hat{W}^t$ of system \ref{4.12dw} becomes in terms of $\tilde{V}$ and $\tilde{W}$, denoting \[\tilde{V}'=\frac{\partial \tilde{V}}{\partial y}\ ,\  
\tilde{W}'=\frac{\partial \tilde{W}}{\partial y}\ ,\]
\bea
&&\theta_1\tilde{V}'=H(y)q(\tilde{V},\tilde{W})(36\tilde{V}'+20\tilde{W}'+c)\nn\\
&&\tilde{W}'=H(y)q(\tilde{V},\tilde{W})(36\tilde{V}'+1)
\eea
where $q(\tilde{V},\tilde{W})$ is a polynomial in $\tilde{V},\tilde{W}$ with positive coefficients; therefore
\bea
&&(\theta_1-36Hq-20\c36(Hq)^2)\tilde{V}'=Hq(20Hq+1)\nn\\
&&\tilde{W}'=Hq+36Hq\frac{Hq(20Hq+c1)}{(\theta_1-36Hq-20\c36(Hq)^2)}\nn\\
&&\ \ \ \ \ =\frac{1}{(\theta_1-36Hq-20\c36(Hq)^2)}\left[\theta_1Hq\right]
\eea
and the two equations can be rewritten as
\bea
&&(\theta_1-36Hq-20\c36(Hq)^2)\tilde{V}'=Hq(20Hq+1)\nn\\
&&(\theta_1-36Hq-20\c36(Hq)^2)\tilde{W}'=\theta_1Hq\ .
\eea
Initial conditions for $\hat{\bf V}_0=\hat{\bf W}_0$ such that \ref{4.16t} holds, are satisfied if we require as initial conditions for $\tilde{V}$
and $\tilde{W}$
\bea
\tilde{V}(0)=0\ \ ,\ \ \tilde{W}(0)=0\ , \eql{4.21}
\eea
therefore we are reduced to prove that the system of equations
\bea
&&(\theta_1-36Hq-20\c36(Hq)^2)\tilde{V}'=Hq(20Hq+1)\nn\\
&&(\theta_1-36Hq-20\c36(Hq)^2)\tilde{W}'=\theta_1Hq\ .\eql{4.22}
\eea
with initial conditions \ref{4.21} has a real analytic solution in a neighborhood of the origin. Choosing $\theta_1$ such that
\bea
\theta_1-36H(0)q(0,0)-20\c36(H(0)q(0,0))^2>0
\eea
equations \ref{4.22} can be written in the form
\bea
&&\tilde{V}'=H(y,\tilde{V},\tilde{W})\nn\\
&&\tilde{W}'=J(y,\tilde{V},\tilde{W})
\eea
where $H$ and $J$ are real analytic functions having power series expansions at $y=0,\tilde{V}=0,\tilde{W}=0$ with non negative expansion
coefficients. Therefore a real analytic solution of the system \ref{4.12dw} exists and there is a neighborhood of $S_0=S(\la_0,\nu_0)$
where the real analytic solution of the system \ref{4.3d} does exists. This completes the analytic part of our result giving a concrete meaning
to Theorem \ref{T3.1}.
\medskip

\NI{\bf Remark:} {\em Recall that the local analytic solution whose proof has been sketched now is a solution for ${\bf V}-{\bf V}_0, {\bf W}-{\bf W}_0$. Then ${\bf V},{\bf W}$ are a local solution of the Einstein equations provided ${\bf V}_0$ and ${\bf W}_0$ are analytic functions satisfying the initial data constraints \ref{5.37la}, \ref{2.84q}. }
\section{A common region of existence for the real analytic solutions of the non linear characteristic problem.}\label{S4}

We look for an ``Alhinac type" result, \cite{S.Ali}, proving that, if some Sobolev norms, $H^s$, of a real analytic solution are controlled up to a certain $s$,\footnote{For our problem they are explicitely defined in Subsection \ref{SS4.2}.} then its ``Cauchy-Kowalevski existence region" can be extended to a larger region whose size depends only on these Sobolev norms. This kind of results have been started by Lax \cite{Lax}, extended by Nirenberg \cite{Nir} and proved, in the main lines, for the Burger equation by S.Klainerman and one of the authors (F.N.) \cite{Kl-Ni:rew}. These results requires the hyperbolicity of the partial differential equations we are considering which is evident in the case of the Burger equation, but more hidden in the case of the Einstein equations.\footnote{The  hyperbolicity for the Einstein equations is explicit in the harmonic gauge, in a more general setting, for instance in the geometric gauge we are considering here, it expresses itself in the existence of a-priori estimates for some energy-type norms, see also \cite{C-K:book}, \cite{Kl-Ni:book}.}

\NI It is appropriate, before discussing our result for the Einstein equations, to look in a detailed way what has been obtained in the case of the Burger equation,\cite{Kl-Ni:rew}.
\smallskip

\subsection{A summary of the analytic extended solution approach to the Burger equation in \cite{Kl-Ni:rew}.}\label{SS4.1}
We give a complete survey of the various steps of this approach in the case of the Burger equation
\bea
\frac{\pr u}{\pr t}+u\frac{\pr u}{\pr x}=0\ .\eql{Burg}
\eea
\smallskip

\NI{\bf Step 1:} We introduce a Banach space $B_{\a,\ro}$ defined by the norm
\bea
||f(\cdot,t)||_{B_{\a,\ro}}=\sum_{n=0}^{\infty}\frac{||D^nf(\cdot,t)||_{L^2}}{n!}n^{\a}\ro^n\ ,
\eea
where $L^2=L^2(R^{n_0})$ and, defining the multiindex $\b=(\b_1,\b_2,...,\b_{n_0})$,
\bea
\frac{||D^nf(\cdot,t)||_{L^2}}{n!}\equiv \sum_{\b;|\b|=n}\frac{||D^\b f(\cdot,t)||_{L^2}}{\b!}\eql{Multi1}
\eea
In the Burger equation case, $n_0=1$, but we keep $n_0\geq 1$ as we are interested to extend this result to a more general case. 
We assume that the initial data of  the hyperbolic equation we are considering are given by an analytic function $u^{(0)}$ which belongs to $B_{\a,\ro_0}\ $ for a certain $\a>2$ and $\ro_0>0$.
\smallskip

\NI Next lemma proves that if   $f(\c,t)\in B_{\a,\ro}$ then, as a function of the ``spatial" variables $\{x^i\}$, $f(\c,t)$ is real analytic in $R^{n_0}$.
\begin{Le}\label{L4.1}
Let $f$ be a function belonging to the Banach space $B_{\a,\ro}$ with \[\a\geq\left[\frac{n_0}{2}\right]\ ,\]
then $f(\c,t)$ is real analytic in $B(x)_{\ro}$.
\end{Le}
{\bf Proof:} If $||f(\cdot,t)||_{B_{\a,\ro}}<\infty$ this implies that there exists 
a numerical sequence $\{h_n\}$, depending on $f$, such that, for any $n>0$, we have
\[{||D^nf(\cdot,t)||_{L^2}}\leq \frac{h_nn!}{n^\a}\frac{1}{\ro^n},\]
where $\{h_n\}$ is such that $h_n\geq 0$ and
\bea
\sum_{n=0}^\infty h_n<\infty\ .
\eea
Therefore if $f$ is a function $\in R^{n_0}$ then by Sobolev Lemma, for $|\b|=n$, 
\bea
&&\ML\ML\sup_{x\in R^{n_0}}|D^\b f(x,t)|\leq c_0\sum_{k=n}^{n+1+[\frac{n_0}{2}]}\sum_{\b;|\b|=k}||D^\b f(\cdot,t)||_{L^2}\nn\\
&&\ML\ML
\leq c_0\sum_{k=n}^{n+1+[\frac{n_0}{2}]}h_k\frac{k!}{k^\a}\frac{1}{\ro^k}
\leq c_0\frac{n!}{\ro^n}\left[\sum_{k=n}^{n+1+[\frac{n_0}{2}]}\frac{k!}{n!}\frac{h_k}{k^\a}\frac{1}{\ro^{k-n}}\right]\\
&&\ML\ML
\leq c_0\frac{n!}{\ro^n}{\overline h}_n\frac{\left[{(n+1)...(n+1+[\frac{n_0}{2}])}\right]}{n^\a}\nn\\
&&\ML\ML\leq c_0{\overline h}_n\frac{n!}{\ro^n}\frac{n^{([\frac{n_0}{2}])}}{n^\a}
\leq c_0{\overline h}_n\frac{n!}{\ro^n}\frac{1}{n^{\a-[\frac{n_0}{2}]}} \ .\nn
\eea
where
${\overline h}_n=\sum_{k=n}^{n+1+[\frac{n_0}{2}]}h_k$ and, if $\ro<1$, $c_0$ depends also on $\ro$. Therefore 
\bea
\sup_{x\in R^{n_0}}|D^\b f(x,t)|\leq c_0(n_0,\ro)\frac{{\overline h}_n}{n^{(\a-[\frac{n_0}{2}])}}\frac{n!}{\ro^n}\ .\eql{Sob1}
\eea
Given an arbitrary $x\in R^{n_0}$ and $y\in B(x)_{\ro}$  the formal power series
\bea
\sum_{n=0}^\infty\sum_{\b; |\b|=n}\frac{D^\b f(x,t)}{\b!}(y-x)^\b\ \ ,
\eea 
is convergent implying that in $B(x)_{\ro}$,  $f(\c,t)$ is a real analytic function. In fact
\bea
&&\bigg|\sum_{n=0}^\infty\sum_{\b; |\b|=n}\frac{D^\b f(x,t)}{\b!}(y-x)^\b\bigg|\leq\sum_{n=0}^\infty\sum_{\b; |\b|=n}\frac{|D^\b f(x,t)|}{\b!}|(y-x)^\b|\nn\\
&&\leq c_0(n_0,\ro)\sum_{n=0}^\infty \frac{{\overline h}_n}{n^{(\a-[\frac{n_0}{2}])}}\sum_{\b; |\b|=n}\frac{n!}{\b_1!\c\c\b_{n_0}!}\frac{\Pi_{i=1}^{n_0}|y^i-x^i|^{\b_i}}{\ro^n}\nn\\
&&\leq  c_0(n_0,\ro)\sum_{n=0}^\infty\frac{{\overline h}_n}{n^{(\a-[\frac{n_0}{2}])}}<\infty\ .
\eea
As the choice of $x$ was arbitrary and the estimate \ref{Sob1} is uniform in $x$ it follows immediately that by unique continuation $f(\c, t)$ is real analytic in $R^{n_0}$. In this case the neighborhoods $B(x)_{\ro}$ where the function can be expanded around $x$ have also a common radius which is a stronger property.

\NI The previous estimate can be done in a better way we use later on, writing
\bea
||D^nf(\cdot,t)||_{L^2}=\sum_{j=1}^\infty||D^nf(\cdot,t)||_{L^2(Q(x_j))}
\eea 
where $R^{n_0}=\cup_{j=1}^\infty Q(x_j)$, where $Q(x_j)$ is a finite size square with ``center" $x_j$. The following estimates hold
\bea
||D^nf(\cdot,t)||_{L^2(Q(x_j))}\leq \frac{h_{n,j}n!}{n^\a}\frac{1}{\ro^n}
\eea
with $h_{n,j}>0$ such that
\bea
\sum_{n=0}^{\infty}\sum_{j=1}^{\infty}h_{n,j}=\sum_{j=1}^{\infty}\sum_{n=0}^{\infty}h_{n,j}<\infty\ .
\eea
Repeating the previous argument we can conclude that if $f(\c,t)\in B_{\a,\ro}$ then
\bea
\sup_{x\in Q(x_j)}|D^\b f(x,t)|\leq c_1C{\overline h}_{n,j}\frac{n!}{\ro^n}\ .
\eea
Next lemma says just the opposite of the previous one and its proof is immediate,
\begin{Le}\label{L4.2}
Let $f(t,x)$ be a real analytic function in a strip around the real axis of uniform height $2\ro$, then for any $T_a<\ro$ $f(t,\c)$ belongs to the Banach space $B_{\a,\ro'}$ with appropriate $\ro'<\ro$ and $\a$.
\end{Le}
\medskip

\NI{\bf Step 2:} The following theorem holds:\footnote{Although this theorem is valid for a larger class of partial differential equations, we look at its result applied to an evolution equation of hyperbolic type.}
\begin{theorem}\label{4.1a} 
Let us consider a p.d.e. whose coefficients are real analytic functions, let $u^{(0)}\in B_{\a,\ro_0}$ be the initial data defined on $R^{n_0}$;  given $\ro\leq \ro_0$ there exists a time $T_a(\ro,\ro_0,u^{(0)})$ and a solution $u(t,x)\in C^1([0,T_a],B_{\a,\ro})$ satisfying the initial conditions. 
\end{theorem} 
\NI This result follows more or less immediately recalling how Cauchy-Kowalevski theorem works; given (real) analytic initial data, using the p.d.e. equation it is possible to control all the derivatives in space and time of the (formal) series which defines the function $u(t,x)$, analytic solution of our equation with initial data $u^{(0)}$ provided the series is convergent. Once this is proved the solution $u(t,x)$ is real analytic with a convergence radius $\ro$, a priori depending on the point $x\in R^{\n_0}$; If $u^{(0)}$ belongs to $B_{\a,\ro_0}$ this implies that the convergence radius of its power series is uniform in $x$. If the real analytic coefficients of the p.d.e. are defined, for instance, in the whole $R^{n_0+1}$ with a common convergence radius, it follows that $u(t,x)$ can be defined in a strip around the real axis of uniform width $2\De_1$. Therefore the solution exists for all $|t|<\De_1$. This does not prevent the possibility that choosing as initial data  on the hypersurface $\Si_{\overline t}$ with $|{\overline t}|=\De_1-\ep$ the function $u({\overline t},x)$ it could be possible, again using the  Cauchy-Kowalevski theorem, to extend the solution to a larger strip and then iterating this procedure forever or up to a moment when this has to end. The first case can happen, for instance, if  the p.d.e. equation is linear as it follows that the strip width does not depend on $t$, while the second case takes place if in the iteration steps the width becomes smaller and smaller so that the strip where the analyticity is proved cannot be extended anymore.

\NI Nevertheless it has to be pointed out that in the statement of the theorem the time  $T_a(\ro,\ro_0,u^{(0)})$ does not define the larger possible existence region, but only the region where the function $u(t,x)$, thought as a function of the space variables only is real analytic, with convergence radius $\ro$. In other words $T_a$ finite does not prevent the possibility that $u(t,x)$ can be extended as real analytic function in a larger region, but only that for $t>T_a$ the function $u(t,\c)$ does not belong to $B_{\a,\ro}$ but, possibly, to a $B_{\a,\ro'}$ with $\ro'<\ro$. To consider the (strip) largest possible analyticity region, possible infinite,  we define
\bea
{\overline T}\equiv \sup\{T_a>0|\ro(T_a,\ro_0,u^{(0)})>0\}\ .\eql{overlineTdef}
\eea
If ${\overline T}<\infty$, it must happen that
\bea
\lim_{T_a\rightarrow {\overline T}}\ro(T_a,\ro_0)=0\ 
\eea
 and therefore in the limit $t\rightarrow{\overline T}$ the function $\lim_{t\rightarrow{\overline T}}u(t,\c)$ is not  real analytic anymore in the space variables.
  
\NI The main conclusion from this discussion is that this approach does not provide a real control of the largest existence region. This can be a problem, for instance, if we are considering a sequence of real analytic solutions $u^{(n)}(t,x)$ associated to a sequence of real analytic initial data converging to a function belonging to a Sobolev space. In fact based on the previous theorem we cannot make a statement on the existence region of the limit solution as the $T_a^{(n)}$  can shrink to zero as $n\rightarrow\infty$.
 \smallskip
 
\NI In the case of the hyperbolic equation the situation is different. While in the general case, as discussed, the control of the time ${\overline T}$ is based, via the  Cauchy-Kowalevski theorem, on the estimates of the coefficients of the p.d.e. equation and on the initial data $u^{(0)}$,  when the equation we are considering is hyperbolic one can have a better control of the solution existence time, namely one can prove that the analytic solution can be extended beyond $\overline T$ up to a time $T$ which depends only on the suitable Sobolev $H^s$ norm of the initial data. 
In the following of this section we discuss in some detail the Klainerman, Nicol\`o result for the Burger equation, \cite{Kl-Ni:rew}.
\smallskip

\NI We start considering the Burger equation solution $u(t,\c)\in C^1([0,T_a],B_{\a,\ro})$ with initial data $u^{(0)}\in B_{\a,\ro_0}$ and with, as discussed before, 
\[\ro=\ro(T_a,\ro_0,u^{(0)})<\ro_0\ .\]
Energy conservation and Sobolev inequalities imply that
\bea
\|u(\c,t)\|_{H^2}\leq \|u^{(0)}\|_{H^2}\exp{c\int_0^t\|u(\c,s)\|_{H^2}}ds
\eea
and
\bea
\sup_{t\in[0,T]}\|u(\c,t)\|_{H^2}\leq \|u^{(0)}\|_{H^2}\exp{c\int_0^T\|u(\c,s)\|_{H^2}}ds
\eea
which implies
\bea
\sup_{t\in[0,T]}\|u(\c,t)\|_{H^2}\leq \|u^{(0)}\|_{H^2}\exp{(cT\sup_{t\in[0,T]}\|u(\c,t)\|_{H^2})}\ .
\eea
Therefore we have the following lemma,
\begin{Le}\label{L.1}
Let $u(t,\c)\in C^1([0,T_a],B_{\a,\ro})$ be  solution of the Burger equation with initial data $u^{(0)}\in B_{\a,\ro_0}$, then there exists a time $T>0$, depending only on $\|u^{(0)}\|_{H^2}$ such that for any $T'\leq min\{T,T_a\}$
\bea
\sup_{t\in[0,T']}\|u(\c,t)\|_{H^2}\leq c_0\|u^{(0)}\|_{H^2}\ .\eql{ineqsob1}
\eea
\end{Le}
Inequality \ref{ineqsob1} is satisfied if  $T$ verifies,
\bea
T\leq \frac{\log c_0}{c_0c\|u^{(0)}\|_{H^2}}\ .\eql{Test}
\eea
{\bf Remark:} {\em In the non linear hyperbolic equations it is possible that the apriori estimates do not hold beyond a certain time value due to the fact that the equations do not have global solutions. In this case there will be a time value $T^*>T$ where some norm blows up. It could also happen that with more refined apriori estimates inequality like \ref{ineqsob1} can be extended for all $t$ values. In the case of the Burger equation we are in the first situation.\footnote{For the vacuum Einstein equations with small data we are in the second case, see \cite{C-K:book}.}.}
\smallskip

\NI{\bf Step 3:} Next goal is to prove the following fundamental theorem:
\begin{theorem}\label{Mainth1}
Let $u^{(0)}\in B_{\a,\ro_0}$ be the initial data, then for $t\in [0,T']$  there exists an analytic solution, $u(\c,t)\in B_{\a,\ro''}$ where $\ro''<\ro_0$  and, moreover, $\ro''$ does not depend on $T_a$, but only on 
$\ro_0$, $\|u^{(0)}\|_{B_{\a,\ro_0}}$. Finally in the time interval $[0,T']$ the following inequality holds with a constant $c_4$,
\bea
\sup_{t\in[0,T']}\|u(\c,t)\|_{B_{\a,\ro''}}\leq c_4\|u^{(0)}\|_{B_{\a,\ro_0}}\ .
\eea
\end{theorem}
{\bf Remark:} {\em The central part of this result is that $\ro''$ appearing in the analytic solution
 \[u\in C^1([0,T'],B_{\a,\ro''})\ ,\]
does not depend on $T_a$,  \[\ro''=\ro''(\ro_0,\|u^{(0)}\|_{B_{\a,\ro_0}})\ ,\] while before, from the Cauchy-Kowalevski theorem, we had 
 \[u\in C^1([0,T_a],B_{\a,\ro})\] 
 with $\ro=\ro(T_a,\ro_0,u^{(0)})$. It is this fact that allows to extend the solution beyond the time $\overline T$ defined in \ref{overlineTdef}, see the next theorem.}
\smallskip

\NI{\bf Proof:} The proof of the theorem is based on the following lemma,
\begin{Le}\label{L.2a}
Let $u(t,\c)\in C^1([0,T_a],B_{\a,\ro})$ be the analytic solution of the Burger equation with initial data $u^{(0)}\in B_{\a,\ro_0}$, with $\a>2$.
Then, for any $J$, the following estimates hold,\footnote{We use the simplified notation
$||D^Ju(t)||_{L^2}\equiv ||D^Ju(\c,t)||_{L^2}$. moreover with $D^J$ we indicate the spatial derivatives, in the Burger case $D^J=\frac{d^J}{dx^J}$.}
\bea
||D^Ju(t)||_{L^2}\leq C_0e^{(J-2)\ga t}\frac{J!}{J^{\a}}\frac{1}{\ro_0^J}\ ,\eql{7.101a}
\eea
where $C_0$ is a constant satisfying
\bea
c_0||u^{(0)}||_{B_{\a,\ro_0}}\leq C_0
\eea
and $c_0$ is defined in the inequality \ref{ineqsob1}.
$\ga>\ga_0$, where  
\bea
\ga_0=\sup_{t\in[0,T']}\|u(\c,t)\|_{H^2}\ \eql{defga0}
\eea 
and $\ga-\ga_0$ depends on $C_0$.
\end{Le}
\NI{\bf Sketch of the proof of the lemma:}
\NI The proof of Lemma \ref{L.2a}, see (Kl-Ni), is inductive, therefore one starts proving that inequality \ref{7.101a} is true for $J=1,2$.\footnote{In the General Relativity case the induction will start at a different value of $J$.} This is obtained from the apriori estimate  \ref{ineqsob1}, in fact, for $J=1,2$ and $t\leq T'$
\bea
&&\ML||D^{J=1,2}u(t)||_{L^2}\leq ||u(\c,t)||_{H^2}\leq c_0||u(0)||_{H^2}=c_0\sum_{J=0}^2||D^Ju^{(0)}||_{L^2}
\eea
Therefore it is enough to choose $C_0$ such that 
\bea
&&||D^{J=1}u(t)||_{L^2}\leq c_0||u(0)||_{H^2}\leq C_0\frac{1}{\ro_0}\nn\\
&&||D^{J=2}u(t)||_{L^2}\leq c_0||u(0)||_{H^2}\leq C_0\frac{1}{2^{\a}}\frac{1}{\ro_0^2}\ .\eql{condC01}
\eea
Both inequalities are satisfied if, with a constant $c_1\geq c_0(2^{\a}\ro_0^2)$, $C_0$ satisfies:
\bea
C_0\!&\geq&\!c_1\|u^{(0)}\|_{H^2}\ .\eql{condC02a}
\eea
We do not repeat here the inductive estimates, see pages 97,98 of \cite{Kl-Ni:rew}, but we recall that in the case of the
Burger equation the structure of the estimate is
\bea
||D^{N}u(t)||_{L^2}\leq e^{(N+2)\ga_0t}||D^{N}u(0)||_{L^2}+\int_0^tdt F(\{D^{J<N}u(t)\})\eql{2.86b}
\eea
where $F$ is a complicated expression, see eqs. (7.20),...,(7.23) of \cite{Kl-Ni:rew}, which, nevertheless, depends only on the partial derivatives of $u$ of order lower than $N$ and is estimated using the inductive assumption. To estimate $||D^{N}u(0)||_{L^2}$ for an arbitrary $N>2$ we can only use the fact that $u(0,\c)=u^{(0)}$ belongs to $B_{\a,\ro_0}$ and therefore satisfies
\bea
||D^{N}u(0)||_{L^2}\leq ||u^{(0)}||_{B_{\a,\ro_0}}\frac{N!}{N^{\a}}\frac{1}{\ro_0^N}\ .\eql{2.86ab}
\eea
Therefore inequality \ref{2.86b} becomes, for $N>2$ and $\ga>5\ga_0$,
\bea
\ML\ML||D^{N}u(t)||_{L^2}\!&\leq&\! e^{(N+2)\ga_0t} ||u^{(0)}||_{B_{\a,\ro_0}}\frac{N!}{N^{\a}}\frac{1}{\ro_0^N}
+\int_0^tdt F(\{D^{J<N}u(t)\})\eql{2.86c}\\
\ML\ML\!&\leq&\!e^{(N-2)\ga t} ||u^{(0)}||_{B_{\a,\ro_0}}\frac{N!}{N^{\a}}\frac{1}{\ro_0^N}
+\int_0^tdt F(\{D^{J<N}u(t)\})\nn\\
\ML\ML\!&\leq&\!e^{(N-2)\ga t}\frac{C_0}{2}\frac{N!}{N^{\a}}\frac{1}{\ro_0^N}+\int_0^tdt F(\{D^{J<N}u(t)\})
\eea
provided we choose $C_0$ satisfying, beside \ref{condC02a},
\bea
C_0\!&\geq&\!2\|u^{(0)}\|_{B_{\a,\ro_0}}\ .\eql{condC03a}
\eea
Using the inductive assumption the integration part in \ref{2.86c} can be estimated, see \cite{Kl-Ni:rew}, as
\bea
\int_0^tdt F(\{D^{J<N}u(t)\})\leq e^{(N-2)\ga t}\frac{C_0}{2}\frac{N!}{N^{\a}}\frac{1}{\ro_0^N}\left[
\frac{2C_0c_1c_2k_1k_2}{\ro_0(\ga-\ga_0)}\right]\ ,
\eea
where $c_1,c_2,k_1,k_2$ are constants which do not depend on the solution norms.
It follows that the lemma is proved if 
\bea
\left[\frac{2C_0c_1c_2k_1k_2}{\ro_0(\ga-\ga_0)}\right]\leq 1\eql{condC04a}
\eea
 Inequality \ref{condC04a} gives a lower bound on $\ga-\ga_0$. Therefore Lemma \ref{L.2a} is proved provided $C_0, (\ga-\ga_0),\a$ satisfy
\bea
\ML C_0\geq (c_1+2)\|u^{(0)}\|_{B_{\a,\ro_0}} \ , \ (\ga-\ga_0)\geq \max\left(\frac{2c_1c_2k_1k_2}{\ro_0}C_0\ ,\ 4\ga_0\right)\ ,\ \a>2\ .
\eql{condC05a}
\eea
Defining
\bea
\ro'=\ro_0e^{-\ga T'}
\eea
it follows that the previous estimates for the $J$ derivative become
\bea
sup_{t\in[0,T']}||D^Ju(t)||_{L^2}\leq C_0\frac{J!}{J^{\a}}\frac{1}{\ro'^J}\ .\eql{7.101b}
\eea
and defining $c_3$ such that
\bea
c_3\|u^{(0)}\|_{B_{\a,\ro_0}}\geq C_0\geq c_1\|u^{(0)}\|_{B_{\a,\ro_0}}
\eea
we obtain, with $\ro''<\ro'<\ro_0$,
\bea
\!\!\!\sup_{t\in[0,T']}\!||u(t)||_{B_{\a,\ro''}}\!\leq C_0\!\!\sum_{N=0}^\infty\left(\frac{\ro''}{\ro'}\right)^{\!N}\!\!\leq c_3\frac{\ro'}{\ro'-\ro''}\|u^{(0)}\|_{B_{\a,\ro_0}}\leq \!c_4\|u^{(0)}\|_{B_{\a,\ro_0}} ,\ \ \  
\eea
proving Theorem \ref{Mainth1}\ .
\smallskip

\NI{\bf Step 4:} We are left to prove that we can extend the solution beyond $T_a$, this is the content of the following theorem: 
\begin{theorem}\label{Mainth2}
Let $u$ be an analytic solution of the Burger equation with initial data $u(\c,0)=u^{(0)}\in B_{\a,\ro_0}$ with $\a>2$ then the solution can be extended to an analytic solution $\in B_{\a,\ro'''}$ with 
\bea
\ro'''<\ro_1=\ro_0e^{-\ga T}\ ,
\eea
 up to a time $T$ which depends only on the Sobolev norm of the initial data $\|u^{(0)}\|_{H^2}$, see \ref{Test}. Moreover
 \bea
\sup_{t\in[0,T]}\|u(\c,t)\|_{B_{\a,\ro'''}}\leq c_5\|u^{(0)}\|_{B_{\a,\ro_0}}\ .
\eea
\end{theorem}
\NI{\bf Proof:}
Given the initial data $u^{(0)}\in B_{\a,\ro_0}$ we know that given $T_a$ there exists an analytic solution $u$ such that for any $t\in [0,T_a]$
\[u(\c,t)\in  C^1([0,T_a]; B_{\a,\ro})\]
with $\ro=\ro(T_a,\ro_0,u^{(0)})$. The previous theorem, Theorem \ref{Mainth1} says something much stronger, namely that there exists $\ro_1=\ro_0e^{-\ga T}$ such that for any $T_a<T$,\footnote{Recall that in the previous theorem $\ro''<\ro'=\ro_0e^{-\ga T'}$ and $T'=min\{T_a,T\}$, therefore if $T_a<T$ we can choose $\ro''>\ro_1=\ro_0e^{-\ga T}$.}
\bea
u\in C^1([0,T_a]; B_{\a,\ro_1})
\eea
where $\ga$ and $T$ depend only on $\|u^{(0)}\|_{B_{\a,\ro_0}}$ and $\|u^{(0)}\|_{H^2(R^{n_0})}$ respectively.
Let us define $T^*$ in the following way:
\bea
T^*=\sup\{T_a\in [0,T]\ | \mbox{for any}\ t\in [0,T_a]\  u(T_a,\c)\in B_{\a,\ro_1} \}\ .\eql{supa}
\eea
As in the interval $[0,T_a]$, $u$ is an analytic solution in all the variables, it follows that taking the limit $t\rightarrow T^*$ also $u(\c,T^*)$ is an analytic solution $\in B_{\a,\ro'''}$ with $\ro'''<\ro_1$. In fact, as discussed before, $T_a$ does not define the largest region of the Cauchy Kowalevski solution, but the largest region where $u(\c,t)\in B_{\a,\ro_1}$. 

\NI If $T^*=T$ we have proved our result. 
Let us, therefore, assume that $T^*<T$,\ 
we can apply again the Cauchy-Kowalevski theorem defining the real analytic function $u^{(1)}\equiv u(\c,{T^*})$ as the initial data on $\Si_{T^*}$  and proving that there exists an interval $[T^*,T^*+\De]$ with $T\geq T^*+\De$ such that, for $t$ in this interval, there is a function, solution of the Burger equation with this initial data,
\[v(t,\c)\in B_{\a,\tilde{\ro}}\ ,\]
with $\tilde{\ro}>0$.\footnote{It follows simply applying Lemma \ref{L4.2}.}

\NI Observe that $\De$ depends only on $\tilde{\ro}$, $\ro_1$ and $\|u^{(1)}\|_{B_{\a,\ro_{1}}}$ or, in other words, only $\tilde{\ro}$, $\ro_0,T, \|u^{(0)}\|_{B_{\a,\ro_0}}$. Let us consider now the function $w$ defined in the following way, for an arbitrary $k\geq 1$,
\bea
&&\ML\ML\ML w(t,\c)=u(t,\c)\in C^k([0,T^*], B_{\a,\ro'''})\ , \  \mbox{for}\ t\in [0,T^*]\ \ \nn\\
&&\ML\ML\ML w(t,\c)=v(t,\c) \in C^k([T^*,T^*+\De], B_{\a,\tilde{\ro}})\ , \ \mbox{for}\ t\in [T^*,T^*+\De]\  .
\eea
It is clear that $w(t,\c)$ is $C^k$ in the time variable for $t\in [0,T^*+\De]$ and we can conclude that ($\tilde{\ro}<\ro'''$),
\[w(t,\c)\in C^k([0,T^*+\De], B_{\a,\tilde{\ro}})\]
and is a solution of the Burger equation in this time interval with initial data $w(0,\c)=u^{(0)}(\c)$. 

\NI Let $\De$ be such that $T^*+\De\leq T$, we can apply again Theorem \ref{Mainth1} to this function,  prove that exactly the same inductive estimates hold\footnote{In this case $T_a=T^*+\De$.} and conclude that
\[w(t,\c)\in C^k([0,T^*+\De], B_{\a,{\ro''}})\subset C^k([0,T^*+\De], B_{\a,{\ro'''}})\ .\]
As $T^*+\De>T^*$, it follows that $T^*$ cannot be the $\sup$ as defined in \ref{supa} unless $T^*=T$. Therefore we conclude that with initial data $u^{(0)}\in B_{\a,\ro_0}$ there exists a solution $u$ in $[0,T]$ such that
\[u(t,\c)\in C^k([0,T],B_{\a,\ro'''})\ ,\]
which means that the size of the spacetime region where the Cauchy-Kowalevski solution can be extended, depends only on the $H^{s=2}$ norm of $u^{(0)}$.
\smallskip

\NI{\bf Remarks:} {\em

\NI 1) Observe that all the previous results are valid if the initial data belong to a different Banach space $B_{\a,\ro'_0}$ with $\ro'_0<\ro_0$, the time $T$ and the fact that the solution can be extended at least up to $T$ does not change.
\smallskip

\NI 2) The fact that $C_0$ is lower bounded by $||u^{(0)}||_{B_{\a,\ro_0}}$ and not only on $||u^{(0)}||_{H^2}$ is not
harmful and, on the other side, is what has to be expected. In fact, at the end of the whole proof, the various real analitic solutions $u_n$ (associated
to the sequence of real analytic initial data $\{u^{(0)}_n\}$) will have the same existence time $T$, but their higher derivatives will have
increasing bounds as $n\rightarrow\infty$. Therefore denoting $C^{(n)}_0$ the constant $C_0$ ``associated" to $u^{(0)}_n$ it will follow that
$C^{(n)}_0\rightarrow\infty$ as $n\rightarrow\infty$\ .}
\smallskip

\NI {\bf Step 5:} {\bf the construction of a Sobolev solution.}
In this case we have a sequence of real analytic functions $\{u^{(0)}_n\}$ converging in the $H^2$ norms to $u^{(0)}\in H^2$. It is clear that due to this convergence the $H^2$ norms of all the $u^{(0)}_n$ functions are bounded by $c\|u^{(0)}\|_{H^2}$. On the other side the $L^2$ norms of the higher tangential derivatives will in general diverge as $n\rightarrow\infty$,  therefore we also have
\bea
\lim_{n\rightarrow\infty}\|u^{(0)}_n(\c)\|_{B_{\a,\ro^{(n)}_0}}=\infty
\eea
Then we have a sequence of real analytic solutions, $\{u_n(t,x)\}$ defined in the same interval of time $[0,T]$ and the $L^2$ norms of the higher, $>2$, derivatives satisfy
\bea
||D^Ju_n(t)||_{L^2}\leq C^{(n)}_0\frac{j!}{j^{\a}}\frac{1}{{\ro_{(n)}}^j} \eql{6.90}
\eea 
and as $C^{(n)}_0\geq c'\|u_n^{(0)}(\c)\|_{B_{\a,\ro}}$,
\bea
 \lim_{n\rightarrow\infty}C^{(n)}_0=\infty\ .
\eea
This is exactly what we expect as the sequence of real analytic functions converge to a function which is not real analytic.
The fact that this sequence has a radius $\ro$ depending on $n$, $\ro^{(n)}$ follows immediately repeating the procedure to obtain the estimate \ref{6.90} and looking at the needed estimate \ref{condC04a} which now reads
\bea
\left[\frac{2C^{(n)}_0c_1c_2k_1k_2}{\ro_0(\ga^{(n)}-\ga_0)}\right]\leq 1\ .\eql{condC04aa}
\eea
As $\ga^{(n)}$ has to increase it follows that\footnote{In the same way also in the construction of the initial data approximating Sobolev data we expect that $\ro_0$ depends on $n$, $\ro_0=\ro_0^{(n)}$.}
\bea
\ro_{(n)}\leq\ro_1^{(n)}=\ro_0e^{-\ga^{(n)}T}\ .
\eea
Therefore, as already said, although the radius of convergence of the sequence of the analytic solutions of the Burger equation, with initial data approximating Sobolev data, tends to shrink, nevertheless the time interval $[0,T]$ where the analytic solutions do exist does not change.

\subsection{The extension of the previous result to the Einstein vacuum equations in the characteristic case}\label{SS4.2}

The extension of the previous result to the Einstein vacuum equations in the characteristic case requires a detailed discussion, let us indicate all the differences, with respect to the Burger case, we have to deal with.
\smallskip

\NI{\bf 1) The initial data:} The analogous of data on $\Si_0$ for the Burger equation are the data given on ${\cal C}=C_0\cup\Cb_0$. The discussion on how to
obtain real analytic data satisfying the constraints on this null hypersurface requires again the use of a ``Cauchy-Kowalevski-type" argument for the local part and a ``Burger-type" argument for the global one.  
More precisely, let us consider first the initial data on $C_0$, first we have to specify on $S_0$ all the tangential derivatives for the ${\cal O}$ and the $\underline{\cal O}$ quantities, where only some constraints have to be fulfilled, see \cite{Ca-Ni:char}. 
Then we use the transport equations, on (a portion of) $C_0$, for the $\nabb^N{\cal O}$ variables for any $N$. Next, once we have these estimates, we can use the transport equations for the  $\nabb^N\underline{\cal O}$ variables for any $N$. Finally we use the transport equations just as relations expressing derivatives with respect to $\nu$ in terms of lower order $\nu$ derivatives (in the case of the ${\cal O}$ quantities) or lower order $\nu$ derivatives and tangential derivatives for the  $\underline{\cal O}$ quantities. Once all these estimates have been done we know that  the ${\cal O}$ and $\underline{\cal O}$ quantities satisfy the transport equations along (a portion of) $C_0$ which are exactly the ``initial condition equations" of Theorem \ref{T3.1}, equations \ref{5.37la}, \ref{2.84q}. 

\NI Exactly the same has to be done on (a portion of) $\Cb_0$ inverting the role of the ${\cal O}$ and $\underline{\cal O}$ quantities. This completes the local construction of the analytic initial data.

\NI Applying now a ``Burger type argument",  we obtain estimates for all the derivatives in $\{\nu,\om^b\}$ of the ${\cal O}$ and $\underline{\cal O}$ quantities proving that, being these quantities in a ${\cal B}_{\a,\ro_0}$-type Banach space, they can be extended as real analytic functions on the whole $C_0\cup\Cb_0$. The details of this construction are given in the following paper.\footnote{The Burger argument developed before requires a bootstrap argument and, before, an apriori estimate. In the case of the initial data the mechanism is slightly different and although also in the initial data case, we need a recursive mechanism and a bootstrap argument, the fact that we can assign in a free way some connection coefficients makes the control of the initial data along the whole $C_0\cup\Cb_0$ easier.}
\smallskip

\NI{\bf 2) The local existence for real analitic solutions:} Next step is to prove a local existence for our problem. This is done using the initial data on a local neighborhood of $S_0$ and the version of the Cauchy-Kowalevski theorem adapted to this non linear characteristic problem as discussed in Section \ref{S3}. This is the preliminary step needed to prove a larger existence region as discussed in Point 3.
\smallskip

\NI{\bf 3) The largest existence region:} This is the central part of the proof. One has to show, as in the Burger case, that the existence region of an analytic solution depends only on some appropriate Sobolev norms of the analytic initial data. Arguing as in the previous discussion of the Burger equation this implies that we can have  a sequence of real analytic data $\{\Psi_n^{(0)}\}$ converging in some Sobolev norms to a ``Sobolev" initial data whose corresponding solutions have a common region of analiticity. This would allow to construct $H^s$ solutions as a limit of a sequence of analytic solutions.

\NI The proof that the existence region of an analytic solution depends only on some appropriate Sobolev norms is obtained, as in the case of the Burger equation, using a contradiction argument, namely we show that we can define a Banach space, ${\cal B}_{\a,\ro}$, where the real analytic solution belongs (as function of space variables), when restricted to a certain region ${\cal K}_a$, 
and prove that the solution remains in this Banach space even when the region is extended to a larger one, ${\cal K}$, whose size depends only on some Sobolev norms of the initial data. Of course, again, the choice of the Banach space must be such that if a function belongs to it and solve the (``hyperbolic") Einstein equations then it is real analytic and can be used as a real analytic initial data on the upper boundary of the largest ${\cal K}_a$ region to extend the solution to ${\cal K}$. Moreover this argument requires that the set of regions ${\cal K}_a$ is not empty which requires preliminary the local existence result discussed in {\bf 2)}. Therefore next problem is that of defining the Banach spaces ${\cal B}_{\a,\ro}$. 
 \smallskip
 
 \NI{\bf 4) The Banach spaces:}  To state precisely the analogy with the Burger equation, the Banach spaces ${\cal B}_{\a,\ro}$, has to be the analogous ones of $B_{\a,\ro}$, the region ${\cal K}_a$ the analogous of the region $R\times[0,T_a]$, the region  ${\cal K}$ of the region $R\times[0,T]$.

\NI Let us recall the definition of the Banach space norm in the case of the Burger equation:
\bea
||u(t,\c)||_{B_{\a,\ro}}=\sum_{n=0}^{\infty}\frac{||D^nf(t,\c)||_{L^2}}{n!}n^{\a}\ro^n
\eea
In the present case let $\Psi=\Psi_{0}+\Psi_{1}$ be a real analytic solution, where
\bea
&&\ML\ML\Psi=(V;W)=(\ga,\oom,v,\psi,\om,\ze,\chi\ ;\ X,w,\omb,\chib)\equiv (\ga,\oom,v,\psi,{\cal O}\ ;\ X,w,\underline{\cal O})\nn\\
&&\ML\ML\Psi_{0}=(\ga,\oom,v,\psi\ ;\ X,w)\ \ ,\ \ \Psi_{1}=({\cal O}\ \! ;\ \! \underline{\cal O})\ ,
\eea
 in a region,
$\{(\la,\nu)|(\la,\nu)\in [0,{\Lambda_a}]\times[0,{\Pi_a}]\}$, of the characteristic problem for the Einstein vacuum equations. Recalling that $V$ denotes the part of the Cauchy-Kowalevski solution whose equations are in the incoming $\la$-direction and $W$ the opposite one, we define
\bea
&&\ML||{\cal O}(\la,\c)||_{B_{\a,\ro}}=\sum_{n=0}^{\infty}\frac{\left(\sup_{p\in\{2,4\}}\sup_{\nu\in C_0}||r^{n+\psi(n)-\frac{2}{p}}\nab^n{\cal O}(\la,\c)||_{L^p(S_0)}\right)}{n!}n^{\a}\ro^n\ \ \ \ \ \ \nn \\
&&\ML||\underline{\cal O}(\nu,\c)||_{B_{\a,\ro}}=\sum_{n=0}^{\infty}\frac{\left(\sup_{p\in\{2,4\}}\sup_{\la\in\Cb_0}
||f_n(r,\la;p)\nab^n\underline{\cal O}(\nu,\c)||_{L^p(S_0)}\right)}{n!}n^{\a}\ro^n\ ,\ \ \ \ \ \ \ \ \ \ \eql{Barodef}
\eea
where $||\nab^n{\cal O}(\la,\c)||_{L^p(S_0)}$, $||\nab^n\underline{\cal O}(\nu,\c)||_{L^p(S_0)}$  are slightly symbolic expressions for
\bea
&&||\nab^n{\cal O}(\nu,\c)||_{L^p(S_0)}:=\sum_{\ p,q;p+q=n}\|\ddb_4^p\nabb^q{\cal O}(\nu,\c)||_{L^p(S_0)}\nn\\
&&||\nab^n\underline{\cal O}(\la,\c)||_{L^p(S_0)}:=\sum_{\ p,q;p+q=n}\|\ddb_3^p\nabb^q\underline{\cal O}(\la,\c)||_{L^p(S_0)}\ ,
\eea
$\psi(n)$ depends on which element of ${\cal O}$ is considered, $f_n(r,\la)$ depends on which element of $\underline{\cal O}$ is considered and has the structure $f_n(r,\la)=r^{\a(n)-\frac{2}{p}}|\la|^{\b(n)}$, see for instance the analogous definitions for the first derivatives in \cite{Kl-Ni:book}. Finally
\bea
||f(\la;\c)||_{L^p(S_0)}=\left(\int_{S_0}\!\!d\mu_{\hat{\ga}_0}|f(\la;\nu,\om^a)|^p\!\right)^{\!\frac{1}{p}}\ ,\eql{6.95}
\eea
where the measure is done  with respect to a given metric tensor $\hat{\ga}_0$, analogous to the  metric $\ga_0$ on $S_0$,
but with a different radial factor
\bea
&&\hat{r}_0(\la,\nu)\equiv r_0+\frac{1}{2}(\nu-\la)\ ,\eql{6.96}\\
&&{\hat{\ga}}_{0(\la,\nu)}(\c,\c)=\frac{\hat{r}_0(\la,\nu)}{r_0}\ga_0(\c,\c)\ .\eql{6.97}
\eea
Finally if $q$ is an $h$-covariant tensor on $S_0$ its pointwise norm $|q|$, defined with respect to the metric tensor ${\hat{\ga}}_{0}$, is
\footnote{We use the metric $\hat{\ga}_0$ instead of the metric $\ga_0$
as we want a ``background" metric ``near" to $\ga(\la,\nu)$ for all the values of $\la,\nu$.}
\bea
|{q}|^2={q}_{a_1a_2...a_s}{q}_{b_1b_2...b_s}{\hat{\ga}}_{0}^{a_1b_1}{\hat{\ga}}_{0}^{a_2b_2}...{\hat{\ga}}_{0}^{a_sb_s}
\equiv |q|^2_{\hat{\ga}_{0}}\ .\eql{6.98}
\eea
The Banach space ${\cal B}_{\a;\ro}$ is defined through the norm
\bea
||\Psi_{(1)}(\la,\nu;\c)||_{{\cal B}_{\a;\ro}}=||{\cal O}(\la,\c)||_{B_{\a,\ro}}
+||\underline{\cal O}(\nu,\c)||_{B_{\a,\ro}}\ .\eql{6.94a}
\eea
\nn{\bf Remarks:} {\em 
\smallskip

\NI { i)} The norms
\[\sup_{p\in\{2,4\}}\sup_{\nu\in C_0}||r^{n+\psi(n)-\frac{2}{p}}\nab^n{\cal O}(\la,\c)||_{L^p(S_0)}\ \ ,\ \sup_{p\in\{2,4\}}\sup_{\la\in\Cb_0}||f_n(r,\la;p)\nab^n\underline{\cal O}(\nu,\c)||_{L^p(S_0)}\] 
play the role of the $\|D^nu(t,\c)\|_{L^2(R^n)}$ norm in the Burger case. The main difference is that, in that case, $D=\frac{\partial}{\partial x}$ involves all the variables in $R^n$ (more specifically the only existing one, but in the case of $R^n$ with $n>1$ all the derivatives $\{\pr_{x^i}\}$ would be present). In the Einstein case the derivatives involved are those ``tangent" to $C_0$ and $\Cb_0$ repectively, namely $\{\ddb_4,\nabb\}$ on $C_0$, $\{\ddb_3,\nabb\}$ on $\Cb_0$.\footnote{Applying $\ddb_4,\ddb_3$ implies deriving with $\dd_4,\dd_3$ and projecting on the $S$-tangent space.} 
\smallskip

\NI { ii)} The previous norms are relative to the connection coefficients ${\cal O}$ and $\underline{\cal O}$  while the
Cauchy-Kowalevski solution $\Psi$ involves also the functions $\ga,\oom,v,\psi; X,w$. Also for these quantities we have to define analogous norms; nevertheless once $\Psi_{1}$ belongs to the Banach space ${\cal B}_{\a;\ro}$ we can easily prove that the analogous norms for these quantities are bounded.} 
\medskip

\NI {\bf 5) The main result:} The core result we have proved and we present in the subsequent paper is that our solution belongs to a ${\cal B}_{\a;\ro}$ Banach space, with appropriate $\ro$ and $\a$, with $(\la,\nu)$ (coordinates of points) in a region $\cal K$ whose size is determined only from some Sobolev norms.\footnote{The region  $\cal K$ is a region of the spacetime, but is completely defined once we give $(\la,\nu)$.} From this it follows immediately that in the same region the solution is real analytic.\footnote{In fact in a slightly larger region.} 
Looking at the discussion of the similar result for the Burger equation we state, first of all, Lemma \ref{L5.1} and Theorem \ref{T5.1} the analogous of Lemma \ref{L4.1} and Theorem \ref{4.1a} respectively. 
\begin{Le}\label{L5.1}
Let ${\cal O}(\la,\c)$, $\underline{\cal O}(\nu,\c)$ be  functions belonging to the Banach space ${\cal B}_{\a;\ro}$ with \[\a\geq \left[\frac{3}{2}\right],\]
then, for any $x=\{\nu,\om^a\}$, ${\cal O}(\la,\c)$ and, for any $y=\{\la,\om^a\}$, $\underline{\cal O}(\nu,\c)$, are real analytic in $B(x)_{\ro}$, $B(y)_{\ro}$ respectively\ .
\end{Le}
\NI{\bf Proof:} The proof goes exactly as the proof of Lemma \ref{L4.1} with the obvious modifications and we do not repeat it here. Next result, the analogous of Theorem \ref{4.1a}, specifies the existence region of  the analytic solution of the characteristic problem solved via the characteristic Cauchy-Kowalevski theorem as discussed in Section \ref{S3}.
\begin{theorem}\label{T5.1}
Let the ``initial data" $\Psi^{(0)}\in {\cal B}_{\a;\ro_0}$ then, given $\ro\leq\ro_0$, there exists a solution of the system of equations \ref{5.37l} and \ref{2.70gql}, $\Psi(\la,\nu,\om^a)\in  {\cal B}_{\a;\ro}$  satisfying the initial conditions in a region,
\[{\cal K}(\Lambda_a,\Pi_a)\equiv\{(\la,\nu)\in[0,{\Lambda_a}]\times[0,{\Pi_a}]\}\ ,\]
where ${\Lambda_a},{\Pi_a}$ depend on $\ro,\ro_0,\Psi^{(0)}$ and are such that
\[\lim_{\ro\rightarrow\ro_0}{\Lambda_a}(\ro,\ro_0,\Psi^{(0)})=\lim_{\ro\rightarrow\ro_0}{\Pi_a}(\ro,\ro_0,\Psi^{(0)})=0\ .\]
\end{theorem}
\NI{\bf Proof:} The proof goes exactly as the proof of Theorem \ref{4.1a} with the obvious modifications and we do not repeat it here. 
\smallskip

\NI Theorem \ref{T5.1} tells us, exactly as in the Burger case, that if we define ${\overline\Lambda},{\overline\Pi}$ such that
\bea
{\cal K}({\overline\Lambda},{\overline\Pi})=\sup\{{\cal K}(\Lambda_a,\Pi_a)| \Psi(\la,\nu,\om^a)\in  B_{\a,\ro} \ \mbox{with}\ \ro>0\}
\eea
then
\bea
\lim_{\Lambda_a,\rightarrow{\overline\Lambda}}\ro(\Lambda_a,\Pi_a;\ro_0,\Psi^{(0)})=\lim_{\Pi_a\rightarrow{\overline\Pi}}\ro(\Lambda_a,\Pi_a;\ro_0,\Psi^{(0)})=0\ .\eql{CKest}
\eea
\smallskip

\NI Next step is to prove Lemma \ref{P5.1} and Theorem \ref{P.2} which correspond to Lemma \ref{L.1} and Theorem \ref{Mainth1}. We first define the analogous of the $H^s$ norm for the Burger equation; we denote
\[\ggg\!=\!(\ga,\oom,X)\ ,\ \cal O\!=\!(\chi,\eta,\om)\ \ ,\ \underline{\cal O}\!=\!(\chib,\etab,\omb)\ \]
and define, \footnote{$$\|{\cal O}\|_{H^s_p}=\left(\sum_{k=0}^s\int_{S_0}\!\!d\mu_{\hat{\ga}_0}|r^{k+\psi(k)-\frac{2}{p}}\nabb^kf(\la;\nu,\om^a)|^p\!\right)^{\!\frac{1}{p}}\ .$$
$$\|\underline{\cal O}\|_{H^s_p}=\left(\sum_{k=0}^s\int_{S_0}\!\!d\mu_{\hat{\ga}_0}|f_k(r,\la;p)\nabb^kf(\la;\nu,\om^a)|^p\!\right)^{\!\frac{1}{p}}\ .$$}
\bea
||\Psi(\la,\nu;\c)||_{H^s_p(S_0)}=\|\ggg\|_{H_p^{s}(S_0)}(\la,\nu)+\|{\cal
O}\|_{H_p^{s}(S_0)}(\la,\nu)+\|\underline{\cal O}\|_{H_p^{s}(S_0)}(\la,\nu)\nn\eql{6.99}
\eea
\[\|\ggg\|_{H_p^{s}(S_0)}\equiv\|\ga\|_{H_p^{s}(S_0)}(\la,\nu)+\|\oom-\frac{1}{2}\|_{H_p^{s}(S_0)}(\la,\nu)+\|X\|_{H_p^{s}(S_0)}(\la,\nu)\ .\]
For the initial data we define analogous norms,
\bea
&&\ML\ML|||\Psi^{(0)}|||_{s\!,p}\!\equiv\!\sup_{\nu\in C_0}\left(\|\ggg\|_{H_{p}^{s}(S_0)}(0,\nu)
+\|{\cal O}\|_{H_p^{s}(S_0)}(0,\nu)+\|\underline{\cal O}\|_{H_p^{s}(S_0)}(0,\nu)\right)\nn\\
&&\ \ \ \ +\sup_{\la\in\Cb_0}\left(\|\ggg\|_{H_p^{s}(S_0)}(\la,0)
+\|{\cal O}\|_{H_p^{s}(S_0)}(\la,0)+\|\underline{\cal O}\|_{H_p^{s}(S_0)}(\la,0)\right)\ .\nn
\eea
\begin{Le}\label{P5.1}
Let $\Psi=(\ggg;{\cal O},\underline{\cal O})=(\ga,\oom,X;\ \om,\ze,\chi,\omb,\chib)$ be a solution of the characteristic problem for the Einstein vacuum equations, in a region,
\[{\cal K}(\Lambda_a,\Pi_a)\equiv\{(\la,\nu)\in[0,{\Lambda_a}]\times[0,{\Pi_a}]\}\ ,\]
 with initial data satisfying $|||\Psi^{(0)}|||_{s\!,p}\leq c$, with $p\in\{2,4\}$. Then there exists a region ${\cal K}(\Lambda,\Pi)$,
whose size depends only on the norm  $|||\Psi^0|||_{s\!,p}$ such that for $s\geq 7$,  $p\in\{2,4\}$, there exists a constant $c_0$ such that 
\bea
\sup_{(\la,\nu)\in {\cal K}(\Lambda',\Pi')}\|\Psi(\la,\nu;\c)\|_{H^s_p(S_0)}\leq c_0|||\Psi^{(0)}|||_ {s\!,p}\eql{6.99a}
\eea
for any region ${\cal K}(\Lambda',\Pi')\subset\ min({\cal K}(\Lambda,\Pi),{\cal K}(\Lambda_a,\Pi_a))$.
\end{Le}
\NI{\bf Remark:} {\em Lemma \ref{P5.1} states, basically, the apriori estimates for the Einstein equations. Differently from the simple Burger equation these estimates are much more involved  as the hyperbolicity of the Einstein equations is more difficult to exploit. }
\smallskip

\NI Next theorem, the analogous of Theorem \ref{Mainth1}, is the basic ingredient to prove our result.
\begin{theorem}\label{P.2}
Let $\Psi=(\ggg;{\cal O},\underline{\cal O})=(\ga,\oom,v,\psi, w, X;\ \om,\ze,\chi,\omb,\chib)$ be a real analytic solution in the region,
\[{\cal K}(\Lambda_a,\Pi_a)\equiv\{(\la,\nu)\in[0,{\Lambda_a}]\times[0,{\Pi_a}]\}\ ,\] of the characteristic
problem for the Einstein vacuum equations with analytic initial data on $C_0\cup\Cb_0$, belonging to the Banach space
${\cal B}_{\a;\ro_0}$.

\NI Then in any region ${\cal K}\subset{\cal K}(\Lambda',\Pi')\subset{\cal K}(\Lambda_a,\Pi_a)$ the following relation holds
\bea
\sup_{(\la,\nu)\in {\cal K}}\|\Psi_{1}(\la,\nu;\c)\|_{{\cal B}_{\a;{\ro''}}}\leq c_1\|\Psi_{1}^{(0)}\|_{{\cal B}_{\a;\ro_0}}
\eea
for some ${\ro''}$ which depends only on $\ro_0$ and on the  $||\Psi_{1}^{(0)}||_{{\cal B}_{\a;\ro_0}}$ norm, but does not depend on the region ${\cal K}(\Lambda_a,\Pi_a)$ .
\end{theorem} 
\NI{\bf Remark:} {\em
Observe that $\ro''$ and $\ro_0$ are the analogous of the same quantities defined in Theorem \ref{Mainth1} in the Burger's equation case. Moreover $(\Lambda_a,\Pi_a)$ are the analogous of $T_a$, $(\Lambda',\Pi')$ are the analogous of $T'$ and $(\Lambda,\Pi)$ are the analogous of $T$ and, exactly in the same way, they depend only on the apriori estimates which means that the size of the region ${\cal K}(\Lambda,\Pi)$ depends only on $\|\Psi^{(0)}\|_{H^{s'}_p(S_0)}$.}

\NI Lemma \ref{P5.1} and Theorem \ref{P.2} are the core of the result we want to obtain, exactly as the proof of  Theorem \ref{Mainth1} is based on Lemma \ref{L.2a}, the proof of Theorem \ref{P.2} is based on the following lemma,
\begin{Le}\label{L6.1a} 
Let $\Psi=(\ggg;{\cal O},\underline{\cal O})=(\ga,\oom,v,\psi, w, X;\ \om,\ze,\chi,\omb,\chib)$ be a real analytic solution in the region,
\[{\cal K}(\Lambda_a,\Pi_a)\equiv\{(\la,\nu)\in[0,{\Lambda_a}]\times[0,{\Pi_a}]\}\ ,\] of the characteristic
problem for the Einstein vacuum equations with analytic initial data on $C_0\cup\Cb_0$, sufficiently small and belonging to the Banach space ${\cal B}_{\a;\ro_0}$, with $\a>3$.
Then, for the generic connection coefficient we indicate with $\cal U$,  the following estimates hold for any $J$ \footnote{The notation of inequality \ref{est45ax} is a bit symbolic, $J$ is an integer except that in $\nabb^J$ where it has to be cosidered a multiindex with $|J|=J$, moreover if with $\cal U$ we denote a $\underline{\cal O}$ connection coefficient then the weight factor $r^{2+J-\frac{2}{p}}$ has to be modified, see equations \ref{Barodef}. }
\bea
|r^{2+J-\frac{2}{p}}\nabb^J{\cal U}|_{p,S}(\la,\nu)\leq C_0\frac{J!}{J^\a}\frac{e^{(J-2)\de}e^{J\Ga(\la,\nu)}}{\ro^J}\eql{est45ax}
\eea
where $\de>\de_0>0$, $C_0$ is a constant satisfying
\bea
C_0\geq c_1||\Psi^{(0)}||_{B_{\a,\ro_0}}
\eea
and 
\bea
\Ga(\la,\nu)=\Ga(\nu)+\underline{\Ga}(\la)\leq \ga
\eea
where, with an appropriate $\hat{C}>0$,
\bea
\Ga(\nu)=\hat{C}\frac{(\nu-\nu_0)}{\nu\nu_0}\ \ ,\ \ \underline{\Ga}(\la)=\hat{C}\frac{(\la-\la_0)}{\la\la_0}\ .\eql{Gadef2}
\eea
$\ga >\ga_0$ where $\ga-\ga_0$ depends on $C_0$ and 
\bea
\Ga_0(\nu)+\underline{\Ga_0}(\la)\leq \ga_0\ .\eql{defga0b}
\eea 
$\ga_0,\de_0$, $\Ga_0(\nu)+\underline{\Ga_0}(\la)$ are relative to the initial data which are assumed to belong to ${\cal B}_{\a;\ro_0}$ and, moreover their angular covariant derivatives satisfy analogous bounds to \ref{est45ax}, with $\ga_0,\de_0$, $\Ga_0(\nu)+\underline{\Ga_0}(\la)$ instead of $\ga,\de$, $\Ga(\nu)+\underline{\Ga}(\la)$.
\end{Le}
\NI{\bf Remarks:} {\em 
\smallskip

\NI a) The  crucial fact in this lemma is that the estimates \ref{est45ax} do not depend on $\Lambda_a,\Pi_a$. Therefore the limit \ref{CKest} is not anymore true and this basically allows to extend the region.
\smallskip

\NI b) Differently from Burger, when initial data are small a factor $|\nu|+|\la|$ in the exponential factor (the analogous of $T$ in the Burger case) is not needed (if we do not assume initial data ``small" we expect nevertheless that the the $|\nu|+|\la|$ has to be present).
\smallskip

\NI c) The goal of this lemma is to provide the appropriate estimates for the covariant derivatives of $\Psi$ to conclude that $\Psi\in B_{\a,\ro''}$. Neverteheless to obtain this result we need to control $the |\c|_{p,s}$ norms not only for the angular derivatives, but also for the $\ddb_4$ derivatives for the $V$ components and the the $\ddb_3$ derivatives for the $W$ ones and also for the mixed derivatives. Nevertheless  we first prove Lemma \ref{L6.1a} and obtain the appropriate estimates for the angular derivatives, then using again the structure equations we control all the remaining mixed derivatives.}
\smallskip

\NI Once Lemma \ref{P5.1} and Theorem \ref{P.2} are proved, next step is the proof of the final theorem, Theorem \ref{T5.3} , which is the analogous of Theorem \ref{Mainth2} for the Burger equation and which we state here. Nevertheless it has be remarked that its proof is significantly different as we are considering now a characteristic problem. 
\begin{theorem}\label{T5.3}
Let $\Psi=(\ggg;{\cal O},\underline{\cal O})=(\ga,\oom,v,\psi, w, X;\ \om,\ze,\chi,\omb,\chib)$ be a real analytic solution  problem for the Einstein vacuum equations with analytic initial data on $C_0\cup\Cb_0$, belonging to the Banach space ${\cal B}_{\a;\ro_0}$.
Then this solution can be extended to an analytic solution $\in{\cal B}_{\a;{\ro'''}}$ in the region ${\cal K}(\Lambda,\Pi)$ with
\bea
\ro'''<\ro_1:=\ro_0e^{-\ga(|\Lambda|+|\Pi |)}
\eea 
where $\ga>0$ depends on the ${\cal B}_{\a;\ro_0}$ norms and $\Lambda,\Pi$ depend only on the Sobolev norms of the initial data.
Moreover if the initial data are ``sufficiently small" then the region where the analytic solution exists is unbounded (in the $\nu$ variable, ${\cal K}({\Lambda,\infty})$).
\end{theorem}
\NI {\bf Proof:}  The first step to prove the theorem is the proof of the following lemma:
\begin{Le}\label{L5.2}
If the results of Theorem \ref{T5.1} and Lemma \ref{P5.1} are true then it is possible to extend this solution to a new solution $\tilde{\Psi}$ real
analytic in the region ${\cal K}(\Lambda_a+\De,\Pi_a+\De)$ and such that, for each $\la,\nu$ in this region, it belongs to the Banach space ${\cal B}_{\a;\ro'''}$ with, depending on the initial conditions,
\bea
\ro'''<\ro_0e^{-\ga(|\Lambda|+|\Pi |)}\ .
\eea
\end{Le}\smallskip
\NI{\bf Proof of Lemma \ref{L5.2}:}
First we prove, using these propositions, that assuming as initial data those on $C_0\cup\Cb_0$ and the real analytic solution on
$\Cb(\Pi_a;[0,\Lambda_a])$ and on $C(\Lambda_a;[0,\Pi_a])$ we can extend the solution to a real analytic solution in

\NI ${\cal K}(\Lambda_a,\Pi_a+\De)\cup{\cal K}(\Lambda_a+\De,\Pi_a)$ and also on the diamond region
\[{\cal K}(\Lambda_a+\De,\Pi_a+\De)/{\cal K}(\Lambda_a,\Pi_a+\De)\cup{\cal K}(\Lambda_a+\De,\Pi_a)\ .\]
The proof of the existence of a real analytic solution in the two strips ${\cal K}(\Lambda_a,\Pi_a+\De)$ and 
${\cal K}(\Lambda_a+\De,\Pi_a)$ mimicks the analogous proof made in \cite{Ca-Ni:exist}, with
the difference that, there, we were building an $H^s$ solution while here we prove the existence of a real analytic solution. The problem is nevertheless of the same type: as these stripes in one direction cannot have an uniformily bounded length we have to use a ``sub-bootstrap" mechanism to prove the existence of the solution there. In the real analytic case the existence of the solution in a strip of a width which does not tend to zero is provided again by the ``a priori estimates" which allow to show that the solution exists along the whole strip.  
\smallskip

\NI Once Lemma \ref{L5.2} has been proved, the remaining steps to prove Theorem \ref{T5.3} are basically identical to those used for the Burger equation, namely Theorem \ref{Mainth2}. Observe that
instead of $[0,T^*]$ we have ${\cal K}({\Lambda^*},{\Pi^*})$ and the definition of $T^*$
\beaa
T^*=\sup\{T_a\in [0,T]\ | \mbox{for any}\ t\in [0,T_a]\  u(T_a,\c)\in B_{\a,\ro_1} \}\ .\eql{sup2}
\eeaa
is substituted by
\beaa
{\cal K}({\Lambda^*},{\Pi^*})\!=\!\!\sup\!\left\{{\cal K}(\Lambda_a,\Pi_a)
\underline{\subset}{\cal K}(\Lambda,\Pi)\big|\forall\ (\la,\nu)\in{\cal K}(\Lambda_a,\Pi_a); \Psi(\la,\nu;\c)\in{\cal
B}_{\a;\ro_1}\right\}.\ \ \ \ \  
\eeaa
Lemma \ref{L5.2} shows that, unless
\bea
{\cal K}({\Lambda^*},{\Pi^*})={\cal K}({\Lambda},{\Pi})\ ,\eql{6.104a}
\eea
the region ${\cal K}({\Lambda^*},{\Pi^*})$ can be extended contradicting its definition. Therefore \ref{6.104a} is proved.

\section{Conclusions}

\NI The results presented in this paper concern the existence region of the real analitic solutions of a class of characteristic problems for the vacuum Einstein equations. As a byproduct to obtain it, we have discussed and developed a gauge associated to the double null cone foliation, first introduced in \cite{Kl-Ni:book}. We consider this gauge very appropriate to this kind of problems and in particular we believe that to obtain the same global result proved here in a gauge like the harmonic gauge could be much more difficult, even if possible.

\NI The choice of the ``double null cone gauge" has also the advantage, in our opinion, that it allows writing the Einstein equations as first order equations in a very natural and geometric way\footnote{This is certainly not new, see for instance \cite{NewPena}, \cite{NewmanPenrose}, but we believe it is presented here in a more complete way.}  because this foliation, differently from any spacelike hypersurfaces foliation, is more intrinsic due to the physical meaning of the null cones. Moreover in this approach the distinction, in the characteristic case, between those Einstein equations which can be considered ``evolution equations" and those which have to be interpreted as ``constraint equations" is completely clear. We also believe that our approach could be used to deal with the characteristic problem with ``initial data" on a null outgoing cone hypersurface, but this has not yet been worked.
\section{Appendix}\label{App 2a}
\subsection{General aspects of the structure equations}
\NI{\bf Proof of Proposition \ref{P2.1}:}
We have
\bea
(\theta^{\ga}\wedge\omm^{\a}_{\ga})(e_{\b},e_{\de})=
\omm^{\a}_{\b}(e_{\de})-\omm^{\a}_{\de}(e_{\b})
\eea
on the other side, recalling how $d$ operates,
\bea
d\theta^{\a}(e_{\b},e_{\de})\!&=&\!e_{\b}(\theta^{\a}(e_{\de}))-e_{\de}(\theta^{\a}(e_{\b}))
-\theta^{\a}([e_{\b},e_{\de}])=-\theta^{\a}([e_{\b},e_{\de}])\nn\\ 
\!&=&\!-\theta^{\a}(\dd_{e_{\b}}e_{\de})+\theta^{\a}(\dd_{e_{\de}}e_{\b})\nn\\
\!&=&\!-\GGa^{\ga}_{\b\de}\theta^{\a}(e_{\ga})+\GGa^{\ga}_{\de\b}\theta^{\a}(e_{\ga})=
-\GGa^{\a}_{\b\de}+\GGa^{\a}_{\de\b}\nn\\
\!&=&\!-\omm^{\a}_{\de}(e_{\b})+\omm^{\a}_{\b}(e_{\de})=
\theta^{\ga}\wedge\omm^{\a}_{\ga}(e_{\b},e_{\de})\ ,
\eea
which proves the first structure equation. To prove the second structure equation we write
\bea
\rr^{\de}_{\ga\a\b}e_{\de}\!&=&\!\rr(e_{\a},e_{\b})e_{\ga}=
\dd_{e_{\a}}(\dd_{e_{\b}}{e_{\ga}})-\dd_{e_{\b}}(\dd_{e_{\a}}{e_{\ga}})
-\dd_{[e_{\a},e_{\b}]}{e_{\ga}}\nn\\
\!&=&\!\dd_{e_{\a}}(\GGa^{\la}_{\b\ga}e_{\la})-\dd_{e_{\b}}(\GGa^{\la}_{\a\ga}e_{\la})-
\dd_{[e_{\a},e_{\b}]}{e_{\ga}}\\
\!&=&\!e_{\a}(\GGa^{\la}_{\b\ga})e_{\la}-e_{\b}(\GGa^{\la}_{\a\ga})e_{\la}+
\GGa^{\la}_{\b\ga}\GGa^{\ep}_{\a\la}e_{\ep}-\GGa^{\la}_{\a\ga}\GGa^{\ep}_{\b\la}e_{\ep}
-\dd_{[e_{\a},e_{\b}]}{e_{\ga}}\nn
\eea
therefore
\bea
\rr^{\de}_{\ga\a\b}\!&=&\!e_{\a}(\GGa^{\de}_{\b\ga})-e_{\b}(\GGa^{\de}_{\a\ga})+
\GGa^{\la}_{\b\ga}\GGa^{\de}_{\a\la}-\GGa^{\la}_{\a\ga}\GGa^{\de}_{\b\la}-
\theta^{\de}(\dd_{[e_{\a},e_{\b}]}{e_{\ga}})\nn\\
\!&=&\!e_{\a}(\omm^{\de}_{\ga}(e_{\b}))-e_{\b}(\omm^{\de}_{\ga}(e_{\a}))+
\omm^{\la}_{\ga}(e_{\b})\omm^{\de}_{\la}(e_{\a})-
\omm^{\la}_{\ga}(e_{\a})\omm^{\de}_{\la}(e_{\b})
-\theta^{\de}(\dd_{[e_{\a},e_{\b}]}{e_{\ga}}).\nn
\eea
On the other side
\bea
(d\omm^{\de}_{\ga}+\omm^{\de}_{\si}\wedge\omm^{\si}_{\ga})(e_{\a},e_{\b})\!&=&\!
d\omm^{\de}_{\ga}(e_{\a},e_{\b})+\omm^{\de}_{\si}(e_{\a})\omm^{\si}_{\ga}(e_{\b})-
\omm^{\de}_{\si}(e_{\b})\omm^{\si}_{\ga}(e_{\a})\nn\\
\!&=&\!e_{\a}(\omm^{\de}_{\ga}(e_{\b}))-e_{\b}(\omm^{\de}_{\ga}(e_{\a}))-
\omm^{\de}_{\ga}([e_{\a},e_{\b}])\nn\\
\!&+&\!\omm^{\de}_{\si}(e_{\a})\omm^{\si}_{\ga}(e_{\b})-
\omm^{\de}_{\si}(e_{\b})\omm^{\si}_{\ga}(e_{\a})
\eea
and
\bea
\omm^{\de}_{\ga}([e_{\a},e_{\b}])\!&=&\!\omm^{\de}_{\ga}(c^{\si}_{\a\b}e_{\si})=
c^{\si}_{\a\b}\omm^{\de}_{\ga}(e_{\si})=c^{\si}_{\a\b}\GGa^{\de}_{\si\ga}\nn\\
\!&=&\!c^{\si}_{\a\b}\theta^{\de}(\dd_{e_{\si}}e_{\ga})=
\theta^{\de}(\dd_{c^{\si}_{\a\b}e_{\si}}e_{\ga})=\theta^{\de}(\dd_{[e_{\a},e_{\b}]}e_{\ga})
\eea
and the thesis follows.
\subsubsection{The general structure equations: $d\om^{\de}_{\ga}=-\om^{\de}_{\si}\wedge\om^{\si}_{\ga}+\Omega^{\de}_{\ga}$}

\bea
\left\{
\begin{array}{llll}
&\dddd_3\chibh+tr\chib\chibh+2\omb\chibh-\nabb\hot\xib+
(2\zeta-\etab-\eta)\hot\xib=-\aa\nn\\
&\dd_3 tr\chib+\frac{1}{2}(tr\chib)^2+2\omb tr\chib+|\chibh|^2-2\divv\xib
-2\xib\cdot(\eta+\etab-2\zeta)=0\nn\\
&-\curll\xib+(2\zeta+\etab-\eta)\wedge\xib=0\nn\\
&\ \ \ \ \ \ \ \ \ \ \ \ \ \ \ \ \ \ \ \ \ \ \ \ \ \ \ \ \ \ \ \ \ \ \ \ \ \ \ \ \\ 
&\ \ \ \ \ \ \ \ \ \ \ \ \ \ \ \ \ \ \ \ \ \ \ \ \ \ \ \ \ \ \ \ \ \ \ \ \ \ \ \ \\
&\dddd_4\chibh+\frac{1}{2}tr\chi\chibh+\frac{1}{2}tr\chib\chih-2\om\chibh
-\nabb\hot\etab-\etab\hot\etab-\xi\hot\xib=0\nn\\
&\dd_4 tr\chib+\frac{1}{2}tr\chi tr\chib-2\om tr\chib+\chih\cdot\chibh-2\divv\etab-2|\etab|^2-
2\xi\cdot\xib=2\ro\nn\\
&\curll\etab + \xi\wedge\xib-\frac{1}{2}(\chih\wedge\chibh)=-\si\nn\\
&\ \ \ \ \ \ \ \ \ \ \ \ \ \ \ \ \ \ \ \ \ \ \ \ \ \ \ \ \ \ \ \ \ \ \ \ \ \ \ \ \\ 
&\ \ \ \ \ \ \ \ \ \ \ \ \ \ \ \ \ \ \ \ \ \ \ \ \ \ \ \ \ \ \ \ \ \ \ \ \ \ \ \ \\
&\dddd_3\chih+\frac{1}{2}tr\chib\chih+\frac{1}{2}tr\chi\chibh-
2\omb\chih-\nabb\hot\eta-\eta\hot\eta-\xib\hot\xi=0\nn\\
&\dd_3 tr\chi+\frac{1}{2}tr\chib{tr\chi}-2\omb tr\chi+\chibh\cdot\chih-2\divv\eta-2|\eta|^2
-2\xib\cdot\xi=2\ro\nn\\
&\curll\eta-\frac{1}{2}\chibh\wedge\chih+\xib\wedge\xi=\si\\
&\ \ \ \ \ \ \ \ \ \ \ \ \ \ \ \ \ \ \ \ \ \ \ \ \ \ \ \ \ \ \ \ \ \ \ \ \ \ \ \ \\ 
&\ \ \ \ \ \ \ \ \ \ \ \ \ \ \ \ \ \ \ \ \ \ \ \ \ \ \ \ \ \ \ \ \ \ \ \ \ \ \ \ \\
&\dddd_4\chih+tr\chi\chih+2\om\chih-\nabb\hot\xi-
(2\zeta+\eta+\etab)\hot\xi=-\a\nn\\
&\dd_4 tr\chi+\frac{1}{2}(tr\chi)^2+2\om tr\chi+|\chih|^2-2\divv\xi
-2\xi\cdot(\etab+\eta+2\zeta)=0\nn\\
&\curll\xi-(-2\zeta+\eta-\etab)\wedge\xib=0\\
&\ \ \ \ \ \ \ \ \ \ \ \ \ \ \ \ \ \ \ \ \ \ \ \ \ \ \ \ \ \ \ \ \ \ \ \ \ \ \ \ \\ 
&\ \ \ \ \ \ \ \ \ \ \ \ \ \ \ \ \ \ \ \ \ \ \ \ \ \ \ \ \ \ \ \ \ \ \ \ \ \ \ \ \\
&\curll\chib-\zeta\wedge\chib={^{*}}\bb\\
&\nabb\tr\chib-\divv\chib+\zeta\cdot\chib-\zeta\tr\chib=-\bb\\
&\ \ \ \ \ \ \ \ \ \ \ \ \ \ \ \ \ \ \ \ \ \ \ \ \ \ \ \ \ \ \ \ \ \ \ \ \ \ \ \ \\ 
&\ \ \ \ \ \ \ \ \ \ \ \ \ \ \ \ \ \ \ \ \ \ \ \ \ \ \ \ \ \ \ \ \ \ \ \ \ \ \ \ \\
&\curll\chi+\zeta\wedge\chi=-{^{*}}\b\\
&\nabb tr\chi-\divv\chi-\zeta\cdot\chi+\zeta\tr\chi=\b\\
&\ \ \ \ \ \ \ \ \ \ \ \ \ \ \ \ \ \ \ \ \ \ \ \ \ \ \ \ \ \ \ \ \ \ \ \ \ \ \ \ \\ 
&\ \ \ \ \ \ \ \ \ \ \ \ \ \ \ \ \ \ \ \ \ \ \ \ \ \ \ \ \ \ \ \ \ \ \ \ \ \ \ \ \\
&\dddd_4\xib-\dddd_3\etab-(\etab-\eta)\cdot\chib-4\om\xib=-\bb\\
&\dddd_3\xi-\dddd_4\eta-(\eta-\etab)\cdot\chi-4\omb\xi=\b\\
&\ \ \ \ \ \ \ \ \ \ \ \ \ \ \ \ \ \ \ \ \ \ \ \ \ \ \ \ \ \ \ \ \ \ \ \ \ \ \ \ \\ 
&\ \ \ \ \ \ \ \ \ \ \ \ \ \ \ \ \ \ \ \ \ \ \ \ \ \ \ \ \ \ \ \ \ \ \ \ \ \ \ \ \\
&\dddd_3\zeta+2\nabb\omb+\chib\cdot(\eta+\zeta)-\chi\cdot\xib+2\omb(\eta-\zeta)-
2\omb\xi=-\bb\\
&\dddd_4\zeta-2\nabb\om-\chi(\etab-\zeta)+\chib\xi-2\om(\etab+\zeta)+
2\om\xib=-\b\\
&\ \ \ \ \ \ \ \ \ \ \ \ \ \ \ \ \ \ \ \ \ \ \ \ \ \ \ \ \ \ \ \ \ \ \ \ \ \ \ \ \\ 
&\ \ \ \ \ \ \ \ \ \ \ \ \ \ \ \ \ \ \ \ \ \ \ \ \ \ \ \ \ \ \ \ \ \ \ \ \ \ \ \ \\
&\curll\zeta-\frac{1}{2}\chibh\wedge\chih=\si\\
&\dd_3\om+\dd_4\omb-4\om\omb+(\etab\eta-\xib\cdot\xi+\etab\cdot\zeta-\eta\cdot\zeta)=\ro\\
&\frac{1}{2}^{(2)}\rr_{dcab}\de_{da}\de_{cb}+\frac{1}{4}tr\chi\tr\chib
-\frac{1}{2}\chibh\cdot\chih=-\ro
\end{array}
\right.
\eea

\NI 
\subsection{The connection coefficients as functions on $S_0\times R\times R$}
To express the tensorial equations \ref{subsetstrinc}, \ref{subsetstrout} as a system of p.d.e. equations we have to transform the covariant derivatives relative to the null directions
into partial derivatives. This can be done expressing everything in a generic set of coordinates. We do it projecting all the connection coefficients on the two dimensional surface $S_0\equiv S(0,0)$ through the pullback associated to the
diffeomorphism, \ref{omdef}, which specifies a coordinate set adapted to the double null foliation.

\NI More precisely assuming the spacetime $(\M,\ggg)$ foliated by a double null foliation,\footnote{In \cite{Kl-Ni:book} the double null foliation had to be ``canonical", see its definition in Chapter 3. This is not needed here.} whose leaves we call, with a slight abuse of notation, outgoing or incoming cones and denote by $C(\la)$ and $\Cb(\nu)$ respectively, the two dimensional surfaces $S(\la,\nu)=C(\la)\cap\Cb(\nu)$ produce a
foliation of each outgoing (incoming) cone, $C(\overline{\la})$,  for instance, being foliated by the leaves $\{S(\overline{\la},\nu)\}$. 
We recall that the vector fields
\bea
N=2\oom^2L\ \ ,\ \ \Nb=2\oom^2\Lb\ ,
\eea
are equivariant vector fields with respect to the leaves $S(\la,\nu)$.\footnote{Obviously they do not commute and their commutator is
$[N,\Nb]=-4\oom^2\ze(e_A)e_A$\ .} This means that the diffeomorphism $\Phi_{\nu}$, generated by the vector field $N$, sends $S(\la, 0)$
onto $\Phi_{\nu}[S(\la,0)]=S(\la,\nu)$, and $\underline{\Phi}_{\la}$, the diffeomorphism generated by the equivariant vector field $\Nb$,
sends $S(0,\nu)$ onto $\underline{\Phi}_{\la}[S(0,\nu)]=S(\la,\nu)$. We have previously defined, see \ref{omdef}, the  diffeomorphism,
\bea
\Psi(\la,\nu):S_0\ni p_0\rightarrow q=\Psi(\la,\nu)(p_0)\in S(\la,\nu)
\eea
\bea
\mbox{where}\ \ \ \
\Psi(\la,\nu)(p_0)=\Phi_{\nu}\circ{\underline\Phi}_{\la}(p_0)={\Phi}_{\nu}(\underline\Phi_{\la}(p_0))={\Phi}_{\nu}(p)\ ,\  
p=\underline\Phi_{\la}(p_0)\ .\ \eql{2.9u}
\eea
From $\Psi(\la,\nu)$, we derive the pullback $\Psi^*(\la,\nu)$ which sends the metric components and the various connection
coefficients, tensors belonging to $T^*S(\la,\nu)\otimes T^*S(\la,\nu)\otimes.....\otimes T^*S(\la,\nu)$, to tensors belonging to
$T^*S_0\otimes T^*S_0\otimes.....\otimes T^*S_0$ and depending on the parameters $\la,\nu$. We define:
\bea
\tilde{\ze}(\la,\nu;p_0)&=&(\Psi^*(\la,\nu)\ze)(p_0)=\Psi^*(\la,\nu)(\ze\circ\Psi(\la,\nu)(p_0))\nn\\
&=&\underline{\Phi}^*_{\la}\Phi^*_{\nu}\ze(\Phi_{\nu}\circ\underline{\Phi}_{\la}(p_0))=\underline{\Phi}^*_{\la}\Phi^*_{\nu}\ze(q)\nn\\
&&\nn\\
\tilde{\chi}(\la,\nu;p_0)&=&(\Psi^*(\la,\nu)\chi)(p_0)=\Psi^*(\la,\nu)(\chi\circ\Psi(\la,\nu)(p_0))\nn\\
&=&\underline{\Phi}^*_{\la}\Phi^*_{\nu}\chi(\Phi_{\nu}\circ\underline{\Phi}_{\la}(p_0))=\underline{\Phi}^*_{\la}\Phi^*_{\nu}\chi(q)\nn\\ 
\tilde{\chib}(\la,\nu;p_0)&=&(\Psi^*(\la,\nu)\chib)(p_0)=\Psi^*(\la,\nu)(\chib\circ\Psi(\la,\nu))(p_0)\eql{2.9tt}\\
&=&\underline{\Phi}^*_{\la}\Phi^*_{\nu}\chib(\Phi_{\nu}\circ\underline{\Phi}_{\la}(p_0))=\underline{\Phi}^*_{\la}\Phi^*_{\nu}\chib(q)\nn\\
&&\nn\\
\tilde{\om}(\la,\nu;p_0)&=&(\om\circ\Psi(\la,\nu))(p_0)=(\om\circ\Phi_{\nu}\circ\underline{\Phi}_{\la})(p_0)=\om(q)\nn\\
\tilde{\omb}(\la,\nu;p_0)&=&(\omb\circ\Psi(\la,\nu))(p_0)=(\omb\circ\Phi_{\nu}\circ\underline{\Phi}_{\la})(p_0)=\omb(q)\nn
\eea
where $p_0\in S_0\!\equiv\!S(0,0)$, is specified by its coordinates $(\theta,\phi)$. Therefore
$\tilde{\chi}$, $\tilde{\chib}$, $\tilde{\ze}$, $\tilde{\om},\tilde{\omb}$ are covariant tensors defined on $S_0$ and
depending on the parameters $\la,\nu$,
\beaa
\tilde{\chi}(\la,\nu;\theta,\phi)\ ,\ \tilde{\chib}(\la,\nu;\theta,\phi)\ ,\ \tilde{\ze}(\la,\nu;\theta,\phi)\ ,\ 
\tilde{\om}(\la,\nu;\theta,\phi)\ ,\ \tilde{\omb}(\la,\nu;\theta,\phi)\ .
\eeaa
The Einstein equations will take the form of a system of first order partial differential equations, with respect to the variable
$\{\la,\nu,\theta,\phi\}$, for the components of these covariant tensors and of the metric tensor. Defining on
$S_0$ an orthonormal basis $\{\tilde{e}_{\theta},\tilde{e}_{\phi}\}$ with respect to a fixed metric $\ga_0$ assigned on $S_0$, 
the explicit expression for $\ze$ is:
\bea
&&\ML\ML\tilde{\ze}(\la,\nu;\theta,\phi)(\c)=\sum_A\tilde{\ze}_A(\la,\nu;\theta,\phi)\tilde{\theta}_A(\c)
=\tilde{\ze}_{\theta}(\la,\nu;\theta,\phi)d\theta(\c)+\tilde{\ze}_{\phi}(\la,\nu;\theta,\phi)d\phi(\c)\ \ \ \ \ \ \ \ \ 
\eea
where
\bea
\tilde{\ze}_A(\la,\nu;\theta,\phi)\equiv\tilde{\ze}(\la,\nu;\theta,\phi)(\tilde{e}_A)\ 
\eea
and exactly analogous expressions hold for $\chi,\chib,\om,\omb$. Finally writing in $\M$ the metric $\ggg(\c,\c)$ as
\bea
{\ggg}=|{X}|^2d\la^2\!-\!2{{\oom}}^2(d\la d\nu\!+\!d\nu d\la)\!-\!{{X}}_a(d\la
d\om^a\!+\!d\om^ad\la)\!+\!{\ga}_{ab}d\om^a d\om^b\ ,\eql{met}
\eea 
we denote the pull back of its components, $\tilde{X}, {\tilde{\oom}}, {\tilde\ga}$. In conclusion we have defined a
diffeomorphism from
$(R^2\times S_0,\tilde{\ggg})$ to $(\M,\ggg)$
\bea
R^2\times S_0\ni p=(\la,\nu,\theta,\phi)\rightarrow q=\Psi(\la,\nu)(p_0)
\eea
 where $p_0=(\theta,\phi)\in S_0$
and
\bea
\tilde{\ggg}=|\tilde{X}|^2d\la^2\!-\!2{\tilde{\oom}}^2(d\la d\nu\!+\!d\nu d\la)\!-\!{\tilde{X}}_a(d\la
d\om^a\!+\!d\om^ad\la)\!+\!{\tilde\ga}_{ab}d\om^a d\om^b\ .\eql{2.9pk}
\eea
The goal of this appendix is to rewrite equations \ref{2.1ff}, as a set of first order partial differential
equations in the variables $\la,\nu,\theta,\phi$. Following the discussion of Section 3 choosing an appropriate subset of equations  \ref{2.1ff}, supplemented with some other equations for the metric
components we obtain equations \ref{subsetstrinc}, \ref{subsetstrout} written (for the various components) in the $(\la,\nu,\theta,\phi)$ coordinates. These are the Einstein equations written in the ``double null foliation" gauge and Proposition \ref{Prop2.2} is, therefore, proved. 

\subsection{The structure equations for the connection coefficients in a vacuum Einstein manifold in the $\{\la,\nu,\theta,\phi\}$
coordinates}
To write the previous equations in terms of the pulled back quantities we look first at those satisfied by $\ze$, corresponding to
${\bf R}(e_4,e_A)\!=0$ and ${\bf R}(e_3,e_A)\!=0\ $,
\bea
&&\dddd_4\zeta+\frac{3}{2}\tr\chi\zeta+\zeta\chih-\divv\chih+\frac{1}{2}\nabb\tr\chi+\ddb_4\nabb\log\oom=0\nn\\
&&\dddd_3\zeta+\frac{3}{2}\tr\chib\zeta+\zeta\chibh+\divv\chibh-\frac{1}{2}\nabb\tr\chib-\ddb_3\nabb\log\oom=0\ .\eql{5.8}
\eea
The derivative with respect to the parameter $\nu$ of $\tilde{\ze}$ is:
\bea
\frac{\partial\tilde{\ze}}{\partial\nu}(\la,\nu;p_0)=(\underline{\Phi}^*_{\la}\Phi^*_{\nu}\Lie_N\ze)(q)\ .
\eea
{In fact}  
\bea
&&\ML\ML\frac{\partial\tilde{\ze}}{\partial\nu}(\la,\nu;p_0)\!=\!\underline{\Phi}^*_{\la}
\lim_{h\rightarrow 0}\frac{1}{h}\big[\Phi^*_{\nu+h}\ \!\ze(\Phi_{\nu+h}\circ\underline{\Phi}_{\la}(p_0))\!-\!(\Phi^*_{\nu}\
\!\ze)(\Phi_{\nu}\circ\underline{\Phi}_{\la}(p_0))\big]\nn\\
&&\ML\ML\ =\underline{\Phi}^*_{\la}\Phi^*_{\nu}\lim_{h\rightarrow0}\frac{1}{h}\big[\Phi^*_{h}\ \!\ze(\Phi_h(q))\!-\!\ze(q)\big]
=(\underline{\Phi}^*_{\la}\Phi^*_{\nu}\Lie_N\ze)(\Psi(\la,\nu)(p_0))=(\underline{\Phi}^*_{\la}\Phi^*_{\nu}\Lie_N\ze)(q)\ .\ \ \ \ \ \ \ \ \ \ 
\eea
A simple computation gives
\bea
(\Lie_N\ze)(\c)=\oom[\dddd_4\zeta(\c)+(\zeta\!\cdot\!\chi)(\c)]=\oom\!\left[-\tr\chi\zeta(\c)+\divv\chi(\c)-\nabb\tr\chi(\c)
-\ddb_4\nabb\log\oom(\c)\right]\nn
\eea
which implies 
\bea
\frac{\partial\tilde{\ze}}{\partial\nu}(\la,\nu;\theta,\phi)(\c)=\big(\underline{\Phi}^*_{\la}\Phi^*_{\nu}\oom\!
\left[-\tr\chi\zeta+\divv\chi\!-\!\nabb\tr\chi\!-\!\ddb_4\nabb\log\oom\right]\!\big)\!(\la,\nu;\theta,\phi)(\c)\ .\ \ \ \ \ \eql{5.10}
\eea
Looking at the r.h.s. of \ref{5.10} it is easy to recognize that, denoting $p_0$ a point of $S_0$ the following relationships hold, whose proof is given later on:
\bea
&&(\underline{\Phi}^*_{\la}\Phi^*_{\nu}\oom\tr\chi\zeta)(p_0)=\tilde{\oom}(\la,\nu;p_0)\widetilde{\tr\chi}(\la,\nu;p_0)\tilde{\ze}(\la,\nu;p_0)\nn\\
&&(\underline{\Phi}^*_{\la}\Phi^*_{\nu}\oom\zeta\!\c\!\chi)(p_0)=\tilde{\oom}(\la,\nu;p_0)\tilde{\ze}(\la,\nu;p_0)\!\c\!\widetilde{\chi}(\la,\nu;p_0)\nn\\
&&(\underline{\Phi}^*_{\la}\Phi^*_{\nu}\oom\divv\chi)(p_0)=\tilde{\oom}(\la,\nu;p_0)\widetilde{\divv}\tilde{\chi}(\la,\nu;p_0)\eql{2.15k}\\
&&(\underline{\Phi}^*_{\la}\Phi^*_{\nu}\oom\nabb\tr\chi)(p_0)=\tilde{\oom}(\la,\nu;p_0)\widetilde{\nabb}\tr\tilde{\chi}(\la,\nu;p_0)\nn\\
&&\underline{\Phi}^*_{\la}\Phi^*_{\nu}\oom\ddb_4\nabb\log\oom(p_0)
=\frac{\partial\tilde{\nabb}{\log{\tilde\oom}}}{\partial\nu}(\la,\nu;p_0)
-\tilde{\oom}\tilde{\nabb}\log\tilde{\oom}\!\c\!\tilde{\chi}(\la,\nu;p_0)\ .\nn
\eea
In conclusion the pullback on $S_0$ of the first equation of \ref{5.8} is
\bea
\frac{\partial\tilde{\ze}}{\partial\nu}+\tilde{\oom}\
\!\widetilde{\tr\chi}\tilde{\ze}-\tilde{\oom}\widetilde{\divv}\tilde{\chi}+\tilde{\oom}\tilde{\nabb}\widetilde{\tr\chi}
+\frac{\partial\tilde{\nabb}\widetilde{\log\oom}}{\partial\nu}-\tilde{\oom}\tilde{\nabb}\log\tilde{\oom}\!\c\!\tilde{\chi}=0\ .\eql{2.30uut}
\eea
Equation \ref{2.30uut} has to be compared with the second equation of \ref{2.1ff}. The remaining equations along the outgoing cone are obtained in
the same way. We recall that to write the pull back of the evolution equation along $C(\la)$ for $\chibh$ the following relation has been used
\footnote{The notation here can be misleading. With $\chibh\!\c\!\chi$ we mean $(\chibh\c\chi)_{ab}=\chibh_{ac}\ga^{cd}\chi_{db}$, while with
$(\chibh\c\chih)$ we indicate $(\chibh\c\chih)=\chibh_{ab}\ga^{ac}\ga^{bd}\chih_{cd}$.} 
\bea
(\Lie_N\chibh)(\c,\c)\!=\!\oom[\dddd_4\chibh\!+\!\chibh\!\cdot\!\chi\!+\!\chi\!\c\!\chibh](\c,\c)
\!=\!\oom[\dddd_4\chibh\!+\!(\chibh\!\cdot\!\chih)\ga\!+\!\tr\chi\chibh](\c,\c)\ .\nn 
\eea
Finally the remaining ``outgoing equations" have the following expressions:
\bea
&&\ML\frac{\partial\widetilde{\tr\chi}}{\partial\nu}+\frac{\tilde{\oom}\widetilde{\tr\chi}}{2}\widetilde{\tr\chi}+2\tilde{\oom}\tilde{\om}
\widetilde{\tr\chi}+\tilde{\oom}|\hat{\tilde{\chi}}|^2\!=\!0\nn\\
&&\ML\frac{\partial\widetilde{\tr\chib}}{\partial\nu}+\tilde{\oom}\widetilde{\tr\chi}\widetilde{\tr\chib}
-2\tilde{\oom}\tilde{\om}\widetilde{\tr\chib}-2\tilde{\oom}\widetilde{\divv}\tilde{\etab}-2\tilde{\oom}|\tilde{\etab}|^2
+2\tilde{\oom}\tilde{\bf K}\!=\!0\nn\\
&&\ML\frac{\partial\hat{\tilde{\chib}}}{\partial\nu}-\frac{\tilde{\oom}\widetilde{\tr\chi}}{2}\hat{\tilde{\chib}}
+\frac{\tilde{\oom}\widetilde{\tr\chib}}{2}\tilde{\chih}
+\frac{\partial\log\tilde{\oom}}{\partial\nu}\hat{\tilde{\chib}}-\tilde{\oom}(\hat{\tilde{\chib}}\c\hat{\tilde{\chi}})\ga
-\tilde{\oom}\tilde{\nabb}\hot\tilde{\etab}-\tilde{\oom}(\tilde{\etab}\hot\tilde{\etab})\!=\!0\nn\\
&&\ML\frac{\partial{\tilde{\omb}}}{\partial\nu}\!-\!2{\tilde\oom}\ \!\!{\tilde{\om}}\
\!{\tilde{\omb}}\!-\!\frac{3}{2}{\tilde{\oom}}|\tilde{\ze}|^2
\!+{\tilde{\oom}}\tilde{\ze}\!\c\!\tilde{\nabb}\log\tilde{\oom}\!+\!\frac{1}{2}\tilde{\oom}|\tilde{\nabb}\log\tilde{\oom}|^2
\!+\!\frac{1}{2}\tilde{\oom}\!\left(\!{\tilde{\bf K}}\!+\!\frac{1}{4}\widetilde{\tr\chib}\widetilde{\tr\chi}
\!-\!\frac{1}{2}\hat{\tilde{\chib}}\c\hat{\tilde{\chi}}\!\right)\!=\!0\ \ .\nn 
\eea

\NI As expected from the fact that the vector fields $N$ and $\Nb$ do not commute, the projection on $S_0$ produces non equivalent expressions
when applied to the equations along the incoming cones.\footnote{Apart from the obvious substitution of $\la$ with $\nu$.}
Let us consider again the equation for $\ze$. The derivative with respect to $\la$ of $\tilde{\ze}$ is:
\bea
\frac{\partial\tilde{\ze}}{\partial\la}(\la,\nu;p_0)=\underline{\Phi}^*_{\la}\Phi^*_{\nu}(\Lie_V\ze)(q)\eql{2.17y}
\eea
where 
\bea
V={\Phi_*}_{\nu}\Nb\ \eql{5.26}
\eea
is the vector field generating the diffeomorphism ${\Phi}_{\nu}\!\circ\!\underline{\Phi}_{\la}\!\circ\!\Phi^{-1}_{\nu}$.
In fact, 
\bea
&&\ML\ML\frac{\partial\tilde{\ze}}{\partial\la}(\la,\nu;p_0)=\underline{\Phi}^*_{\la}
\lim_{h\rightarrow0}\frac{1}{h}\big[\underline{\Phi}^*_{h}\Phi^*_{\nu}\ \!\ze(\Phi_{\nu}\circ\underline{\Phi}_{\la+h}(p_0))\!-\!\Phi^*_{\nu}\
\!\ze(\Phi_{\nu}\circ\underline{\Phi}_{\la}(p_0))\big]\nn\\
&&\ML\ML=\underline{\Phi}^*_{\la}\Phi^*_{\nu}\lim_{h\rightarrow0}\frac{1}{h}\big[{\Phi^*}^{-1}_{\nu}\underline{\Phi}^*_{h}\Phi^*_{\nu}\
\!\ze(\Phi_{\nu}\circ\underline{\Phi}_{\la+h}(p_0))\!-\!\ze(\Phi_{\nu}\circ\underline{\Phi}_{\la}(p_0))\big]\\
&&\ML\ML=\underline{\Phi}^*_{\la}\Phi^*_{\nu}\lim_{h\rightarrow0}\frac{1}{h}\big[({\Phi^*}^{-1}_{\nu}\underline{\Phi}^*_{h}\Phi^*_{\nu}\
\!\ze)(q)\!-\!\ze(q)\big]=\underline{\Phi}^*_{\la}\Phi^*_{\nu}(\Lie_V\ze)(\Psi(\la,\nu)(p_0))=\underline{\Phi}^*_{\la}\Phi^*_{\nu}(\Lie_V\ze)(q)\ .\ \ \ \ \ \nn
\eea 
The following relation holds:
\bea
(\Lie_V\ze)(q)=(\Lie_{\Nb}\ze)(q)-(\Lie_X\ze)(q)\eql{5.34}
\eea
and $\tilde{\ze}$ satisfies the following equation:
\bea
\frac{\partial\tilde{\ze}}{\partial\la}\!-\!\frac{\partial\tilde{\nabb}\widetilde{\log\oom}}{\partial\la}\!+\!\tilde{\oom}\
\!\widetilde{\tr\chib}\tilde{\ze}\!+\!\tilde{\oom}\widetilde{\divv}\tilde{\chib}\!-\!\tilde{\oom}\tilde{\nabb}\widetilde{\tr\chib}
\!+\!\tilde{\nabb}\log\tilde{\oom}\c\tilde{\chib}\!+\!{\Lie_{\tilde X}\tilde{\ze}}\!-\!\Lie_{\tilde X}\tilde{\nabb}\log\tilde{\oom}=0\ .\ \ \
\eql{2.21yh}
\eea

\NI To express all the remaining equations on the incoming cones pulled back to $S_0$ we use also the following equation, whose proof follows in the next subsection:
\bea
&&\underline{\Phi}^*_{\la}\Phi^*_{\nu}(\oom\ddb_3f)(q)
=\frac{\partial \tilde{f}}{\partial\la}(p_0)+{\tilde X}|_{p_0}({\tilde f})\eql{2.22wy}\ .
\eea
Using also equation \ref{2.22wy} we obtain
\bea
&&\ML\frac{\partial{\tilde{\om}}}{\partial\la}\!+\!\tilde{\nabb}_{\!\tilde X}\tilde{\om}\!-\!2{\tilde\oom}\ \!\!{\tilde{\omb}}\
\!{\tilde{\om}}\!-\!\frac{3}{2}{\tilde{\oom}}|\tilde{\ze}|^2\!-\!{\tilde{\oom}}\tilde{\ze}\!\c\!\tilde{\nabb}\log\tilde{\oom}
\!+\!\frac{1}{2}\tilde{\oom}|\tilde{\nabb}\log\tilde{\oom}|^2\!+\!\frac{1}{2}\tilde{\oom}\!\left({\tilde{\bf
K}}\!+\!\frac{1}{4}\widetilde{\tr\chi}\widetilde{\tr\chib}\!-\!\frac{1}{2}\hat{\tilde{\chi}}\!\c\!\hat{\tilde{\chib}}\!\right)=0\ \nn\\ 
&&\ML\frac{\partial\widetilde{\tr\chi}}{\partial\la}+\tilde{\oom}\widetilde{\tr\chib}\widetilde{\tr\chi}
-2\tilde{\oom}\tilde{\omb}\widetilde{\tr\chi}+\tilde{\nabb}_{\!\tilde X}\widetilde{\tr\chi}
-2\tilde{\oom}\widetilde{\divv}\tilde{\eta}-2\tilde{\oom}|\tilde{\eta}|^2 +2\tilde{\oom}\tilde{\bf K}\!=\!0\eql{5.37a}\\
&&\ML\frac{\partial\hat{\tilde{\chi}}}{\partial\la}\!-\!\frac{\tilde{\oom}\widetilde{\tr\chib}}{2}\hat{\tilde{\chi}}
\!+\!\frac{\tilde{\oom}\widetilde{\tr\chi}}{2}\hat{\tilde{\chib}}
\!+\!\frac{\partial\log\tilde{\oom}}{\partial\la}\hat{\tilde{\chi}}\!+\!(\tilde{\nabb}_{\!\tilde X}\log\tilde{\oom})\hat{\tilde{\chi}}
\!-\!\tilde{\oom}(\hat{\tilde{\chi}}\c\hat{\tilde{\chib}})\ga-\tilde{\oom}\
\!\tilde{\nabb}\hot\tilde{\eta}\!-\!\tilde{\oom}(\tilde{\eta}\hot\tilde{\eta})\!+\!\Lie_{\tilde{X}}\hat{\tilde{\chi}}\!=\!0\nn\\
&&\ML\frac{\partial\tilde{\ze}}{\partial\la}+\tilde{\oom}\
\!\widetilde{\tr\chib}\tilde{\ze}+\tilde{\oom}\widetilde{\divv}\tilde{\chib}-\tilde{\oom}\tilde{\nabb}\widetilde{\tr\chib}
-\frac{\partial\tilde{\nabb}\widetilde{\log\oom}}{\partial\la}+\tilde{\oom}\tilde{\nabb}\log\tilde{\oom}\!\c\!\tilde{\chib}
+{\Lie_{\tilde X}\tilde{\ze}}-\Lie_{\tilde X}\tilde{\nabb}\log\tilde{\oom}\!=\!0\ .\nn\\
&&\ML\frac{\partial\widetilde{\tr\chib}}{\partial\la}+\frac{\tilde{\oom}\widetilde{\tr\chib}}{2}\widetilde{\tr\chib} +\tilde{\nabb}_{\!\tilde
X}\widetilde{\tr\chib}+2\tilde{\oom}\tilde{\omb}\widetilde{\tr\chib}+\tilde{\oom}|\hat{\tilde{\chib}}|^2\!=\!0\ .\nn
\eea

\subsection{The structure equations in a vacuum Einstein spacetime for the metric components in the $\{\la,\nu,\theta,\phi\}$ coordinates.}
To complete the set of p.d.e. equations  representing the Einstein equations
in this ``double null foliation gauge" we have still to add some other equations. The previous equations are
the analogue of the first order equations for the second fundamental form $k_{ij}$ associated to the spacelike foliation of the Einstein spacetime. We need the analogue of the equations for the three dimensional metric $g_{ij}$. Observe that in this foliation the metric written in the ``adapted" coordinates is given in \ref{met} and the six quantity associated to the metric are $\oom, X_a, \ga_{ab}$.
The corresponding quantities, pulledback to $TS_0$, are
\bea
&&\!\!\!\!\!\!\!\!\!\!\tilde{\ga}(\la,\nu;p_0)
=\!\underline{\Phi}^*_{\la}\Phi^*_{\nu}\ga(\Phi_{\nu}\circ\underline{\Phi}_{\la}(p_0))=\underline{\Phi}^*_{\la}\Phi^*_{\nu}\ga(q)\nn\\
&&\!\!\!\!\!\!\!\!\!\!\tilde{X}(\la,\nu;p_0)
=\underline{\Phi}^*_{\la}\Phi^*_{\nu}X(\Phi_{\nu}\circ\underline{\Phi}_{\la}(p_0))=\underline{\Phi}^*_{\la}\Phi^*_{\nu}X(q)\nn\\ 
&&\!\!\!\!\!\!\!\!\!\!\tilde{\oom}(\la,\nu;p_0)
=\oom(\Phi_{\nu}\circ\underline{\Phi}_{\la}(p_0))=\oom(q)\ ,\nn
\eea
and, proceeding as before, their partial derivatives with respect to $\nu$ and $\la$ are
\bea
&&\ML\frac{\partial\tilde{\ga}}{\partial\nu}(\c,\c)(\la,\nu;\theta,\phi)
=\underline{\Phi}^*_{\la}\Phi^*_{\nu}(\Lie_N\ga)(\c,\c)(q)=2\tilde{\oom}\ \!\tilde{\chi}(\c,\c)(\la,\nu;\theta,\phi)\eql{2.49xz}\\
&&\ML\frac{\partial\tilde{\ga}}{\partial\la}(\c,\c)(\la,\nu;\theta,\phi)
\!=\!\underline{\Phi}^*_{\la}\Phi^*_{\nu}(\Lie_V\ga)(\c,\c)(q)\!=\!2\tilde{\oom}\ \!\tilde{\chib}(\c,\c)(\la,\nu;\theta,\phi)
\!-\!(\Lie_{\tilde X}\tilde{\ga})(\c,\c)(\la,\nu;\theta,\phi)\ .\nn
\eea
Beside these equations we need, in these coordinates, the equation connecting $X$ to the connection coefficients and  analogous equations for $\oom$. From the definition of $\om$ and $\omb$, in subsection \ref{SS2.3}, it follows
\bea
&&\tilde{\om}(\la,\nu;p_0)=-\frac{1}{2\tilde{\oom}}\frac{\partial\log\tilde{\oom}}{\partial\nu}(\la,\nu;p_0)\eql{2.51xz}\\
&&\tilde{\omb}(\la,\nu;p_0)
=-\frac{1}{2\tilde{\oom}}\bigg(\frac{\partial\log\tilde{\oom}}{\partial\la}+\tilde{\nabb}_{\tilde{X}}\log\tilde{\oom}\bigg)\!(\la,\nu;p_0)\ .\nn
\eea 
The equation for $X$ is derived from the commutation relation of $N$ with $\Nb$,
\bea
&&[N,\Nb]=-4\oom^2Z\ ,\\
&&[N,\Nb]=\Lie_N(\frac{\partial}{\partial \la}+X)=\Lie_{\frac{\partial}{\partial\nu}}(\frac{\partial}{\partial \la}+X)=\Lie_NX\ .\nn
\eea
It follows, in the $\{\la,\nu,\om^a\}$ coordinates, 
\bea
&&\ML\ML\frac{\partial{\tilde X}}{\partial\nu}(\la,\nu;p_0)
=-{\underline{\Phi}_*}^{-1}_{\la}{\Phi_*}^{-1}_{\nu}4\oom^2Z=4{\tilde\oom}^2{\tilde Z}(\la,\nu;p_0)\ ,\eql{2.56xz}
\eea
where ${\tilde Z}(\c)=\sum_{A\in\{1,2\}}{\tilde\ze}({\tilde e}_A){\tilde e}_A(\c)$\ \ .

\subsection{The equations  \ref{2.1ff} in the $\{\la,\nu,\theta,\phi\}$ coordinates.}
We collect all the equations \ref{2.1ff}, written now as p.d.e. equations for covariant tensors on $S_0$, depending on
the parameters $\la,\nu$.\footnote{It is clear that this equations can be immediately written as equations for the various components of these tensors. In fact the orthonormal frame introduced in $S_0$ does not depend on $\la,\nu$. } They are, indicating also the Ricci component to which they are associated,
\bea
&&{\bf R}(e_4,e_4)\!=\!0:\nn\\
&&\frac{\partial\widetilde{\tr\chi}}{\partial\nu}+\frac{\tilde{\oom}\widetilde{\tr\chi}}{2}\widetilde{\tr\chi}+2\tilde{\oom}\tilde{\om}
\widetilde{\tr\chi}+\tilde{\oom}|\hat{\tilde{\chi}}|^2\!=\!0\nn\\
&&{\bf R}(e_4,e_A)\!=\!0:\nn\\
&&\frac{\partial\tilde{\ze}}{\partial\nu}+\tilde{\oom}\
\!\widetilde{\tr\chi}\tilde{\ze}-\tilde{\oom}\widetilde{\divv}\tilde{\chi}+\tilde{\oom}\tilde{\nabb}\widetilde{\tr\chi}
+\frac{\partial\tilde{\nabb}\widetilde{\log\oom}}{\partial\nu}-\tilde{\oom}\tilde{\nabb}\log\tilde{\oom}\c\tilde{\chi}\!=\!0\nn\\
&&\de_{AB}{\bf R}(e_A,e_B)\!=\!0:\nn\\
&&\frac{\partial\widetilde{\tr\chib}}{\partial\nu}+\tilde{\oom}\widetilde{\tr\chi}\widetilde{\tr\chib}
-2\tilde{\oom}\tilde{\om}\widetilde{\tr\chib}-2\tilde{\oom}\widetilde{\divv}\tilde{\etab}-2\tilde{\oom}|\tilde{\etab}|^2+2\tilde{\oom}\tilde{\bf
K}\!=\!0\nn\\
&&\widehat{{\bf R}(e_A,e_B)}\!=\!0:\nn\\
&&\frac{\partial\hat{\tilde{\chib}}}{\partial\nu}-\frac{\tilde{\oom}\widetilde{\tr\chi}}{2}\hat{\tilde{\chib}}
+\frac{\tilde{\oom}\widetilde{\tr\chib}}{2}\tilde{\chih}
+\frac{\partial\log\tilde{\oom}}{\partial\nu}\hat{\tilde{\chib}}-\tilde{\oom}(\hat{\tilde{\chib}}\c\hat{\tilde{\chi}})\ga
-\tilde{\oom}\tilde{\nabb}\hot\tilde{\etab}-\tilde{\oom}(\tilde{\etab}\hot\tilde{\etab})\!=\!0\nn\\
&&\widehat{{\bf R}(e_3,e_4)}\!=\!0:\nn\\
&&\frac{\partial{\tilde{\omb}}}{\partial\nu}\!-\!2{\tilde\oom}\ \!\!{\tilde{\om}}\
\!{\tilde{\omb}}\!-\!\frac{3}{2}{\tilde{\oom}}|\tilde{\ze}|^2
\!+{\tilde{\oom}}\tilde{\ze}\!\c\!\tilde{\nabb}\log\tilde{\oom}\!+\!\frac{1}{2}\tilde{\oom}|\tilde{\nabb}\log\tilde{\oom}|^2
\!+\!\frac{1}{2}\tilde{\oom}\!\left(\!{\tilde{\bf K}}\!+\!\frac{1}{4}\widetilde{\tr\chib}\widetilde{\tr\chi}
\!-\!\frac{1}{2}\hat{\tilde{\chib}}\c\hat{\tilde{\chi}}\!\right)\!=\!0\nn\\ 
&&\widehat{{\bf R}(e_3,e_4)}\!=\!0:\nn\\
&&\frac{\partial{\tilde{\om}}}{\partial\la}\!+\!\tilde{\nabb}_{\!\tilde X}\tilde{\om}\!-\!2{\tilde\oom}\ \!\!{\tilde{\omb}}\
\!{\tilde{\om}}\!-\!\frac{3}{2}{\tilde{\oom}}|\tilde{\ze}|^2\!-\!{\tilde{\oom}}\tilde{\ze}\!\c\!\tilde{\nabb}\log\tilde{\oom}
\!+\!\frac{1}{2}\tilde{\oom}|\tilde{\nabb}\log\tilde{\oom}|^2\!+\!\frac{1}{2}\tilde{\oom}\!\left({\tilde{\bf
K}}\!+\!\frac{1}{4}\widetilde{\tr\chi}\widetilde{\tr\chib}\!-\!\frac{1}{2}\hat{\tilde{\chi}}\!\c\!\hat{\tilde{\chib}}\!\right)=0\ \nn\\ 
&&\de_{AB}{{\bf R}(e_A,e_B)}\!=\!0:\nn\\
&&\frac{\partial\widetilde{\tr\chi}}{\partial\la}+\tilde{\oom}\widetilde{\tr\chib}\widetilde{\tr\chi}
-2\tilde{\oom}\tilde{\omb}\widetilde{\tr\chi}+\tilde{\nabb}_{\!\tilde X}\widetilde{\tr\chi}
-2\tilde{\oom}\widetilde{\divv}\tilde{\eta}-2\tilde{\oom}|\tilde{\eta}|^2 +2\tilde{\oom}\tilde{\bf K}\!=\!0\eql{5.37b}\\
&&\widehat{{\bf R}(e_A,e_B)}\!=\!0:\nn\\
&&\frac{\partial\hat{\tilde{\chi}}}{\partial\la}\!-\!\frac{\tilde{\oom}\widetilde{\tr\chib}}{2}\hat{\tilde{\chi}}
\!+\!\frac{\tilde{\oom}\widetilde{\tr\chi}}{2}\hat{\tilde{\chib}}
\!+\!\frac{\partial\log\tilde{\oom}}{\partial\la}\hat{\tilde{\chi}}\!+\!(\tilde{\nabb}_{\!\tilde X}\log\tilde{\oom})\hat{\tilde{\chi}}
\!-\!\tilde{\oom}(\hat{\tilde{\chi}}\c\hat{\tilde{\chib}})\ga-\tilde{\oom}\
\!\tilde{\nabb}\hot\tilde{\eta}\!-\!\tilde{\oom}(\tilde{\eta}\hot\tilde{\eta})\!+\!\Lie_{\tilde{X}}\hat{\tilde{\chi}}\!=\!0\nn\\
&&{\bf R}(e_3,e_A)\!=\!0:\nn\\
&&\frac{\partial\tilde{\ze}}{\partial\la}+\tilde{\oom}\
\!\widetilde{\tr\chib}\tilde{\ze}+\tilde{\oom}\widetilde{\divv}\tilde{\chib}-\tilde{\oom}\tilde{\nabb}\widetilde{\tr\chib}
-\frac{\partial\tilde{\nabb}\widetilde{\log\oom}}{\partial\la}+\tilde{\oom}\tilde{\nabb}\log\tilde{\oom}\!\c\!\tilde{\chib}
+{\Lie_{\tilde X}\tilde{\ze}}-\Lie_{\tilde X}\tilde{\nabb}\log\tilde{\oom}\!=\!0\ .\nn\\
&&\widehat{{\bf R}(e_3,e_3)}\!=\!0:\nn\\
&&\frac{\partial\widetilde{\tr\chib}}{\partial\la}+\frac{\tilde{\oom}\widetilde{\tr\chib}}{2}\widetilde{\tr\chib} +\tilde{\nabb}_{\!\tilde
X}\widetilde{\tr\chib}+2\tilde{\oom}\tilde{\omb}\widetilde{\tr\chib}+\tilde{\oom}|\hat{\tilde{\chib}}|^2\!=\!0\ ,\nn
\eea
where \ $\eta=\ze+\nabb\log\oom$ and $\etab=-\ze+\nabb\log\oom$ and $\widehat{{\bf R}(\c,\c)}$ is the traceless part of the Ricci tensor.
To have a closed set of equations we have to add the equations at the level of the metric components, \ref{2.49xz}, \ref{2.51xz} and
\ref{2.56xz},
\bea
&&\ML\frac{\partial\tilde{\ga}}{\partial\nu}
-2\tilde{\oom}\ \!\tilde{\chi}=0\ \ \ \ ,\ \ \ \ \ \ \frac{\partial\tilde{\ga}}{\partial\la}
-2\tilde{\oom}\ \!\tilde{\chib}+\Lie_{\tilde X}\tilde{\ga}=0\eql{2.59y}\\
&&\ML\frac{\partial\log\tilde{\oom}}{\partial\nu}+{2\tilde{\oom}}\ \!\tilde{\om}=0\ \ \ ,\ \ 
\frac{\partial\log\tilde{\oom}}{\partial\la}+{\tilde{\nabb}}_{X}\log\tilde{\oom}
+{2\tilde{\oom}}\ \!\tilde{\omb}=0\nn\\
&&\ML\frac{\partial{\tilde X}}{\partial\nu}-4{\tilde\oom}^2{\tilde Z}=0\ .\nn
\eea

\subsection{The technical proofs.}
\subsubsection{Proof of equations \ref{2.15k}}
$(\underline{\Phi}^*_{\la}\Phi^*_{\nu}\tr\chi\zeta)(p_0)$ is the pullback of
$\tr\chi(q)\ze(q)$. Therefore, denoting $Y_{p_0}$ a vector $\in {TS_0}|_{p_0}$ we have 
\bea
&&\ML(\underline{\Phi}^*_{\la}\Phi^*_{\nu}\oom\tr\chi\zeta)(Y_{p_0})=\oom(q)(\tr\chi\zeta)(q)({\Phi_*}_{\nu}
{\underline{\Phi}_*}_{\la}Y_{p_0})\\
&&\ML=(\oom\tr\chi)(\Psi(\la,\nu)(p_0))(\underline{\Phi}^*_{\la}\Phi^*_{\nu}\zeta)(p_0)(Y_{p_0})
=\tilde{\oom}(\la,\nu;p_0)\widetilde{\tr\chi}(\la,\nu;p_0)\tilde{\ze}(\la,\nu;p_0)(Y_{p_0})\ .\nn
\eea
Proceeding in the same way we obtain
\bea
(\underline{\Phi}^*_{\la}\oom\Phi^*_{\nu}\zeta\!\c\!\chi)(p_0)=\tilde{\oom}(\la,\nu;p_0)\tilde{\ze}(\la,\nu;p_0)\!\c\!\widetilde{\chi}(\la,\nu;p_0)\ .
\eea
To prove the third relation of \ref{2.15k},
\beaa
(\underline{\Phi}^*_{\la}\Phi^*_{\nu}\oom\divv\chi)(p_0)=\tilde{\oom}(\la,\nu;p_0)\widetilde{\divv}\tilde{\chi}(\la,\nu;p_0)\ ,
\eeaa
observe that $\divv\chi(\c)$ is a one form defined on the spacetime; let $Y_{p_0}$ be an $S$-tangent vector field  at $p_0$, then
\bea
&&(\underline{\Phi}^*_{\la}\Phi^*_{\nu}\divv\chi)(p_0)(Y_{p_0})=\divv\chi(q)({\Phi_*}_{\nu}{\underline{\Phi}_*}_{\la}Y_{p_0})
=\divv\chi(q)(({\Phi_*}_{\nu}{\underline{\Phi}_*}_{\la}Y)_{q})\ \ \ \nn\\
&&=\ga^{ab}(q)(\nabb_a\chi)_{bc}(q)(({\Phi_*}_{\nu}{\underline{\Phi}_*}_{\la}Y)_{q})
=\ga^{ab}(q)(\nabb_a\chi)_{be}(q)(({\Phi_*}_{\nu}{\underline{\Phi}_*}_{\la}Y)_{q})^e\ .\ \ \ \ \ \ \ \ \ \ \ \eql{5.16}
\eea
Immediately, with $p={\underline{\Phi}}_{\la}(p_0)$,
\bea
({\Phi_*}_{\nu}{\underline{\Phi}_*}_{\la}Y)^e(q)=\frac{\partial{\Phi}^e_{\nu}}{\partial\om^d}\bigg|_{p}
\frac{\partial\underline{\Phi}^d_{\la}}{\partial\om^c}\bigg|_{p_0}\!Y^c(p_0)={\Phi_*}_{\nu}{\underline{\Phi}_*}_{\la}Y(p_0)\ .
\eea
Therefore \ref{5.16} becomes
\bea
\ML\ML(\underline{\Phi}^*_{\la}{\Phi}^*_{\nu}\divv\chi)(p_0)(Y_{p_0})\!&=&\!\ga^{ab}(q)(\nabb_a\chi)_{be}(q)\frac{\partial{\Phi}^e_{\nu}}{\partial\om^d}\bigg|_{p}\frac{\partial\underline{\Phi}^d_{\la}}{\partial\om^c}\bigg|_{p_0}\!Y^c(p_0)\nn\\
\!&=&\!\ga^{ab}(q)(\nabb_a\chi)_{be}(q)
\frac{\partial{\Psi}(\la,\nu)^e}{\partial\om^c}\bigg|_{p_0}\!\!Y^c(p_0)\ ,\nn 
\eea
so that,
\bea
(\underline{\Phi}^*_{\la}{\Phi}^*_{\nu}\divv\chi)(p_0)=\left(\!\ga^{ab}(q)(\nabb_a\chi)_{be}(q)
\frac{\partial{\Psi}(\la,\nu)^e}{\partial\om^c}\bigg|_{p_0}\!\right)\!d\tilde{\om}^c|_{p_0}\ .
\eea
The remaining part of the proof is trivial and we do not report it here.\footnote{It is enough to observe that concerning the indices $a$ and $b$
$\ga^{ab}(q)(\nabb_a\chi)_{be}(q)$ is a scalar and therefore it can be rewritten in terms of the tilded quantities.}
To prove the last equation in \ref{2.15k} we observe that
\bea
\ML(\oom\dd_4\nabb\log\oom)(e_A)\!&=&\!(\Lie_N\nabb\log\oom)(e_A)+(\nabb\log\oom)([N,e_A])-\nabb\log\oom(\dd_Ne_A)\nn\\
\ML\!&=&\!(\Lie_N\nabb\log\oom-\oom\nabb\log\oom\!\c\!\chi)(e_A)\ .
\eea 
Therefore
\bea
\ML\underline{\Phi}^*_{\la}\Phi^*_{\nu}\oom\ddb_4\nabb\log\oom\!&=&\!\underline{\Phi}^*_{\la}\Phi^*_{\nu}\Lie_N\nabb\log\oom
-\underline{\Phi}^*_{\la}\Phi^*_{\nu}(\oom\nabb\log\oom\!\c\!\chi)\nn\\
\ML\!&=&\!\frac{\partial\tilde{\nabb}{\log{\tilde\oom}}}{\partial\nu}-\tilde{\oom}\tilde{\nabb}\log\tilde{\oom}\!\c\!\tilde{\chi}\ \ \ .
\eea
\subsubsection{Proof of equation \ref{5.34}.}
We compute $V={\Phi_*}_{\nu}\Nb$ in the following way:
\bea
({\Phi_*}_{\nu}\Nb)|_q(f)=\Nb|_{p}(f\circ{\Phi}_{\nu})
\eea
where $p=\underline{\Phi}_{\la}(p_0)$ and $q=\Phi_{\nu}(p)=\Phi_{\nu}\!\circ\!{\Phi}_{\la}(p_0)$, $f(\c)$ is a scalar function defined on $T\M_q$ and
$f\!\circ\!{\Phi}_{\nu}(\c)$ is defined on $T\M_{p}$. Moreover $p\in S(\la,0)\subset\Cb_0$ and we assume that on $\Cb_0$,\footnote{As part of the
initial conditions.} $\Nb=\frac{\partial}{\partial \la}$. Therefore
\bea
\Nb|_{p}(f\circ{\Phi}_{\nu})=\frac{\partial}{\partial u}\bigg|_{p}(f\circ{\Phi}_{\nu})=\frac{\partial{{\Phi}_{\nu}}^{\ro}}{\partial u}\bigg|_{p}
\frac{\partial f}{\partial x^{\ro}}\bigg|_{q}\ 
\eea
and from it
\bea
({\Phi_{\nu}}_*\Nb)|_q=\frac{\partial{{\Phi}_{\nu}}^{\ro}}{\partial u}\bigg|_{p}\frac{\partial}{\partial x^{\ro}}\bigg|_{q}\ .\eql{5.30}
\eea
As ${\Phi}_{\nu}$ is generated by the vector field $N=\frac{\partial}{\partial\nu}$, 
in the $\{\la,\nu,\om^a\}$ coordinates, a null geodesic along $C(\la)$ starting from the point $p=(0,\la,\theta,\phi)\in S(\la,0)\subset\Cb_0$ is,
\bea
&&\Phi^{\ub}_{\nu}(0,\la,\theta,\phi)=x^{\ub}(\nu;0,\la,\theta,\phi)=\nu\nn\\
&&\Phi^{u}_{\nu}(0,\la,\theta,\phi)=x^{u}(\nu;0,\la,\theta,\phi)=\la\\
&&\Phi^{\theta}_{\nu}(0,\la,\theta,\phi)=x^{\theta}(\nu;0,\la,\theta,\phi)=\theta\nn\\
&&\Phi^{\phi}_{\nu}(0,\la,\theta,\phi)=x^{\phi}(\nu;0,\la,\theta,\phi)=\phi\ .\nn
\eea
This implies ${\partial{{\Phi}_{\nu}}^{\ro}}/{\partial\la}=\de^{\ro}_{\la}$ which, applied to \ref{5.30}, gives
\bea
T\M_q\ni V|_q=({\Phi_{\nu}}_*\Nb)|_q=\frac{\partial}{\partial\la}\bigg|_q=\Nb|_q-X|_q\eql{5.33}
\eea
and, therefore,
\bea
(\Lie_V\ze)(q)=(\Lie_{\Nb}\ze)(q)-(\Lie_X\ze)(q)\ .\eql{5.34w}
\eea
\subsubsection{Proof of equation \ref{2.21yh}.}
\bea
&&\!\!\!\!\!\!\!\!\!\frac{\partial\tilde{\ze}}{\partial\la}(q)(\c)=\underline{\Phi}^*_{\la}\Phi^*_{\nu}
\left[(\Lie_{\Nb}\ze)(q)-(\Lie_X\ze)(q)\right](\c)
=\underline{\Phi}^*_{\la}\Phi^*_{\nu}\left[(\oom\dddd_3\zeta+\oom\zeta\!\cdot\!\chib-\Lie_X\ze)(q)\right](\c)\nn\\
&&\!\!\!\!\!\!\!\!\!=\left(\underline{\Phi}^*_{\la}\Phi^*_{\nu}\oom\left[(-\tr\chib\zeta-\divv\chib+\nabb\tr\chib
+\ddb_3\nabb\log\oom-\oom^{-1}\Lie_X\ze)\right]\right)\!(q)(\c)\ \ \ \ \nn
\eea
where we have applied equation \ref{5.8} satisfied by $\ze$ along the incoming cones.
\smallskip

\NI Let us compute $\underline{\Phi}^*_{\la}\Phi^*_{\nu}\oom\ddb_3\nabb\log\oom$ as we did in the previous case:
\bea
&&(\oom\dd_3\nabb\log\oom)(e_A)=\Lie_{\Nb}((\nabb\log\oom)(e_A))-\nabb\log\oom(\dd_{\Nb}e_A)\nn\\
&&=(\Lie_{\Nb}\nabb\log\oom)(e_A)+(\nabb\log\oom)([{\Nb},e_A])-\nabb\log\oom(\dd_{\Nb}e_A)\nn\\
&&=(\Lie_{\Nb}\nabb\log\oom-\oom\nabb\log\oom\!\c\!\chib)(e_A)\ .
\eea
Observe that
\bea
\widetilde{\Lie_X\ze}={\Lie_{\tilde X}\tilde{\ze}}
\eea
where \footnote{Observe that ${\tilde X}$ belongs to $TS_0$, but has not to be confused with $X$ on the points of $S_0$ which is identically zero.}
\bea
{\tilde X}(p_0)=\underline{\Phi}^{-1}_{*\la}\Phi^{-1}_{*\nu}X(q)\ .\eql{5.38a}
\eea
Recalling equation \ref{5.33},
\bea
&&\underline{\Phi}^*_{\la}\Phi^*_{\nu}\oom\ddb_3\nabb\log\oom=\underline{\Phi}^*_{\la}\Phi^*_{\nu}\Lie_V\nabb\log\oom+
\underline{\Phi}^*_{\la}\Phi^*_{\nu}\Lie_X\nabb\log\oom
-\underline{\Phi}^*_{\la}\Phi^*_{\nu}(\oom\nabb\log\oom\!\c\!\chib)\nn\\
&&=\frac{\partial\tilde{\nabb}\widetilde{\log\oom}}{\partial\la}+\Lie_{\tilde
X}\tilde{\nabb}\log\tilde{\oom}-\tilde{\oom}\tilde{\nabb}\log\tilde{\oom}\!\c\!\tilde{\chib}
\eea
and, therefore the following relation holds:
\bea
&&\!\!\!\!\!\!\!\!\!\left(\underline{\Phi}^*_{\la}\Phi^*_{\nu}\oom\left[(-\tr\chib\zeta-\divv\chib+\nabb\tr\chib
+\ddb_3\nabb\log\oom\right]\right)\!(\la,\nu;p_0)\nn\\
&&\ML =\big(-\tilde{\oom}\
\!\widetilde{\tr\chib}\tilde{\ze}-\tilde{\oom}\widetilde{\divv}\tilde{\chib}+\tilde{\oom}\tilde{\nabb}\widetilde{\tr\chib}
+\frac{\partial\tilde{\nabb}\widetilde{\log\oom}}{\partial\la}+\Lie_{\tilde X}\tilde{\nabb}\log\tilde{\oom}
-\tilde{\oom}\tilde{\nabb}\log\tilde{\oom}\c\tilde{\chib}\big)\!(\la,\nu;p_0)\nn
\eea
so that, finally,
\bea
\frac{\partial\tilde{\ze}}{\partial\la}-\frac{\partial\tilde{\nabb}\widetilde{\log\oom}}{\partial\la}+\tilde{\oom}\
\!\widetilde{\tr\chib}\tilde{\ze}+\tilde{\oom}\widetilde{\divv}\tilde{\chib}-\tilde{\oom}\tilde{\nabb}\widetilde{\tr\chib}
+\tilde{\nabb}\log\tilde{\oom}\c\tilde{\chib}+{\Lie_{\tilde X}\tilde{\ze}}-\Lie_{\tilde X}\tilde{\nabb}\log\tilde{\oom}=0\ .\nn
\eea
\subsubsection{Proof of equation \ref{2.22wy}}
Let us consider $\frac{\partial \tilde{f}}{\partial\la}$ where ${\tilde f}(p_0)=f(\Phi_{\nu}\circ\underline{\Phi}_{\la}(p_0))$
is a scalar function,
\bea
&&\frac{\partial \tilde{f}}{\partial\la}(p_0)=\lim_{h\rightarrow 0}\frac{1}{h}
\left[f(\Phi_{\nu}\circ\underline{\Phi}_{h}\circ\Phi^{-1}_{\nu}\circ\Phi_{\nu}\circ\underline{\Phi}_{\la}(p_0)
-f(\Phi_{\nu}\circ\underline{\Phi}_{\la}(p_0))\right]\nn\\
&&\lim_{h\rightarrow 0}\frac{1}{h}
\left[f(\Phi_{\nu}\circ\underline{\Phi}_{h}\circ\Phi^{-1}_{\nu}(\Phi_{\nu}\circ\underline{\Phi}_{\la}(p_0)))
-f(\Phi_{\nu}\circ\underline{\Phi}_{\la}(p_0))\right]\nn\\
&&=V|_{\Phi_{\nu}\circ\underline{\Phi}_{\la}(p_0)}(f)
=\Nb|_{(\Phi_{\nu}\circ\underline{\Phi}_{\la}(p_0))}(f)-X|_{(\Phi_{\nu}\circ\underline{\Phi}_{\la}(p_0))}(f)\
.\nn
\eea
Therefore
\bea
&&\underline{\Phi}^*_{\la}\Phi^*_{\nu}(\oom\ddb_3f)(q)=({\tilde\oom}\widetilde{\ddb_3f})(p_0)
=(\oom\ddb_3f)(\Phi_{\nu}\circ\underline{\Phi}_{\la}(p_0))\nn\\
&&=\Nb|_{(\Phi_{\nu}\circ\underline{\Phi}_{\la}(p_0))}(f)=\frac{\partial
\tilde{f}}{\partial\la}(p_0)+X|_{(\Phi_{\nu}\circ\underline{\Phi}_{\la}(p_0))}(f)
=\frac{\partial \tilde{f}}{\partial\la}(p_0)+{\tilde X}|_{p_0}({\tilde f})\nn\ .
\eea

\subsubsection{Proof of equations \ref{2.51xz}}
\bea
&&\ML\ML\tilde{\om}(\la,\nu;p_0)={(\om\circ\Phi_{\nu}\circ\underline{\Phi}_{\la})(p_0)}
=-\frac{1}{2}\left((\dd_4\log\oom)\circ\Phi_{\nu}\circ{\underline{\Phi}}_{\la}\right)(p_0)\nn\\
&&\ML\ML=-\frac{1}{2\oom(\Phi_{\nu}\circ{\underline{\Phi}}_{\la}(p_0))}(D_N\log\oom)(\Phi_{\nu}\circ{\underline{\Phi}}_{\la}(p_0))\nn\\
&&\ML\ML=-\frac{1}{2\oom(\Phi_{\nu}\circ{\underline{\Phi}}_{\la}(p_0))}\lim_{h\rightarrow
0}\frac{1}{h}\left[\log\oom(\Phi_h\circ\Phi_{\nu}\circ\underline{\Phi}_{\la})(p_0)-\log\oom(\Phi_{\nu}\circ\underline{\Phi}_{\la})(p_0)\right]\nn\\
&&\ML\ML
=-\frac{1}{2\tilde{\oom}}\frac{\partial\log\tilde{\oom}}{\partial\nu}(\la,\nu;p_0)\ .
\eea
Proceeding in the analogous way for $\tilde{\omb}(\la,\nu;p_0)$ we obtain
\bea
&&\ML\tilde{\omb}(\la,\nu;p_0)={(\omb\circ\Phi_{\nu}\circ\underline{\Phi}_{\la})(p_0)}
=-\frac{1}{2}\left((\dd_3\log\oom)\circ\Phi_{\nu}\circ{\underline{\Phi}}_{\la}\right)(p_0)\nn\\
&&\ML=-\frac{1}{2\oom(\Phi_{\nu}\circ{\underline{\Phi}}_{\la}(p_0))}(D_{\Nb}\log\oom)(\Phi_{\nu}\circ{\underline{\Phi}}_{\la}(p_0))\nn\\
&&\ML=-\frac{1}{2\oom(\Phi_{\nu}\circ{\underline{\Phi}}_{\la}(p_0))}\left[D_{V}\!\log\oom(\Phi_{\nu}\circ{\underline{\Phi}}_{\la}(p_0))
+\nabb_{\!X}\!\log\oom(\Phi_{\nu}\circ{\underline{\Phi}}_{\la}(p_0))\right]\nn\\
&&\ML=-\frac{1}{2\oom(\Phi_{\nu}\circ{\underline{\Phi}}_{\la}(p_0))}\lim_{h\rightarrow
0}\frac{1}{h}\left(\log\oom(\Phi_{\nu}\circ\underline{\Phi}_{h}\circ\Phi^{-1}_{\nu}\circ\Phi_{\nu}\circ\underline{\Phi}_{\la})(p_0)
-\log\oom(\Phi_{\nu}\circ\underline{\Phi}_{\la})(p_0)\right)\nn\\
&&-\frac{1}{2\oom(\Phi_{\nu}\circ{\underline{\Phi}}_{\la}(p_0))}\nabb_{\!X}\!\log\oom(\Phi_{\nu}\circ{\underline{\Phi}}_{\la}(p_0))
=-\frac{1}{2\tilde{\oom}}\frac{\partial\log\tilde{\oom}}{\partial\la}(\la,\nu;p_0)-\tilde{\nabb}_{\tilde{X}}\log\tilde{\oom}(\la,\nu;p_0)\
.\nn
\eea
\subsubsection{Proof of equation \ref{2.56xz}}
\bea
&&\ML\frac{\partial{\tilde
X}}{\partial\nu}(\la,\nu;p_0)=\!{\underline{\Phi}_*}^{-1}_{\la}{\Phi_*}^{-1}_{\nu}\lim_{h\rightarrow
0}\frac{1}{h}\big[{\Phi_*}^{-1}_{h}X(\Phi_{\nu+h}\circ\underline{\Phi}_{\la}(p_0))-X(\Phi_{\nu}\circ\underline{\Phi}_{\la}(p_0))\big]\nn\\
&&\ML=\!{\underline{\Phi}_*}^{-1}_{\la}{\Phi_*}^{-1}_{\nu}\lim_{h\rightarrow
0}\frac{1}{h}\big[({\Phi_*}^{-1}_{h}X)(q)-X(q)\big]
=\!{\underline{\Phi}_*}^{-1}_{\la}{\Phi_*}^{-1}_{\nu}(\Lie_NX)(\Phi_{\nu}\circ\underline{\Phi}_{\la}(p_0))\nn\\
&&\ML=\!-{\underline{\Phi}_*}^{-1}_{\la}{\Phi_*}^{-1}_{\nu}4\oom^2Z=4{\tilde\oom}^2{\tilde Z}(\la,\nu;p_0)\ .\nn
\eea
\subsubsection{Proof of equations \ref{3.18d}.}
\NI{\bf Remark:} Observe that $\ga(\la,\nu)(\c,\c)$ is a symmetric covariant two tensor defined on $S_0$. We can
consider it as a metric tensor, but the natural metric tensor on $S_0$ is $\ga_0(\c,\c)\equiv\ga(0,0)(\c,\c)$.\footnote{With respect to the
metric
$\ga_0(\c,\c)\equiv\ga(0,0)(\c,\c)$, we define the area of $S_0$ as $|S_0|_{\ga_0}$ and  its radius $\sqrt{4\pi}r_0\equiv
|S_0|^{\frac{1}{2}}_{\ga_0}$. As $S_0$ is diffeomorphic to $S^2$, if the initial data on $S_0$ are ``small", see for more details
\cite{Ca-Ni:char}, we can conclude that the metric $\ga_0(\c,\c)$ is ``near" to $r_0(d\theta^2+\sin\theta^2d\phi^2)$.}
On the other side, as discussed before, $\ga(\la,\nu)(\c,\c)$ is the pull back via $\Psi^*(\la,\nu)$ of a metric tensor induced on
$S(\la,\nu)$. The relationship, reintroducing the tildas temporarily, is \ref{2.49xz},
\[\frac{\partial}{\partial\nu}\tilde{\ga}(\la,\nu;p_0)(\c,\c)=(\Psi^*(\la,\nu)\Lie_N\ga)(\Psi(\la,\nu)(p_0))(\c,\c)\ .\]
Denoting $\{e^{(0)}_A\}$ an orthonormal frame on $S_0$ with respect to the metric $\ga_0$ it is clear that, in general,
$\tilde{\ga}(e^{(0)}_A,e^{(0)}_B)\neq\de_{AB}$.
The variables $v_{cba}(\la,\nu,\om)$ are the (non covariant) derivatives with respect to the angular variables of
$\tilde{\ga}(\la,\nu;p_0)=\tilde{\ga}(\la,\nu;\om)$ and do not have to be thought as the transport of partial derivatives of ${\ga}(\la,\nu;p_0)$
on $S(\la,\nu)$. The pullback is defined for the covariant tensors on
$S(\la,\nu)$. This is true also for the covariant derivatives in the sense that denoted $W$
a tensor on $S(\la,\nu)$ it follows that $\widetilde{\nabb W}=\tilde{\nabb}\tilde{W}$ and $\tilde{\nabb}$ is the covariant derivative
done with respect to the metric $\tilde{\ga}$. Therefore if, for instance, $W$ is a one form we have
\bea
\widetilde{\nabb W}_{ab}=\tilde{\nabb}_a\tilde{W}_b=\frac{\partial}{\partial\om^a}\tilde{W}_b-{\tilde{\Ga}}^c_{ab}\tilde{W}_c
\eea
\bea
\ML\ML\ML\ML\ML\mbox{and}\ \ \ \ \ \ \ \ \ \ \ \ \ \
{\tilde{\Ga}}^c_{ab}=\frac{1}{2}\tilde{\ga}^{cd}\left({v}_{adb}+{v}_{bad}-{v}_{dab}\right)\ ,\eql{3.16d}
\eea
where $v_{abc}$ are the unknown functions of the first order system defined in \ref{3.13wq}. 
Omitting the tildas, we have:
\bea
&&\ML\ML\frac{\partial}{\partial\la}\psi_a=\frac{\partial}{\partial\la}\nabb_a\log\oom=\nabb_a\frac{\partial}{\partial\la}\log\oom
=\nabb_a\left(-2\oom\omb-\nabb_X\log\oom\right)\nn\\
&&\ML\ML=-2\oom\nabb_a\omb-2\oom\omb\psi_a-(\nabb_aX)^c\psi_c-X^c\nabb_c\psi_a\ .
\eea
\bea
&&\ML\ML\frac{\partial}{\partial\la}v_{cba}=\frac{\partial}{\partial\la}(\partial\ga)_{cba}=\partial_c\frac{\partial}{\partial\la}\ga_{ba}
=-\partial_c(\Lie_X\ga)_{ab}+\partial_c(2\oom\chib)_{ab}\nn\\
&&\ML\ML=-(\pr_cX^d)\pr_d\ga_{ab}-\pr_X\pr_c\ga_{ab}+\partial_c\left(\partial_aX^d\ga_{db}+\partial_bX^d\ga_{ad}\right)+2\oom\partial_c\chib_{ab}+2\oom(\nabb_c\log\oom)\chib_{ab}\nn\\
&&\ML\ML=-(\pr_cX^d)v_{dab}-\pr_Xv_{cab}+\partial_c\partial_aX_b+\partial_c\partial_bX_a+2\oom\partial_c\chib_{ab}+2\oom\psi_c\chib_{ab}\nn\\
&&\ML\ML=-(\pr_cX^d)v_{dab}-\pr_Xv_{cab}+\partial_cw_{ab}+\partial_cw_{ba}+2\oom\partial_c\chib_{ab}+2\oom\psi_c\chib_{ab}\nn
\eea
To obtain the equation for $w_{ab}$ we write:
\bea
&&\ML\frac{\partial w_{ab}}{\partial\nu}=\frac{\partial{\partial_a}{X_b}}{\partial\nu}
=\partial_a\frac{\partial\ga_{bc}X^c}{\partial\nu}=-4{\partial_a}({{\oom}}^2\ze_b)
+\left(\partial_a\!\frac{\partial\ga_{bc}}{\partial\nu}\right)X^c\!\nn\\
&&\ML=-4{\partial_a}({{\oom}}^2{\ze_b})+\partial_a\left(2\oom\chi_{bc}\right)X^c\nn\\
&&\ML=-8\oom^2(\nabb_a\log\oom)\ze_b-4\oom^2\partial_a\ze_b+2\oom(\nabb_a\log\oom)\chi_{bc}X^c
+2\oom(\partial_a\chi_{bc})X^c\nn\\
&&\ML=-8\oom^2\psi_a\ze_b-4\oom^2\partial_a\ze_b+2\oom\psi_a\chi_{bc}X^c
+2\oom(\partial_a\chi_{bc})X^c\ .\nn
\eea
We collect the three equations obtained,
\bea
&&\ML\frac{\partial}{\partial\la}\psi_a=-2\oom\partial_a\omb-2\oom\omb\psi_a-(\nabb_aX)^c\psi_c-X^c\nabb_c\psi_a\nn\\
&&\ML\frac{\partial}{\partial\la}v_{cba}=-(\pr_cX^d)v_{dab}-\pr_Xv_{cab}+\partial_cw_{ab}+\partial_cw_{ba}+2\oom\partial_c\chib_{ab}+2\oom\psi_c\chib_{ab}\nn\\
&&\ML\frac{\partial w_{ab}}{\partial\nu}=-8\oom^2\psi_a\ze_b-4\oom^2\partial_a\ze_b
+2\oom\psi_a\chi_{bc}X^c+2\oom(\partial_a\chi_{bc})X^c\ .\eql{3.18daa}
\eea

\subsection{The null components of the Riemann tensor on $S_0\times R\times R$ and their Bianchi equations}\label{SS5.2}
In our notation the ten independent components of the (conformal part of the) Riemann tensor are denoted $\a,\b,\ro,\si,\bb,\aa$,
respectively two covariant two-tensors, two covariant vectors and two scalar functions, tangential at each point to $S(\la,\nu)$, the two
dimensional surface containing it. All these components written in terms of the connection coefficients and their derivatives are, see
\cite{C-K:book}, \cite{Kl-Ni:book}:
\bea
&&\a=-\dddd_4\chih-\tr\chi\chih+(\dd_4\log\oom)\chih\nn\\
&&\b=\nabb\tr\chi-\divv\chi-\zeta\c\chi+\zeta\tr\chi\nn\\
&&\ro=-{\bf K}-\frac{1}{4}\tr\chi\tr\chib+\frac{1}{2}\chibh\c\chih\nn\\
&&\si=\curll\zeta-\frac{1}{2}\chibh\wedge\chih\eql{4.2p}\\
&&\bb=-\nabb\tr\chib+\divv\chib-\zeta\c\chib+\zeta\tr\chib\nn\\
&&\aa=-\dddd_3\chibh-\tr\chib\ \!\chibh+(\dd_3\log\oom)\chibh\nn
\eea
We will consider these quantities pulled back on the two dimensional surface $S_0$, as we did before for the connection coefficients. 
Therefore we define
\bea
&&\!\!\!\!\!\!\!\!\!\!\tilde{\a}(\la,\nu;p_0)=(\Psi^*(\la,\nu)\a)(p_0)=\!\underline{\Phi}^*_{\la}\Phi^*_{\nu}
\a(\Phi_{\nu}\circ\underline{\Phi}_{\la}(p_0))=\underline{\Phi}^*_{\la}\Phi^*_{\nu}\a(q)\nn\\
&&\!\!\!\!\!\!\!\!\!\!\tilde{\b}(\la,\nu;p_0)=(\Psi^*(\la,\nu)\b)(p_0)=
\underline{\Phi}^*_{\la}\Phi^*_{\nu}\b(\Phi_{\nu}\circ\underline{\Phi}_{\la}(p_0))=\underline{\Phi}^*_{\la}\Phi^*_{\nu}\b(q)\nn\\ 
&&\!\!\!\!\!\!\!\!\!\!\tilde{\bb}(\la,\nu;p_0)=(\Psi^*(\la,\nu)\bb)(p_0)
=\underline{\Phi}^*_{\la}\Phi^*_{\nu}\bb(\Phi_{\nu}\circ\underline{\Phi}_{\la}(p_0))=\underline{\Phi}^*_{\la}\Phi^*_{\nu}\bb(q)\nn\\
&&\!\!\!\!\!\!\!\!\!\!\tilde{\aa}(\la,\nu;p_0)=(\Psi^*(\la,\nu)\aa)(p_0)=\!\underline{\Phi}^*_{\la}\Phi^*_{\nu}
\aa(\Phi_{\nu}\circ\underline{\Phi}_{\la}(p_0))=\underline{\Phi}^*_{\la}\Phi^*_{\nu}\aa(q)\nn\\
&&\!\!\!\!\!\!\!\!\!\!\tilde{\ro}(\la,\nu;p_0)=(\ro\circ\Psi(\la,\nu))(p_0)=(\ro\circ\Phi_{\nu}\circ\underline{\Phi}_{\la})(p_0)=\ro(q)\nn\\
&&\!\!\!\!\!\!\!\!\!\!\tilde{\si}(\la,\nu;p_0)=(\si\circ\Psi(\la,\nu))(p_0)=(\si\circ\Phi_{\nu}\circ\underline{\Phi}_{\la})(p_0)=\si(q)
\eea
where $p_0\in S_0$ and is specified by its coordinates $(\theta,\phi)$.  In conclusion we have the following covariant tensors defined on $S_0$,
\beaa
\tilde{\a}(\la,\nu;\theta,\phi)\ ,\ \tilde{\aa}(\la,\nu;\theta,\phi)\ ,\ \tilde{\b}(\la,\nu;\theta,\phi)\ ,\ \tilde{\bb}(\la,\nu;\theta,\phi)\ ,\ 
\tilde{\ro}(\la,\nu;\theta,\phi)\ ,\ \tilde{\si}(\la,\nu;\theta,\phi)\ .\ \ \ \ 
\eeaa
\begin{Le}\label{L5.2a}
The  explicit expressions for the pullback of the various null Riemann components  are 
\bea
&&\ML\ML\tilde{\a}=-\frac{1}{\tilde{\oom}}\frac{\partial\hat{\tilde\chi}}{\partial\nu}+|\hat{\tilde\chi}|^2\ga-2{\tilde\om}\hat{\tilde\chi}\nn\\
&&\ML\ML\tilde{\b}=\frac{1}{2}\nabb\widetilde{\tr\chi}-\widetilde{\divv}\hat{\tilde\chi}-\tilde{\zeta}\!\c\hat{\tilde\chi}
+\frac{1}{2}\tilde{\zeta}\widetilde{\tr\chi}\nn\\
&&\ML\ML\tilde{\ro}=-\tilde{\bf K}-\frac{1}{4}\widetilde{\tr\chi}\widetilde{\tr\chib}+\frac{1}{2}\hat{\tilde{\chib}}\!\c\!\hat{\tilde{\chi}}\nn\\
&&\ML\ML\tilde{\si}=\widetilde{\curll}\tilde{\zeta}-\frac{1}{2}\hat{\tilde{\chib}}\wedge\hat{\tilde{\chi}}\eql{5.88c}\\
&&\ML\ML\tilde{\bb}=-\frac{1}{2}\tilde{\nabb}\widetilde{\tr\chib}+\widetilde{\divv}\hat{\tilde{\chib}}
-\zeta\!\c\!\hat{\tilde{\chib}}+\frac{1}{2}\tilde{\zeta}\widetilde{\tr\chib}\nn\\
&&\ML\ML\tilde{\aa}=-\frac{1}{\tilde{\oom}}\bigg(\frac{\partial\hat{\tilde{\chib}}}{\partial\la}+{\Lie_{\tilde X}}\hat{\tilde{\chib}}\bigg)
+|\hat{\tilde{\chib}}|^2\ga-2\tilde{\omb}\hat{\tilde{\chib}}\nn
\eea
\end{Le}
\NI{\bf Proof:} It follows immediately from the explicit expressions of the various Riemann null components, \ref{4.2p}, and the pullback of the
various connection coefficients and their derivatives.53

\medskip

\NI The Bianchi equations take the form of a system of transport equations with respect to the variables $\la$ and $\nu$ for these covariant
tensors. Defining on
$S_0$ an orthonormal basis $\{\tilde{e}_{\theta},\tilde{e}_{\phi}\}$, each of them can be written in the following way:
\bea
&&\tilde{\a}(\c)=\sum_A\tilde{\a}_{AB}(\la,\nu;\theta,\phi)\tilde{\theta}_A\!\otimes\!\tilde{\theta}_B(\c,\c)
\eea
\beaa\ML\ML\ML\ML\mbox{where}\ \ \ \ \ \ \ \ \ \ \ \ \ \ \ \ \ \ \ \ \ \ \ \ 
\tilde{\a}_{AB}(\la,\nu;\theta,\phi)\equiv\tilde{\a}(\la,\nu;\theta,\phi)(\tilde{e}_A,\tilde{e}_B)\ .
\eeaa
We start writing the derivative with respect to $\nu$ of $\tilde{\aa}$ obtaining:
\bea
\frac{\partial\tilde{\aa}}{\partial\nu}(\la,\nu;p_0)=\underline{\Phi}^*_{\la}\Phi^*_{\nu}(\Lie_N\aa)(q)
\eea
The proof goes over exactly as the one for $\frac{\partial\tilde{\ze}}{\partial\nu}(\la,\nu;p_0)$ and we do not repeat it.
A simple computation gives
\bea
&&\ML(\Lie_N\aa)(\c)=\oom\dddd_4\aa(\c)+\oom(\aa\!\cdot\!\chi+\chi\!\cdot\!\aa)(\c)=\oom\left[\dddd_4\aa+\tr\chi\aa+(\aa\!\cdot\!\chih)\ga\right](\c)\nn\\
&&\ML=\oom\bigg[\frac{\tr\chi}{2}\aa+(\aa\cdot\chih)\ga-\nabb\hot\bb
+\left(4\om\aa-3(\hat{\chib}\ro-\dual\hat{\chib}\si)+(\ze-4\etab)\hot\bb\right)\bigg]\ \ \ \ \ \ \ \ \                      
\eea
which implies
\bea
\frac{\partial\tilde{\aa}}{\partial\nu}(\la,\nu;\theta,\phi)(\c)\!\!&=&\!\!\bigg(\underline{\Phi}^*_{\la}\Phi^*_{\nu}\oom
\bigg[\frac{\tr\chi}{2}\aa+(\aa\cdot\chih)\ga-\nabb\hot\bb+\left(4\om\aa-3(\hat{\chib}\ro-\dual\hat{\chib}\si)\right.\nn\\
&+&\!\!\left.(\ze-4\etab)\hot\bb\right)\bigg]\!\bigg)\!(\la,\nu;\theta,\phi)(\c)\ .\ \ \ \ \ \eql{5.74}
\eea
Applying, as before, the pull-back on the right hand side, the final expression of the Bianchi equations, for this component, is
\bea
\frac{\partial\tilde{\aa}}{\partial\nu}(\la,\nu;\theta,\phi)(\c)\!\!&=&\!\!\!\tilde{\oom}
\bigg[\frac{\widetilde{\tr\chi}}{2}\tilde{\aa}+(\tilde{\aa}\cdot\hat{\tilde{\chi}})\tilde{\ga}-\nabb\hot\tilde{\bb}+\left(4\tilde{\om}\tilde{\aa}
-3(\hat{\tilde{\chib}}\tilde{\ro}-\dual\hat{\tilde{\chib}}\tilde{\si})\right.\nn\\
&+&\!\!\!\!\!\left.(\tilde{\ze}-4\tilde{\etab})\hot\tilde{\bb}\right)\bigg]\!(\la,\nu;\theta,\phi)(\c)\ .\ \ \ \ \ \eql{5.75}
\eea
It is easy now to project the remaining Bianchi equations on $S_0$. 
\bea
\frac{\partial\tilde{\bb}}{\partial{\nu}}\!\!\!&=&\!\!\!\underline{\Phi}^*_{\la}\Phi^*_{\nu}\!\big(\oom[\dddd_4\bb+\chi\c\bb]\big)
=\underline{\Phi}^*_{\la}\Phi^*_{\nu}\!\left[\oom\!\left(-\frac{1}{2}\tr\chi\bb\!+\!\chih\!\c\!\bb\!-\!\nabb\ro\!+\!\left[2\om\bb+2\hat{\chib}\!\c\!\b\right.\right.\right.\nn\\
&&\ \ \ \ \ \ \ \ \ \ \ \ \ \ \ \ \ \ \ \ \ \ \ \ \ \ \ \ \ \ \
+\left.\left.\left.\!\!\!\dual\nabb\si\!-\!3(\etab\ro-\dual\etab\si)\right]\right)\right]\nn\\
&=&\!\!\tilde{\oom}\!\left(-\frac{1}{2}\widetilde{\tr\chi}\tilde{\bb}+\hat{\tilde{\chi}}\!\c\!\tilde{\bb}\!-\!\tilde{\nabb}\tilde{\ro}
+\left[2\tilde{\om}\tilde{\bb}+2\hat{\tilde{\chib}}\!\c\!\tilde{\b}+\dual\tilde{\nabb}\tilde{\si}
-3(\tilde{\etab}\tilde{\ro}-\dual\tilde{\etab}\tilde{\si})\right]\right)\nn\\
\frac{\partial\tilde{\ro}}{\partial{\nu}}\!\!&=&\!\!\underline{\Phi}^*_{\la}\Phi^*_{\nu}(\oom\dd_4\ro)
=\underline{\Phi}^*_{\la}\Phi^*_{\nu}\left[\oom\!\left(-\frac{3}{2}\tr\chi\ro+\divv\b
-\left[\frac{1}{2}\hat{\chib}\!\c\!\a+\ze\!\c\!\b-2\nabb\log\oom\!\c\!\b\right]\right)\right]\nn\eql{nullBian}\\
\!\!&=&\!\!\tilde{\oom}\!\left(-\frac{3}{2}\widetilde{\tr\chi}\tilde{\ro}+\widetilde{\divv}\tilde{\b}
-\left[\frac{1}{2}\hat{\tilde{\chib}}\!\c\!\tilde{\a}+\tilde{\ze}\!\c\!\tilde{\b}-2\nabb\log\oom\!\c\!\tilde{\b}\right]\right)\nn\\
\frac{\partial\tilde{\si}}{\partial{\nu}}\!\!&=&\!\!\underline{\Phi}^*_{\la}\Phi^*_{\nu}(\oom\dd_4\si)
=\underline{\Phi}^*_{\la}\Phi^*_{\nu}\left[\oom\!\left(-\frac{3}{2}\tr\chi\si+\divv\dual\b
-\left[\frac{1}{2}\hat{\chib}\!\c\!\dual\a+\ze\!\c\!\dual\b-2\nabb\log\oom\!\c\!\dual\b\right]\right)\right]\nn\\
\!\!&=&\!\!\tilde{\oom}\!\left(-\frac{3}{2}\widetilde{\tr\chi}\tilde{\si}+\widetilde{\divv}\dual\tilde{\b}
-\left[\frac{1}{2}\hat{\tilde{\chib}}\!\c\!\dual\tilde{\a}+\tilde{\ze}\!\c\!\dual\tilde{\b}-2\nabb\log\oom\!\c\!\dual\tilde{\b}\right]\right)\nn\\
\frac{\partial\tilde{\b}}{\partial{\nu}}\!\!&=&\!\!\underline{\Phi}^*_{\la}\Phi^*_{\nu}\big(\oom[\dddd_4\b+\chi\!\c\!\b]\big)
=\underline{\Phi}^*_{\la}\Phi^*_{\nu}\!\left[\!\oom\!\left(\!-\frac{3}{2}\tr\chi\!+\!\divv\a\!-\!2\om\b
\!+\!\chih\!\c\!\b\!+\!(\ze\!+\!\nabb\log\oom)\!\c\!\a\!\right)\right]\nn\\
\!\!&=&\!\!\tilde{\oom}\!\left(-\frac{3}{2}\widetilde{\tr\chi}\tilde{\b}\!+\!\widetilde{\divv}\tilde{\a}\!-\!2\om\b
\!+\!\left[\chih\!\c\!\b\!+\!(\ze\!+\!\nabb\log\oom)\!\c\!\a\right]\!\right)\ \ .
\eea
The Bianchi equations which are transport equations along the incoming cones can be obtained exactly in the same way as done for the connection
coefficients. In this case we have to use the relation, see \ref{5.34}, for a generic Riemann null component $w$,
\bea
&&\!\!\!\!\!\!\!\!\!\frac{\partial\tilde{w}}{\partial\la}(\la,\nu;\theta,\phi)(\c)
=\underline{\Phi}^*_{\la}\Phi^*_{\nu}\left[(\Lie_{\Nb}w)(q)-(\Lie_Xw)(q)\right]\ .
\eea 
\bea
\frac{\partial\tilde{\bb}}{\partial{\la}}\!\!&=&\!\!\tilde{\oom}\!\left(-\frac{3}{2}\widetilde{\tr\chib}\ \!\tilde{\bb}+\hat{\tilde{\chib}}\!\c\!\tilde{\bb}
-\tilde{\divv}\tilde{\aa}-\left[2\tilde{\omb}\tilde{\bb}+(-2\tilde{\ze}+\tilde{\eta})\!\c\!\tilde{\aa}\right]\right)-\Lie_{\tilde{X}}\tilde{\bb}\nn\\
\frac{\partial\tilde{\ro}}{\partial{\la}}\!\!&=&\!\!
\tilde{\oom}\!\left(-\frac{3}{2}\widetilde{\tr\chib}\tilde{\ro}-\widetilde{\divv}\tilde{\bb}
-\left[\frac{1}{2}\hat{\tilde{\chi}}\!\c\!\tilde{\aa}-\tilde{\ze}\!\c\!\tilde{\bb}+2\tilde{\eta}\!\c\!\tilde{\bb}\right]\right)
-\tilde{\nabb}_{\tilde X}\tilde{\ro}\nn\\
\frac{\partial\tilde{\si}}{\partial{\la}}\!\!&=&\!\!
\tilde{\oom}\!\left(-\frac{3}{2}\widetilde{\tr\chib}\tilde{\si}-\widetilde{\divv}\dual\tilde{\bb}
+\left[\frac{1}{2}\hat{\tilde{\chi}}\!\c\!\dual\tilde{\aa}-\tilde{\ze}\!\c\!\dual\tilde{\bb}-2\tilde{\eta}\!\c\!\dual\tilde{\bb}\right]\right)
-{\tilde{\nabb}}_{\tilde X}\tilde{\si}\eql{5.78}\\
\frac{\partial\tilde{\b}}{\partial{\la}}
\!\!&=&\!\!\tilde{\oom}\!\left(-\frac{1}{2}\widetilde{\tr\chib}\tilde{\b}+\hat{\tilde{\chib}}\!\c\!\tilde{\b}+\tilde{\nabb}\tilde{\ro}
+\left[2\tilde{\omb}\tilde{\b}+2\chih\!\c\!\bb+\dual\tilde{\nabb}\tilde{\si}
+3(\tilde{\eta}\tilde{\ro}+\dual\tilde{\eta}\tilde{\si})\right]\right)-\Lie_{\tilde{X}}\tilde{\b}\nn\\
\frac{\partial\tilde{\a}}{\partial\la}\!\!&=&\!\!\!\tilde{\oom}
\bigg[\frac{\widetilde{\tr\chib}}{2}\tilde{\a}+(\tilde{\a}\!\c\!\hat{\tilde{\chib}})\ga+\nabb\hot\tilde{\b}+\left(4\tilde{\omb}\tilde{\a}
-3(\hat{\tilde{\chi}}\tilde{\ro}+\dual\hat{\tilde{\chi}}\tilde{\si})+(\tilde{\ze}+4\tilde{\eta})\hot\tilde{\b}\right)\bigg]-\Lie_{\tilde{X}}\tilde{\a}\ .\nn
\eea

\end{document}